\documentclass[10pt,final,doublecolumn]{IEEEtran}
\usepackage{amsopn,amsmath,amssymb,amsfonts}
\usepackage{cite}
\usepackage{subfigure}
\usepackage{citesort}
\usepackage{balance}
\usepackage{float}
\usepackage{graphics}
\usepackage{array}
\usepackage{epsfig}
\usepackage{epstopdf}
\usepackage{multirow}

\begin{document}

\title{Towards Massive Machine Type Communications in Ultra-Dense Cellular IoT Networks: Current Issues and Machine Learning-Assisted Solutions}
\author{
\authorblockN{Shree Krishna Sharma,  \emph{Senior Member, IEEE}, and Xianbin Wang, \emph{Fellow, IEEE}}
\thanks{This work was supported in part by NSERC Discovery and CREATE programs under project number RGPIN-2018-06254 and 432280-2013.
(\emph{Corresponding author: Xianbin Wang}).}
\thanks{The authors are with the Department of Electrical and Computer Engineering, Western University, London, ON, N6A 3K7, Canada,
Email: sshar323@uwo.ca, xianbin.wang@uwo.ca.}}

\markboth{IEEE Communications Surveys \& Tutorials (Draft)}
{}
\maketitle

\begin{abstract}
The ever-increasing number of resource-constrained Machine-Type Communication (MTC) devices is leading to the critical challenge of fulfilling diverse communication requirements in dynamic and ultra-dense wireless environments. Among different application scenarios that the upcoming 5G and beyond cellular networks are expected to support, such as enhanced Mobile Broadband (eMBB), massive Machine Type Communications (mMTC) and Ultra-Reliable and Low Latency Communications (URLLC), the mMTC brings the unique technical challenge of supporting a huge number of MTC devices in cellular networks, which is the main focus of this paper. The related challenges include Quality of Service (QoS) provisioning, handling highly dynamic and sporadic MTC traffic, huge signalling overhead and Radio Access Network (RAN) congestion. In this regard, this paper aims to identify and analyze the involved technical issues, to review recent advances, to highlight potential solutions and to propose new research directions. First, starting with an overview of mMTC features and QoS provisioning issues, we present the key enablers for mMTC in cellular networks. Along with the highlights on the inefficiency of the legacy Random Access (RA) procedure in the mMTC scenario, we then present the key features and channel access mechanisms in the emerging cellular IoT standards, namely, LTE-M and Narrowband IoT (NB-IoT). Subsequently, we present a framework for the performance analysis of transmission scheduling with the QoS support along with the issues involved in short data packet transmission. Next, we provide a detailed overview of the existing and emerging solutions towards addressing RAN congestion problem, and then identify potential advantages, challenges and use cases for the applications of emerging Machine Learning (ML) techniques in ultra-dense cellular networks. Out of several ML techniques, we focus on the application of low-complexity Q-learning approach in the mMTC scenario along with the recent advances towards enhancing its learning performance and convergence. Finally, we discuss some open research challenges and promising future research directions.
\end{abstract}



\vspace{-10 pt}

\section{Introduction}
\label{sec:_sec1}
The convergence of emerging wireless communication technologies, ubiquitous wireless infrastructure and vertical Internet of Things (IoT) applications such as industrial automation, connected cars and smart-grid is leading to an integrated enabling platform for future smart and connected societies. This platform envisions to
synergistically integrate the ever-increasing number of smart devices (forecasted by IHS Markit to be around $125$ billion by $2030$),
intelligent industry processes, people and societies together to enhance the overall quality of our daily life
\cite{Fuqaha2015IoT}. Towards supporting connected IoT devices, there are several recent developments in the area of licensed cellular technologies such as Long Term Evolution (LTE) for Machine-Type Communications (LTE-M) and Narrow-Band IoT (NB-IoT), and unlicensed technologies such as WiFi, ZigBee and LoRa \cite{Wang2017survey}. Out of these, cellular technologies are considered to be promising due to their several advantages including Quality of Service (QoS) provisioning, wide coverage area and tight coordination, and therefore, cellular IoT is of the main focus in this paper.
\subsection{Recent Developments in Cellular IoT}
\label{sec:_sec11}
In recent years, cellular IoT has gained significant importance from academia, industries, regulators and standardization bodies to enable the incorporation of IoT devices in the existing cellular infrastructures. The ITU-R has categorized the emerging diversified telecommunication services in the upcoming 5G and beyond cellular networks into the following three classes \cite{ITURvision2015}: (i) enhanced Mobile Broadband (eMBB),(ii) massive Machine Type Communications (mMTC), and (iii) Ultra-Reliable and Low Latency Communications (URLLC).

Out of the above-mentioned categories, the eMBB comprises of high data rate services of 5G systems while the mMTC deals with the scalable connectivity to a massive number of devices in the order of $10^6$ devices per square kilometers with diverse QoS requirements \cite{JayawickramaspectrumIoT}. On the other hand, URLLC aims to provide robust connectivity with very low latency. The main challenge in the mMTC case is to support a huge number of devices with the limited radio resources whereas the key challenge for the URLLC scenario is to provide extremely high reliability in the order of $99.999 \%$ within a very short duration in the order of $1$ ms \cite{3GPPtechnical2017}. Among these usage scenarios, the mMTC has to deal with various non-conventional challenges including QoS provisioning, Random Access Network (RAN) congestion, highly dynamic and sporadic traffic, and large signalling overhead. To this end, this paper focuses on the involved issues and the potential enablers of the mMTC scenario with a particular emphasis on the RAN congestion problem and emerging Machine Learning (ML)-based solutions.

In terms of ongoing standardization efforts, the main cellular IoT standards introduced by the 3GPP are LTE-M and NB-IoT. Out of these, LTE-M is intended for mid-range IoT applications which can support voice and video services, while NB-IoT systems target to provide very large coverage and support for ultra-low cost devices \cite{Ghavimi2015M2M}. Since MTC devices usually do not require high channel throughput, the existing LTE-M and NB-IoT standards allocate a small bandwidth for IoT devices, i.e., LTE-M assigns $1.4$ MHz bandwidth while the NB-IoT allocates a significantly lower bandwidth of $180$ kHz \cite{Koseoglu2017}. Despite these recent developments, there are still several challenges to be addressed while supporting MTC devices in cellular systems.

\subsection{Challenges in Cellular IoT}
\label{sec:_sec12}
Although centralized cellular systems provide several advantages in terms of providing large coverage, tight time synchronization and handover operations for mobile users, they are sluggish in terms of handling low-end devices and face several challenges in supporting a large number of MTC devices with diverse QoS requirements. While incorporating MTC devices in the existing LTE/LTE-A based cellular systems, cellular operators have to face a lot of challenges both at the operational and planning levels. More specifically, there arise various issues related to the MTC device deployment, mMTC traffic, energy efficiency of low-cost MTC devices and the network protocol aspects such as signalling overhead \cite{DawymMTC2017}. Furthermore, network congestion may occur in different segments of LTE/LTE-A based cellular network including RAN, core network and signalling network \cite{3GPPMTCservice}. Out of these, RAN congestion problem is crucial in ultra-dense cellular IoT networks due to the limited available radio resources at the access-side and the massive number of sporadic access attempts from heterogeneous MTC devices.

Existing contention-based protocols are effective to support the conventional Human-Type Communications (HTC), however,
their performance significantly degrades in mMTC scenarios due to infrequent and massive number of access requests \cite{Belloeuropean2014}. Also, due to limited available preambles in the existing LTE-based systems, several MTC devices may need to select the same preambles at the same time, resulting in a significantly high probability of collision in the access network. Furthermore, the number of transmission attempts from the massive number of heterogeneous IoT devices could be significantly large \cite{Durisi2016IEEEproc},
and their activation periods and frame sizes could be very different \cite{SingminMDPI}. This sporadic and dynamic nature of mMTC access attempts and data transmissions may result in the peak traffic in both the access and traffic channels well beyond the capacity of the IoT access network, thus leading to the inevitable congestion in an IoT access network \cite{Sharmacommletter2018}. Moreover, although data packets transmitted by IoT devices are relatively short, very high signalling overhead per data packet becomes another critical issue \cite{Bockelmann2016,DawymMTC2017}. To this end, it is significantly important to investigate suitable transmission scheduling and efficient signalling reduction techniques in ultra-dense scenarios by utilizing emerging tools such as ML.

\subsection{Need of Machine Learning and Associated Challenges in IoT/mMTC Networks}
\label{sec:_sec13}
Optimizing the operation of cellular networks in dynamic wireless environments has been challenging over the generations since the number of configurable parameters of a cellular network has been rapidly increasing from one cellular generation to the next one \cite{Imran2014challenges,Liintelligent2017}. The widely-used link adaptation techniques in the existing wireless systems, which adapt different physical layer parameters including transmission power and modulation and coding scheme based on the reliability/link of a communication link, may not be efficient in ultra-dense cellular IoT networks. This adaptation is based on the prediction of reliability of a wireless link in the form of some metrics such as  Packet Error Rate (PER) and this prediction process becomes extremely complex due to the increasing trend of using multiple antennas, wideband signals and a number of advanced signal processing algorithms \cite{Daniels2009}. Furthermore, the prediction of PER with good accuracy becomes difficult in practice by using the conventional signal processing tools. Moreover, due to a significantly large number of environmental parameters such as channel state information, signal power, noise variance, non-Gaussian noise effect and transceiver hardware impairments, it becomes complicated to provide the near-optimal/optimal tuning of the transmission parameters to achieve the efficient link adaptation \cite{Yun2010reinforcement}. The severity of this problem greatly increases in ultra-dense networks due to the involvement of massive number of devices and system parameters.

\begin{table*}
\caption{\small{Classification of survey/overview works in the areas of IoT/mMTC, ML and UDNs.}}
	\centering
\renewcommand{\arraystretch}{0.95}
\begin{tabular}{|l|l|l|}
\hline
Main domain & Sub-topics & References\\ \hline
  & Enabling technologies/protocols, challenges and applications & \cite{Fuqaha2015IoT,Wang2017survey,Ghavimi2015M2M,Durisi2016IEEEproc,Bockelmann2016,Wang2016cellular,Dawy2017towards,Alvarino2016overview,Elsaadany2017cellularLTEA,Hoglund2017overview} \\
& Random access schemes  &  \cite{Hasanrandom2013,Layarandomaccess2014,Yangnarrwoband2017,Ali2017LTE} \\
IoT/mMTC  & Traffic characterization and issues   & \cite{Soltanmohammadi2016} \\
& Transmission scheduling     & \cite{Gotsis2012M2M,Mehaseb2016classification} \\
& IoT big data analytics   & \cite{Marjani2017bigdata,Verma2017survey,SKSIEEE2017} \\
& QoS provisioning & \cite{Elhammouti2017self} \\
\hline
  & Intelligence in 5G networks & \cite{Liintelligent2017,Wang2015IEEE} \\
Machine Learning (ML) & Learning in IoT/sensor networks & \cite{ParkIEEE2016,Mohammadienabling,Sezer2018context} \\
& Reinforcement learning & \cite{Busoniu2008Trans,deepRL2017} \\
\hline
Ultra-Dense Networks (UDNs)  &  & \cite{Baldemair2015ultra,ChenM2Msurvey,Kamel2016ultra} \\
\hline
\end{tabular}
	\vspace{-15 pt}
\label{tab: referencesclassification}
\end{table*}

Understanding the context of the surrounding wireless environment significantly facilitates in developing context-aware adaptive communication protocols and in taking optimized decisions. Nevertheless, handling self-configuration, self-optimization and self-healing operations in the ultra-dense cellular networks becomes challenging since the networks need to observe dynamic environmental variations, learn uncertainties, plan response actions and configure the associated network parameters effectively. To this end, emerging ML-assisted techniques seem promising since they can play significant roles in learning the system variations/parameter uncertainties, classifying the involved cases/issues, predicting the future results/challenges and investigating potential solutions/actions \cite{Liintelligent2017}. Moreover, the conventional link adaptation techniques are more localized to a particular network and a geographical region, and do not usually consider their impacts on the other systems. However, future ultra-dense cellular networks will need to handle mutual impact among the involved entities to maximize the overall system performance. To this end, by utilizing the emerging collaborative edge-cloud processing platform \cite{SKSIEEE2017}, ML-assisted solutions can enable the utilization of global network knowledge at the edge-side, and also facilitate the coordination among different distributed systems. In this direction, the application of ML techniques to address various issues in dynamic wireless environments has recently received an important attention \cite{Liintelligent2017,Wang2015IEEE} and in the context of MTC environments, some existing works have already studied the applications of different ML techniques in learning various system parameters \cite{Belloeuropean2014,Moon2017access,Hasanrandom2013,Mohammed2015base,Park2016resource,Portelli2017leveraging}.

However, the direct application of conventional ML techniques to complex and dynamic wireless IoT environments
is not straight-forward due to several underlying constraints such as low computational capability of MTC devices,
distributed nature and heterogeneous QoS requirements of IoT devices, and the distinct features of mMTC traffic as
compared to the conventional HTC traffic \cite{ParkIEEE2016}. Furthermore, due to the
limited computed power and low memory size of IoT devices, implementing sophisticated learning techniques in IoT devices
becomes challenging \cite{Tang2017computer}. In this regard, this paper identifies the implementation issues of the ML
techniques in ultra-dense IoT scenarios and provides an emphasis on computationally simpler Q-learning based solutions.

\subsection{Review of Related Overview/Survey Articles}
\label{sec:_sec14}
In this subsection, we provide a brief overview of the existing survey works in the main domains related to this paper, namely, IoT/mMTC, ML and Ultra-Dense Networks (UDNs). Also, we present the classification of the existing references related to these domains into different sub-topics which are listed in Table \ref{tab: referencesclassification}.

Several existing papers have provided the review of enabling technologies, protocols, challenges and applications of IoT/mMTC in different contexts \cite{Fuqaha2015IoT,Wang2017survey,Ghavimi2015M2M,Durisi2016IEEEproc,Bockelmann2016,Wang2016cellular,Dawy2017towards,Alvarino2016overview,Elsaadany2017cellularLTEA,Hoglund2017overview}.
Authors in \cite{Fuqaha2015IoT} provided a comprehensive overview of existing IoT protocols including application protocols, service discovery protocols, infrastructure protocols, and discussed some enabling technologies including cloud computing, edge computing and big data analytics for IoT systems. Furthermore, the authors in \cite{Ghavimi2015M2M} presented a detailed survey of MTC systems including its features, requirements and  the required architectural enhancements in LTE/LTE-A based networks. In the context of short packet transmissions in mMTC/IoT environment, the contribution in \cite{Durisi2016IEEEproc} provided a review of the recent advances in information theoretic principles governing the transmissions of short data packets and discussed the applications of these principles to different scenarios including a two-way channel, a downlink broadcast channel and an uplink Random Access Channel (RACH). Besides, the article \cite{Bockelmann2016} highlighted the requirements and design challenges for mMTC systems and discussed various physical and Medium Access Control (MAC) layer solutions for energy-efficient and massive access. Another overview paper \cite{Wang2016cellular} discussed the physical limitations of MTC devices while operating in cellular networks, and then analyzed the impact of these device limitations on the link performance and the link budget design.

In addition, the article \cite{Alvarino2016overview} presented a review on various features defined by the 3GPP to support Machine-to-Machine (M2M) communications in LTE-based cellular systems and discussed recent advances in different layers including the physical layer improvements under the enhanced MTC (eMTC), and MAC and higher layer enhancements brought by the extended Discontinuous Reception (eDRX). Furthermore, the survey article \cite{Wang2017survey} provided a comprehensive survey of three main low power and long range M2M solutions, namely, Low Power Wide Area Network (LPWAN), IEEE 802.11ah-based network and cellular M2M including LTE-M and NB-IoT. Besides, the overview article \cite{Dawy2017towards} presented the new requirements and challenges in large-scale MTC applications, and discussed some enabling techniques including efficient overhead signalling protocols, data aggregation and in-device intelligent processing. Also, the survey article \cite{Elsaadany2017cellularLTEA} provided a comprehensive tutorial on the development of MTC design over different releases of LTE and recent user equipments belonging to the MTC and the NB-IoT categories, called CAT-M and CAT-N, respectively. Moreover, another overview article \cite{Hoglund2017overview} provided a review of various features of NB-IoT introduced in LTE Release $14$ including the increased positioning accuracy, multi-casting, enhanced non-anchor carrier operation and lower device power class, and the applicability of these features for NB-IoT systems.

The design of effective Random Access (RA) schemes in the mMTC environment is an important challenge due to massive access requests and sporadic device transmissions from a huge number of resource-constrained MTC devices. In this regard, authors in \cite{Hasanrandom2013} provided an overview of different RA overload control mechanisms to avoid the RAN congestion caused by the random channel access from the MTC devices. Furthermore, the article \cite{Layarandomaccess2014} presented a comprehensive survey of various RA solutions attempting to enhance the RACH operation of LTE/LTE-A based cellular networks, and carried out the performance evaluation of LTE RACH from the energy efficiency perspective. Moreover, the authors in \cite{Yangnarrwoband2017} provided a review of emerging LPWAN technologies both in the unlicensed band (LoRa and SIGFOX) and in the licensed band (LTE-M and NB-IoT) while considering three common fundamental objectives of these access mechanisms, namely, high system capacity, wide coverage and long battery life. In addition, the article \cite{Ali2017LTE} provided an overview of the existing RA solutions towards supporting MTC devices in LTE/LTE-A based networks and these solutions are compared in terms of five key metrics, namely, access success rate, access delay, QoS guarantee, energy efficiency and the impact on the HTC.

The nature of MTC traffic is significantly different from the HTC traffic as detailed later in Section \ref{sec:_sec24}, however, existing cellular networks are mainly optimized to support the HTC traffic. Therefore, it is important to understand and characterize MTC traffic to facilitate the incorporation of MTC devices in cellular networks. To this end, the authors in \cite{Soltanmohammadi2016} provided a discussion on the traffic issues of MTC and the associated congestion problems on the access channels, traffic channels and core network, and presented a comprehensive review of the existing solutions towards addressing these problems along with their advantages and disadvantages.

One of the promising solutions to handle massive access requests in the mMTC scenarios is to employ suitable transmission scheduling techniques at the distributed MTC devices. In this context, the authors in \cite{Gotsis2012M2M} identified limitations for signalling and scheduling of M2M devices over the existing LTE-based cellular infrastructures and discussed some of the existing proposals. Furthermore, the article \cite{Mehaseb2016classification} provided a detailed survey on the uplink scheduling techniques for M2M devices over LTE/LTE-A based cellular networks by considering various aspects of M2M communications such as scalability, energy efficiency, QoS support and multi-hop connectivity.

Another important aspect in ultra-dense IoT networks is how to handle the massive amount of data generated from the resource-constrained sensors and MTC devices. In this regard, authors in \cite{Marjani2017bigdata,Verma2017survey} discussed the connection between IoT and big data analytics, and provided a survey on the existing research attempts in the domain of big IoT data analytics. Also, authors in \cite{Verma2017survey} provided an overview of the existing network methodologies suitable for real-time IoT data analytics along with the fundamentals of real-time IoT analytics, software platforms and use cases, and highlighted real-time IoT analytics issues related to network scalability, network fault tolerance, spectral efficiency and network delay. Moreover, the article \cite{SKSIEEE2017} presented basic features, challenges and enablers for big data analytics in wireless IoT networks, and discussed the importance of collaborative cloud-edge processing for live data analytics along with the associated challenges and potential enablers.

Besides, providing QoS support in ultra-dense IoT networks is challenging due to massive connectivity, heterogeneity and  the resource constraints of the MTC devices as detailed later in Section \ref{sec:_sec2}. While analyzing from the energy efficiency perspective, maximizing QoS usually becomes energy costly and higher energy efficiency can be achieved by considering satisfactory QoS levels. Motivated by this, authors in \cite{Elhammouti2017self} provided a discussion on the need for QoS satisfaction and the methods to achieve QoS satisfaction efficiently. Also, game theory-based fully distributed algorithms were presented to enhance the energy efficiency of IoT systems while maintaining a desired QoS threshold.

In the direction of incorporating intelligence in 5G and beyond networks, there have been some recent attempts in applying Artificial Intelligence (AI)-based techniques to address various issues in wireless communications. The AI techniques can provide significant benefits in achieving efficient management, organization and optimization of various system resources in emerging ultra-dense 5G Heterogeneous Networks (HetNets). In this regard, the article \cite{Wang2015IEEE} discussed the state-of-the-art AI-based techniques for intelligent HetNet systems by considering the objective of achieving self-configuration, self-optimization and self-healing. Furthermore, authors in \cite{Liintelligent2017} recently introduced the fundamental concepts of AI and its relation with 5G candidate technologies. Also, the challenges and opportunities for the application of AI in managing network resources in intelligent 5G networks were discussed.

In the IoT/mMTC environment, learning techniques need to consider the unique features such as heterogeneity, resource constraints and QoS requirements. In this direction, the article \cite{ParkIEEE2016} discussed the applicability of different types of learning techniques in the IoT scenarios by taking their learning performance, computational complexity and required input information into account. Furthermore, authors in \cite{Mohammadienabling} discussed various aspects of deep Reinforcement Learning (RL) and its application in building cognitive smart cities while considering the use cases in the areas of water consumption, energy and agriculture. Moreover, the recent article \cite{Sezer2018context} provided an overview of existing context-aware computing studies along with the learning and big data related works in the direction of intelligent IoT systems. In the context of RL techniques, a comprehensive survey of multi-agent RL is provided in \cite{Busoniu2008Trans} by considering the aspects of stability of the learning dynamics of the agents and adaptation to the varying behavior of other learning agents. Also, the recent article \cite{deepRL2017} provided a brief survey of the existing deep RL algorithms along with the highlights on current research areas and the associated challenges.

Additionally, there exist a few survey and overview papers in the area of UDNs \cite{Baldemair2015ultra,ChenM2Msurvey,Kamel2016ultra}. The authors in \cite{Baldemair2015ultra} provided an overview of the operation of UDNs in the millimeter-wave band and presented wireless self-backhauling across multiple hops to improve the deployment flexibility. Furthermore, the article \cite{ChenM2Msurvey} highlighted the key issues in incorporating M2M communications in the emerging UDNs and also identified different ways to support M2M communications the UDNs from the perspective of different protocol layers including physical, MAC, network and application. Moreover, authors in \cite{Kamel2016ultra} provided a comprehensive review on the recent advances and enabling technologies for UDNs along with a discussion on the widely-used performance metrics and modeling techniques.

\begin{figure*}
	\begin{center}
		\includegraphics[width=4.5 in]{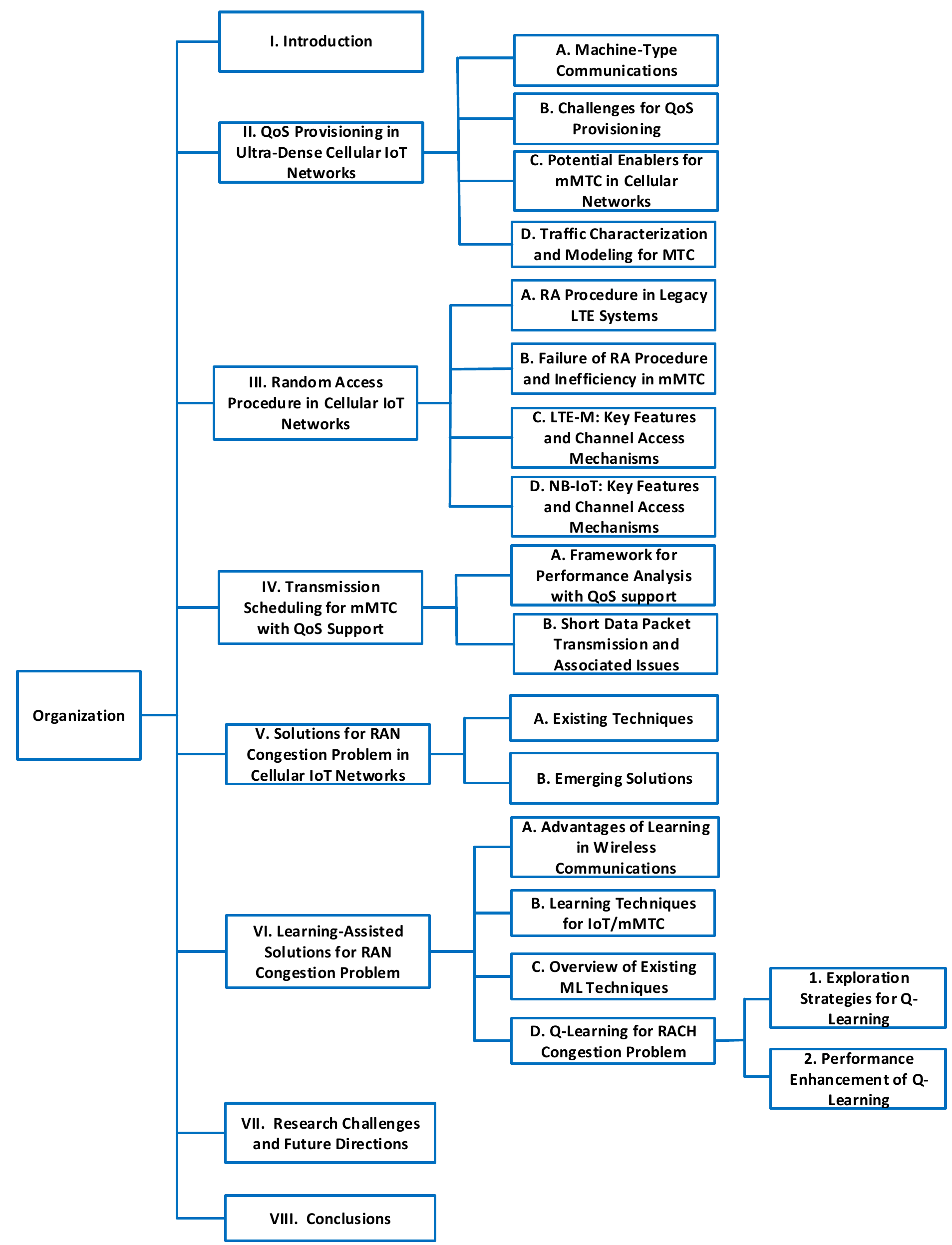}
		\caption{\small{Structure of the Paper}}
		\label{fig: paperstructure}
	\end{center}
	\vspace{-20 pt}
\end{figure*}
\subsection{Contributions}
\label{sec:_sec15}
Although several existing survey/overview articles reviewed in Section \ref{sec:_sec14} have considered different aspects of mMTC systems, ML techniques and UDNs, a comprehensive analysis of the research issues involved in supporting the massive number of MTC devices in ultra-dense cellular IoT networks and a detailed review of the recent advances including ML-assisted solutions attempting to address these challenges are missing in the literature. As highlighted earlier in Section \ref{sec:_sec12}, there arise several challenges while incorporating MTC devices in the existing LTE/LTE-A based cellular systems. The main issues include QoS provisioning to heterogeneous MTC devices, addressing random and dynamic MTC traffic, transmission scheduling with QoS support and RAN congestion. To this end, the overall aim of this paper is to analyze these different issues, to review the existing works attempting to overcome these challenges, to highlight potential enablers, and to propose ML-assisted solutions to address various challenges in ultra-dense cellular IoT networks. In the following, we highlight the main contributions of this survey paper.
\begin{enumerate}
\item The major challenges faced by the existing cellular IoT networks in supporting the massive number of MTC devices are identified and the potential enabling technologies are highlighted along with the key features, traffic characterization and the application scenarios of the mMTC. 
\item The inefficiency of the legacy LTE RA procedure in supporting MTC devices is pointed out and its adaptation for mMTC systems is presented along with the main features and channel access mechanisms of emerging cellular IoT standards (LTE-M and NB-IoT).
\item A mathematical framework for the performance analysis of transmission scheduling with the QoS support in an mMTC system is presented, and several limitations and the design aspects of short data packet transmission are identified.
\item Existing solutions towards addressing the RAN congestion problem in cellular IoT networks are reviewed along with the highlights on three emerging techniques.
\item The potential benefits, challenges and promising use case scenarios for the applications of emerging ML techniques in ultra-dense cellular networks are identified and the existing ML techniques are reviewed by broadly categorizing them into supervised, unsupervised and RL techniques.
\item A framework for the application of low-complexity Q-learning in addressing the RACH congestion problem is presented along with different exploration strategies, and some performance enhancement techniques are suggested in multi-agent and dynamic wireless environments.
\item Various research issues are identified and some interesting future directions are presented to stimulate future research activities in the related domains.
\end{enumerate}

\begin{table*}
\caption{\small{Definitions of Acronyms}}
\centering
\renewcommand{\arraystretch}{0.9}
\begin{tabular}{ll|ll}
  \hline 
 \textbf{Acronyms}  & \textbf{Definitions} & \textbf{Acronyms}  & \textbf{Definitions} \\
 \hline
ACB & Access Class Barring & NPSS & Narrowband Primary Synchronization Signal \\
ACK & Acknowledgement & NSSS & Narrowband Secondary Synchronization Signal  \\
AI & Artificial Intelligence & NPBCH & Narrowband Physical Broadcast Channel  \\
BS & Base Station &  NRS & Narrowband Reference Signal \\
CE & Coverage Enhancement  & NPDCCH & Narrowband Physical Downlink Control Channel  \\
CRA & Coded Random Access & NPDSCH & Narrowband Physical Downlink Shared Channel  \\
CS  & Compressive Sensing &  NOMA & Non-Orthogonal Multiple Access \\
CIoT & Cellular IoT & OFDM & Orthogonal Frequency-Division Multiplexing   \\
DCI & Downlink Control Information & OFDMA & Orthogonal Frequency-Division Multiple Access   \\
DNC  & Deterministic Network Calculus & PSK & Phase-Shift Keying  \\
DQCA & Distributed Queuing Collision Avoidance & PDU & Periodic Update   \\
eDRX & Extended Discontinuous Reception & PDCCH & Physical Downlink Control Channel  \\
EPDCCH & Enhanced PDCCH & PRACH & Physical RACH   \\
eMBB   & Enhanced Mobile Broadband & PUSCH & Physical Uplink Shared Channel  \\
ED & Event Driven & RB  & Resource Block   \\
FDD & Frequency Division Duplex & RRC & Radio Resource Control   \\
HARQ & Hybrid Automatic Repeat Request & RAN & Random Access Network \\
HetNet & Heterogeneous Network & RA & Random Access  \\
HTC   & Human-Type Communications & RACH & Random Access Channel  \\
H2H   & Human to Human & RAR & Random Access Response  \\
IoT    & Internet of Things & RAW & Random Access Window    \\
LSA   & Licensed Shared Access & RL & Reinforcement Learning \\
LTE   & Long Term Evolution & SAS & Spectrum Access System   \\
LTE-A & LTE-Advanced & SCMA & Sparse Code Multiple Access   \\
MAC    & Medium Access Control & SC-FDMA & Single Carrier Frequency Division Multiple Access  \\
MCL & Maximum Coupling Loss &  SDN & Software Defined Networking  \\
MDP & Markov Decision Process & SL & Sequential Learning \\
MTC & Machine-Type Communications & TA & Timing Alignment  \\
mMTC  & Massive MTC & TDD & Time Division Duplex    \\
M2M  & Machine-to-Machine & TTI & Transmission Time Interval \\
ML & Machine Learning & UE & User Equipment  \\
MPRACH & MTC PRACH & UDN & Ultra-Dense Network   \\
MPDCCH & MTC PDCCH & URLLC & Ultra-Reliable and Low-latency Communications  \\
MUD & Multi-User Detection  &  UFMC & Universal Filtered Multi-Carrier \\
NB-IoT & Narrowband IoT & QoS & Quality of Service \\
\hline 
\end{tabular}
\label{tab: Acronyms}
\vspace{-10 pt}
\end{table*}

\subsection{Paper Organization}
\label{sec:_sec16}
The remainder of this paper is organized as follows: Section \ref{sec:_sec2} identifies the main features, application areas and the potential enablers for mMTC in cellular networks along with a discussion on various issues associated with QoS provisioning in ultra-dense IoT networks. Section \ref{sec:_sec3} highlights the inefficiency of the conventional
LTE RA procedure in mMTC systems along with the basics of RA procedure in legacy LTE systems, and then present key features and channel access mechanisms in two emerging cellular IoT standards, namely, LTE-M and NB-IoT. Also, the characterization of MTC traffic is presented along with the 3GPP-based MTC traffic models and related works. Subsequently, Section \ref{sec:_sec4} provides a mathematical framework for the performance analysis of transmission scheduling with QoS support in the mMTC systems, and also present various design aspects and limitations of short data packet transmission in the mMTC environment. Section \ref{sec:_sec5} presents a review of the existing solutions for RAN congestion problem in cellular IoT networks along with some emerging solutions while Section \ref{sec:_sec6} presents the advantages and challenges of ML techniques in wireless IoT systems and also provides an overview of the existing ML techniques. Also, it provides a detailed explanation on the Q-learning mechanism from the perspective of addressing RAN congestion minimization along with some Q-learning performance enhancement techniques. Finally, Section \ref{sec:_sec7} provides some research challenges and future directions, and Section \ref{sec:_sec8} concludes this paper. To improve the flow of this paper, we provide the structure of the paper in Fig. \ref{fig: paperstructure} and the definitions of acronyms in Table \ref{tab: Acronyms}.

\begin{table*}
\caption{\small{Advantages of achieving desired QoS levels instead of maximizing QoS}}
	\centering
\renewcommand{\arraystretch}{0.95}
\begin{tabular}{|l|l|}
\hline
\textbf{Advantages} & \textbf{Related Causes}  \\
\hline
1. Reduction of energy consumption & 1. Extraneous energy will be wasted while maximizing QoS.  \\ \hline
2. Better compliance with the fixed data rate services & 2. No need to maximize data rates for fixed data rate services \\
& such as video surveillance and online gaming \\ \hline
3. A good perceived performance at the users' end & 3. End-users are usually insensitive to small changes. \\
& in QoS levels, allowing the room for energy saving \\ \hline
4. Better support for emerging application-oriented networks & 4. Desired QoS level is required only within a specific coverage area. \\ \hline
5. The set of feasible regions of optimization solutions are enlarged. & 5. Relaxation of global optimum for the QoS maximization makes \\
& the mathematical problems less restrictive. \\ \hline
6. Adaptive resource allocation problem leads to the cost-effective solutions. & 6. No need to waste additional radio resources by considering  \\
& the conventional optimization assumptions such as full-buffer traffic. \\
\hline
\end{tabular}
	\vspace{-15 pt}
\label{tab: referencesFDcapacitygain}
\end{table*}

\section{QoS Provisioning in Ultra-Dense IoT Networks}
\label{sec:_sec2}
In this section, we first discuss various aspects related to QoS provisioning in ultra-dense IoT networks in the general context, and then provide specific details related to cellular IoT scenarios.  The International Mobile Telecommunication (IMT) vision for 2020 and beyond envisions to provide the connection density target of about $10^6$ devices per square kilometers \cite{ITURvision2015}. However, the available radio resources and communication infrastructures are limited and there is a significant amount of cost involved while acquiring new radio
resources such as spectrum and building new infrastructures. Because of this issue, existing communication networks need to
be made as efficient as possible to support the massive number of devices, thus leading to the concept of ultra-dense IoT networks. Due to diverse types of emerging IoT services and heterogeneous capabilities of MTC devices, QoS provisioning in ultra-dense IoT networks is a crucial challenge.

The overall network performance and QoS of emerging Internet protocol-based networks significantly depend on the effective management of instantaneous traffic flowing in the network. In wireless IoT networks, the instantaneous aggregated traffic at the IoT gateway can be bursty and may greatly exceed the average aggregated traffic since it aggregates periodic transmissions from a large number of sensor nodes with different periods and frame sizes \cite{SingminMDPI}. Due to this,
there may occur congestion during possible bursty intervals and much higher backhaul link (from aggregators to the cloud) bandwidth is needed than that required for the non-bursty traffic. In this direction, one of the important research
challenges is how to make the aggregated traffic as close to the average traffic as possible.

In contrast to the conventional HTC traffic, there are several unique features of MTC traffic \cite{SmiljkovicM2M} which need to be considered while devising transmission scheduling and traffic management strategies for wireless IoT networks. The amount of small packets in IoT-type networks could become significantly large due to the resource-constrained sensor devices and the transmission of short packets from mMTC devices \cite{Durisi2016IEEEproc}. As compared to the dominant
downlink traffic in the conventional cellular systems, uplink to downlink ratio for the MTC traffic is much higher. Furthermore, MTC devices usually have limited power budget and the MTC traffic consists of packets with the short payload length. Also, MTC traffic may arrive in the batch-mode due to high density of devices and correlated transmission \cite{LiIoTjournal}. Moreover, the standard Poisson process may not be suitable for modeling the MTC traffic since the MTC transmissions usually exhibit spatial and temporal synchronism. In addition, in contrast to the conventional voice traffic which has a constant sampling rate for a given codec, MTC traffic usually comprises of different packet sizes and inter-arrival patterns \cite{Afrin2014adaptive}. Different MTC applications have distinct characteristics and service requirements such as priority and delay constraints, thus leading to the need of separate traffic modelling and scheduling schemes to incorporate mMTC devices in the current LTE-based cellular networks.

In addition, MTC devices have completely different QoS requirements than that of the conventional HTC devices.
MTC nodes are usually constrained in terms of battery power and the employed protocols need to be as
energy-efficient as possible. Although the previous research in the area of wireless sensor network protocols
mainly focused on monitoring applications based on low-rate delay-tolerant data collection, the current IoT-based
research has moved to several new applications such as eHealthCare, industrial automation, military and smart home.
These applications have different QoS requirements in terms of  delay, throughput, priority, reliability and different traffic patterns such as event-driven, periodic and streaming \cite{Kim2014M2M}. To this end, it is crucial to consider these distinct QoS features while designing transmission and access techniques for the MTC devices.

In the above context, several existing works deal with the maximization of QoS while attempting to minimize the energy consumption. Several techniques such as sleep mode optimization, power control mechanisms, adaptation of the data rates and learning-assisted algorithms have been employed in various settings \cite{Elhammouti2017self}. However, the objective of QoS maximization may lead to unnecessary energy consumption and achieving satisfactory levels of QoS may be sufficient to balance other performance metrics of systems such as energy efficiency. In contrast to the existing works related to the maximization of QoS, the objective of achieving satisfactory QoS levels can provide several benefits as highlighted in Table \ref{tab: referencesFDcapacitygain} \cite{Elhammouti2017self}.

Moreover, in time-critical MTC applications, several requirements in terms of low end-to-end delay,
deterministic delay, bounds on systematic delay variations and linear delay-payload (packet size) dependence need
to be considered \cite{FabiniM2M}. The packet arrival period in the HTC systems such as multimedia ranges from $10$ ms to
$40$ ms whereas this may range from about $10$ ms to several minutes in MTC systems \cite{Lien2011massive}. In addition, some applications such as data reporting in the smart grid have deterministic (hard) timing constraints and serious consequences may occur in the case of violence of these constraints. In this regard, multiplexing massive accesses with these diverse QoS characteristics effectively is a crucial challenge in ultra-dense IoT networks.

In the following subsections, we describe several aspects of MTC systems, highlight existing challenges for QoS provisioning in ultra-dense cellular IoT networks, present potential enablers for the incorporation of MTC devices in cellular IoT systems, and then present the characterization and modeling of the MTC traffic.

\subsection{Machine-Type Communications}
\label{sec:_sec21}
MTC has got a wide variety of application areas ranging from industrial automation and control to environmental monitoring towards building an information ambient society. The main applications of the MTC are listed below \cite{Ghavimi2015M2M}.
\begin{enumerate}
\item \textbf{Industrial automation and control}: This class includes several scenarios such as production on demand, quality control, automatic interactions among machines, optimization of packaging, logistics and supply chain, and inventory tracking.
\item \textbf{Intelligent transportation}: Under this category, MTC finds applications in different scenarios such as logistic services, M2M assisted driving, fleet management, e-ticketing and passenger services, smart parking and smart car counting.
\item \textbf{Smart-grid}: This category include various application scenarios including automatic meter reading, power demand management, smart electricity distribution and patrolling, online monitoring of transmission lines and transmission tower protection \cite{Saleemarxiv2017}.
\item \textbf{Smart environment}: Several application scenarios including smart homes/offices/shops, smart lighting, smart industrial plants, smart water supply, environmental monitoring and green environment can be enabled with MTC.
\item \textbf{Security and public safety}: Several applications such as remote surveillance, personal tracking and public infrastructure protection can be considered under this category.
\item \textbf{e-Health}: In this application area, various scenarios exist such as tracking or monitoring a patient or a segment of an organ in a patient, identification and authentication of patients, diagnosing patient conditions and providing real-time information on patients health related data to the remote monitoring center.
\end{enumerate}

Since MTC applications are significantly different from their HTC counterparts, they have distinct QoS requirements with different service features such as time-controlled, time-tolerant, small data transmission, low or no mobility, group-based connection, priority-based transmissions and low power consumption \cite{Zhengradio2012}. In this regard, the 3GPP has identified the following $14$ features for M2M communications\footnote{For the description of these features, interested readers may refer to \cite{3GPPMTCservice}.}: (i) low mobility, (ii) time-controlled, (iii) time-tolerant, (iv) packet switched only, (v) mobile originated only, (vi) small data transmission, (vii) infrequent mobile terminated, (viii) M2M monitoring, (ix) priority alarm message, (x) secure connection, (xi) location specific trigger, (xii) network-provided destination for uplink, (xiii) infrequent transmission, and (xiv) group-based policing and addressing.

Furthermore, the 3GPP has specified various general requirements for the MTC systems \cite{3GPPMTCservice,Ghavimi2015M2M} in order to effectively operate MTC devices and also to establish successful linkage between an MTC subscriber and the network operator. Some of the main technical requirements include: (i) providing a control mechanism to the network operators for the addition/removal/restriction of individual MTC device features, (ii) exploring a peak reduction mechanism for data and signaling traffic when a  number of MTC devices concurrently attempt for their data transmissions, (iii) yielding a mechanism to restrict downlink data traffic and also limiting access towards a specific access point name in case of network overload, and (iv) investigating techniques to maintain efficient connectivity for a large number of MTC devices and to lower the corresponding energy consumption.

In comparison to the conventional HTC, the emerging MTC has the features of infrequent transmissions and low data rates. Also, the size of signalling data packets can be much larger than the size of user data packets in M2M applications \cite{Gotsis2012M2M}. Furthermore, although M2M devices need to transmit small amounts of data, communication infrastructure may get congested if a huge number of M2M devices attempt to access the network near-simultaneously \cite{Chengoverload}. Hence, the performance of the existing cellular standards has to be evaluated for this emerging type of traffic.

In addition, the 3GPP has identified the following performance objectives to support mMTC in the emerging air interface 5G New Radio (NR) \cite{Ratasuk2017LTEM,3GPPtechnical2017}.
\begin{enumerate}
\item Very high connection density of about $10^6$ devices per $\mathrm{km}^2$ in an urban environment
\item Ultra-low complexity and low-cost IoT devices/networks
\item Battery life in extreme coverage beyond $10$ years with the battery life evaluated at $164$ dB MCL, and a battery capacity of $5$ Wh.
\item Maximum Coupling Loss (MCL) of about $164$dB for a data rate of $160$ bps at the application layer
\item Latency of about $10$ seconds or less on the uplink to deliver a $20$-byte application layer packet (measured at $164$dB MCL)
\end{enumerate}

\subsection{Challenges for QoS Provisioning in Ultra-Dense IoT Networks}
\label{sec:_sec22}
\begin{figure*}
	\begin{center}
		\includegraphics[width=4.0 in]{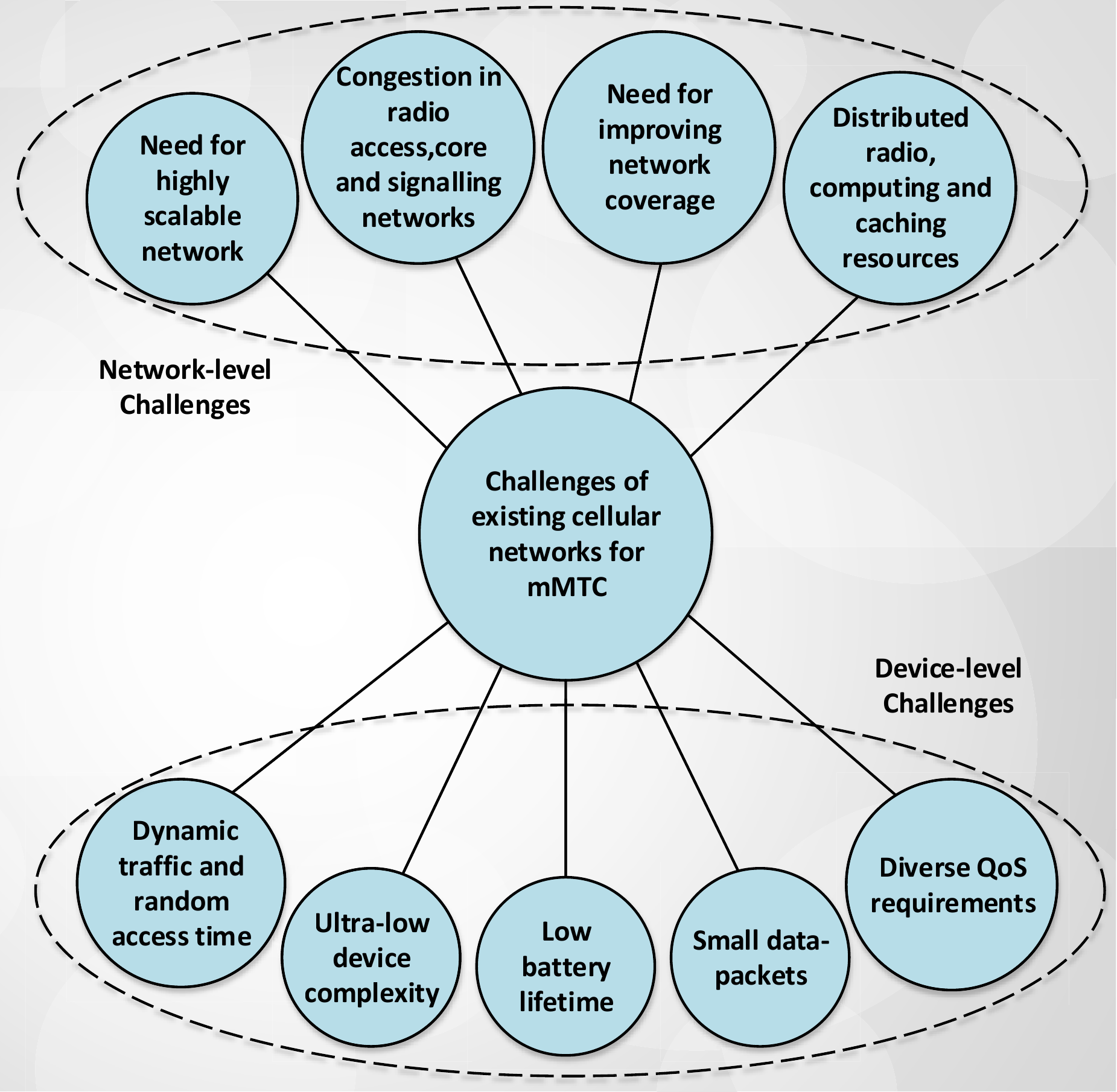}
		\caption{\footnotesize{Challenges of existing cellular networks to support emerging massive machine-type communications.}}
	\label{fig: challengesmMTC}
	\end{center}
\vspace{-15 pt}
\end{figure*}

There arise several challenges in incorporating MTC devices in LTE/LTE-A based cellular networks. First, the massive number of devices try to access the scarce network resources in a short period of time and there may arise the need of either utilizing the available resources efficiently or allocating additional bandwidth
to incorporate these devices \cite{Wali2014}. Secondly, there are significant differences in the transceiver properties and the applications of MTC devices from the existing LTE-based user terminals \cite{Wang2016cellular}.
In most of the applications, MTC devices consume low power and have intermittent low rate transmissions. Furthermore,
due to the need of cost-effective deployment of massive devices, MTC devices have degraded transceiver performance and reduced coverage as compared to the LTE user terminals. Besides, their effects in the communication performance of the existing LTE-A users need to be monitored and mitigated carefully. In this regard, one of the important research questions is
how to provide concurrent access to a large number of MTC devices without degrading the QoS of the existing cellular users.

Since a network interface is fully utilized during the peak time, the devices may not be able to send or
receive data and it is crucial to optimize the peak traffic in the emerging content-centric wireless networks \cite{AlAdwani2012}. Furthermore, as highlighted earlier in Section \ref{sec:_sec21}, the emerging MTC applications are quite different from the traditional HTC applications due to unique features such as group-based communications, time-controlled, small data transmissions, and low or no mobility \cite{Zhengradio2012,Lien2011ubiquitous}. These distinct features of MTC applications result in diverse QoS requirements and it is important to take these QoS requirements into account while devising multiple access techniques for future cellular IoT networks. The main performance indicators of an mMTC system are the number of concurrent connections to be supported, energy efficiency and network coverage \cite{Lienabling2017hetnet}.

Existing cellular networks face the following major problems in supporting MTC devices \cite{3GppTR2015,Maldonado2017NBIoT,3GPPMTCservice}. In Figure \ref{fig: challengesmMTC}, we present the pictorial representation of these issues in the form of device-level and network-level challenges.
\begin{enumerate}
\item \textbf{Highly dynamic traffic and random access time}: The data traffic arisen from the MTC devices is highly dynamic in nature as compared to more predictable HTC traffic. Furthermore, there arises a need to handle the mixed traffic models with the event-driven and periodic traffics. In addition, the existing contention-based radio access schemes will need to coordinate random transmissions from the massive number of devices \cite{Sharmacommletter2018}.
\item \textbf{Ultra-low device complexity}: Due to the requirement of cheap MTC devices for mass deployment, the devices are constrained in terms of computational and memory resources, thus providing the limited performance.
\item \textbf{Low battery lifetime}: Because of the cost and space constraints, MTC devices are limited in their battery capacity. Furthermore, due to distributed nature of IoT devices and the involved cost-issues in replacing the batteries, the battery lifetime of MTC devices is expected to be more than $10$ years with the battery capacity of $5$ Wh, thus leading to the need of investigating power saving methods for ultra-dense cellular IoT networks.
\item \textbf{Small data packet transmissions}: In addition to the huge signalling burden associated with a large number of small packet transmissions from MTC devices, there arise other challenges such as the requirement of higher resource granularity and efficient channel coding for short block lengths in contrast to the channel coding schemes designed for long packets in the conventional cellular systems \cite{Bockelmann2016}.
\item \textbf{Diverse QoS requirements}: MTC devices have diverse QoS requirements in terms of data rate and latency requirements and existing cellular technologies need to adapted to handle these features.
\item  \textbf{Network congestion}: As highlighted earlier in Section \ref{sec:_sec12}, the incorporation of massive MTC devices in the existing LTE/LTE-based cellular network may result in congestion in different segments of the network including RAN, the core network and the signalling network.
\item  \textbf{Highly scalable network}: Because of the need to support a significantly large number of connected devices ranging from a factor of $10 \times$ to $100 \times$ as compared to the cellular devices, it is crucial to maintain the system performance with the increase in the connection density.
\item \textbf{Need for improving network coverage}: There arises significant shrinkage in the link budget due to the reduced capability of MTC devices. In order to increase the coverage to the areas where MTC devices are deployed (such as deep inside a building), LTE release 13 targeted the coverage extension of at least $15$ dB for the MTC devices. This coverage improvement enables the support of the devices in the locations where the conventional cellular networks face difficulty.
\item \textbf{Distributed radio, computing and caching resources}: With the recent trend of migrating communications networks from the connection-oriented to the content-oriented nature, it is important to investigate synergies among communications, computing and caching resources which are distributed across different devices in ultra-dense IoT networks \cite{SKSIEEE2017}. However, the conventional cellular networks based on the centralized management are sluggish in terms of network resource management and they need to evolve to deal with the management of distributed resources.
\end{enumerate}

Towards modeling and analysis of the QoS of wireless networks, one of the important mathematical tools is Deterministic Network Calculus (DNC), which is useful to calculate delay parameters such as delay bound, backlog bound and other service quality parameters by utilizing the traffic/packet arrival and service curves \cite{Huang2017M2M}. This DNC tool enables the determined boundary analysis for the system performance and offers a strict service guarantee by considering the worst-case scenarios. The main QoS metrics that can be evaluated include delay bound and backlog bound. The metric delay bound represents the maximum between arrival and service curves while the backlog bound denotes the maximum vertical deviation between these two curves.

In addition to providing high capacity to the fairly limited number of traditional user equipments to support high data rate services such as video streaming, the air interface of 5G cellular network has to provide connectivity to the massive number of concurrent transmissions coming from the MTC devices \cite{Centenaro2017comparison}. Also, the exchange of signalling information needs to be minimized both in the uplink and downlink due to a large number of MTC devices to be supported with the limited available radio resources. Furthermore, the RA procedure at the device-side should be simplified as much as possible by shifting the burden to the network side/eNodeB due to the resource constrained and low-cost nature of MTC devices.

Moreover, the conventional centralized approaches for congestion management in cellular networks are not scalable as desired by the MTC systems and also the distributed scheduling approaches can not easily acquire the knowledge about the network load and requirements of other applications \cite{UmaIBM}. Furthermore, the conventional congestion management process is mostly a reactive process instead of the proactive one needed for MTC devices. By deferring and shaping transmissions at the source itself in a network and being aware of the underlying application properties, better congestion management can be obtained for MTC devices \cite{UmaIBM}.

Authors in \cite{Dawy2017towards} highlighted the difference between HTC over cellular and MTC over cellular in terms of various parameters such as uplink, downlink, subscriber load, device types and requirements in terms of delay, energy, signalling and cellular architecture. Furthermore, the signalling overhead in the uplink and downlink MTC links are analyzed considering the SMS-type raw data size ($<248$ bytes) and email-type raw data size for smart metering and vehicular sensing applications via experimental measurements. It has been shown that the signalling overhead for the downlink control messages is considerably higher than for the uplink case, and higher signalling overhead occurs in vehicular applications than in the smart metering applications.

\subsection{Potential Enablers for mMTC in Cellular Networks}
\label{sec:_sec23}
Due to some distinct transitions while going from the conventional HTC platform to the emerging mMTC such as from the larger packet sizes to the smaller packet sizes, from the downlink-focused communication scenario to the uplink dominant, and from high data rate to low data rate transmissions from MTC devices, mMTC systems have new design requirements than those of the conventional HTC systems. To this end, it is important to look into the mMTC network design problem from a different perspective than the tradition approach followed for the HTC systems.

\begin{figure}
	\begin{center}
		\includegraphics[width=3.6 in]{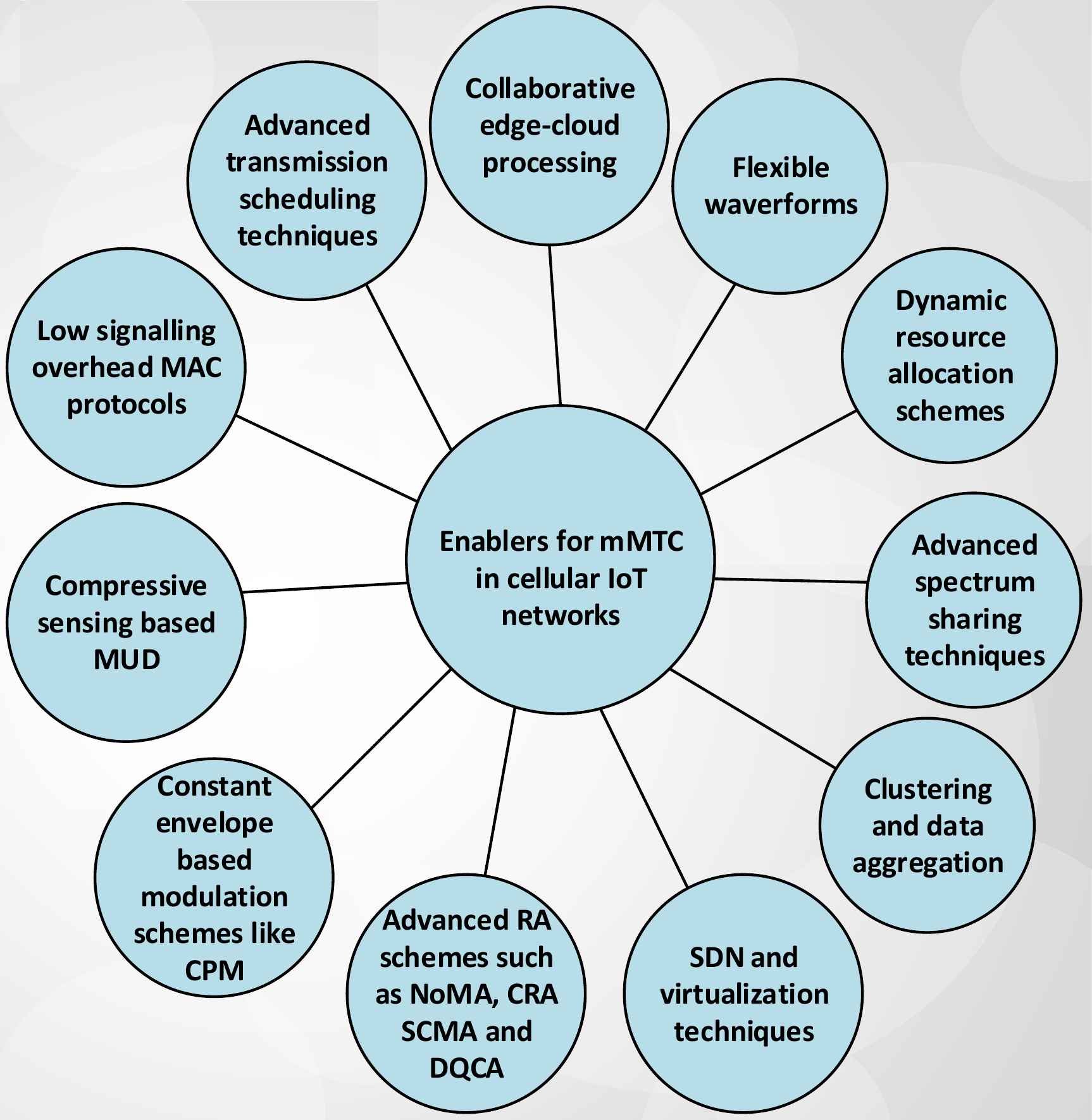}
		\caption{\footnotesize{Potential enabling techniques for mMTC in cellular IoT networks.}}
	\label{fig: mMTCenablers}
	\end{center}
\vspace{-15 pt}
\end{figure}

In Figure \ref{fig: mMTCenablers}, we present the main enabling techniques being considered to facilitate the incorporation of mMTC in the upcoming cellular IoT networks. In the following, we briefly describe these enabling techniques.
\begin{enumerate}
\item \textbf{Flexible waveform design}: The design of flexible waveforms can enable the in-band mMTC channels within the LTE carrier \cite{Bockelmann2016}. In this regard, the traditional waveforms designed for HTC communications need to be adapted to support mMTC while considering various aspects such as end-to-end latency,
    robustness against time and synchronization errors, out-of-band radiations, spectral efficiency and transceiver complexity \cite{Medjahdi2017road}.
\item \textbf{Dynamic resource allocation techniques}: The 3GPP suggested to allocate RACH resources dynamically to address the RAN congestion problem \cite{3GPPRAN}. For example, by deciding the number of preambles adaptively without knowing the number of devices and access probability, the RACH throughput can be maximized \cite{Choi2016commlett}. Furthermore, since this approach can dynamically change the size of the RACH resource pool and other resources, the total data collection time from the resource-constrained MTC devices can be minimized in delay-sensitive/emergency applications \cite{Lee2017MTC}. The two main issues for employing this process are the requirement of estimating the number of contending devices in an RA slot and determining the preamble pool size.
\item \textbf{Advanced spectrum sharing methods}: Although both the licensed and unlicensed bands can be exploited for mMTC applications, lack of QoS guarantees in the unlicensed band becomes highly problematic \cite{JayawickramaspectrumIoT}. In this regard, emerging advanced spectrum sharing techniques such as Licensed Shared Access (LSA) and Spectrum Access System (SAS) \cite{Sharma2018DSS} could be potential solutions for mMTC applications since they can provide better interference characterization.
\item \textbf{Clustering and data aggregation schemes}: By grouping MTC devices into smaller clusters based on some suitable criteria such as geographical locations or QoS requirements and then aggregating the individual device data at the MTC gateway/aggregator, the RAN congestion can be significantly minimized \cite{Lienabling2017hetnet}. Furthermore, the investigation of energy-efficient clustering schemes facilitates the deployment of low-power MTC devices \cite{Lienabling2017hetnet}.
 \item \textbf{Software Defined Networking (SDN) and virtualization techniques}: Based on the functionalities of MTC devices and their QoS requirements, a physical cellular network can be virtualized into different networks such as industrial, vehicular, smart grids and emergency networks, with all these networks sharing the same set of radio, computing and networking resources \cite{Lirandomaccess2017}. The dynamic sharing of resources and the reconfiguration of network elements among thus virtualized networks can be carried out by utlizing an SDN paradigm which decouples the control plane from the data plane and incorporates the capability of programming in the IoT network.
 \item \textbf{Advanced RA schemes}: Several emerging RA schemes such as Non-Orthogonal Multiple Access (NOMA), Sparse Code Multiple Access (SCMA), Coded Random Access (CRA) \cite{Bockelmann2016} and distributed queueing based access protocol \cite{Samir2016} can be considered as the promising enablers for the mMTC in cellular networks.
 \item \textbf{Constant envelope coded-modulation schemes}: Due to space/cost constraints, MTC devices need to use low-cost amplifiers which are prone to non-linearities and hardware imperfections. In this scenario, constant envelope signals can enable the non-linear power-efficient and cost-effective operation at the MTC devices. Therefore, constant envelope coded modulation schemes such as Continuous Phase Modulation (CPM) can be considered as enablers for the mMTC \cite{Bockelmann2016}.
 \item \textbf{Compressed Sensing (CS)-based Multi-User Detection (MUD)}: The amount of collisions in the IoT access network can be further minimized by employing advanced interference cancellation receivers. As an example, the CS-MUD can enhance the resource efficiency and serve higher number of users by using the combination of non-orthogonal RA and joint detection of user data and activity \cite{Bockelmann2016}. In this regard, the combination of advanced MAC protocols with the CS-based MUD can be utilized by exploiting the sparse joint activity in the mMTC environment \cite{Compressivecoded2015}.
\item \textbf{Low signalling overhead MAC protocols}: One of the main technical challenges in an mMTC system is to reduce the amount of signalling overhead generated by the MTC devices and the design of low-signalling overhead protocols will facilitate the deployment of MTC devices in cellular networks \cite{Dawy2017towards}.
\item \textbf{Advanced transmission scheduling techniques}: The transmission scheduling techniques designed for cellular IoT systems should be able to accommodate the MTC devices with heterogeneous QoS requirements in addition to the legacy cellular users. In this regard, advanced scheduling techniques such as latency-aware scheduling \cite{Delgado2012european}, fast uplink grant \cite{AliCoRR2018} and learning-assisted scheduling \cite{Belloeuropean2014} seem promising to schedule the sporadic transmissions from a huge number of MTC devices over limited RACH resources.
\item \textbf{Collaborative cloud-edge processing}: Cloud computing platform has very high computational and storage capacity, and has a global view of the network but is not suitable for delay sensitive applications. On the other hand, edge-computing is suitable for applications demanding low delay and high QoS but has lower computational resources and storage capacity. In this regard, collaborative processing between these two platforms will be a promising approach to address various issues including latency minimization \cite{Sabin2017Globecom}, dynamic spectrum sharing \cite{Sharmacooperative2017}, peak traffic management and data offloading in ultra-dense IoT networks \cite{SKSIEEE2017}.
\end{enumerate}

The ML-assisted techniques, detailed later in Section \ref{sec:_sec6}, can address various issues related to self-configuration, self-optimization and self-healing in emerging wireless networks and seem promising in facilitating the implementation of the most of the technology enablers listed in Fig. \ref{fig: mMTCenablers} towards enhancing the performance of mMTC systems. However, the ML techniques should be as simple as possible to be applied in the MTC devices and the investigation of low-complexity adaptive ML techniques is one of the emerging future research directions as highlighted later in Section \ref{sec:_sec7}.

\subsection{Traffic Characterization and Modeling for mMTC Systems}
\label{sec:_sec24}
The characterization and modeling of mMTC traffic is crucial to support MTC devices in the existing cellular networks due to various reasons specified in the following. The incorporation of MTC devices in cellular networks may cause harmful interference to the existing cellular users and may significantly degrade the system performance of LTE/LTE-A based cellular systems. To this end, it is important to analyze the impact of MTC traffic on the existing cellular users by utilizing suitable interference modeling in realistic wireless environments \cite{Andrade2015impact}. Furthermore, suitable interference mitigation, resource allocation and resource sharing schemes need to be investigated to ensure the sufficient protection of the cellular users against harmful interference caused by the massive number of MTC devices and by utilizing the given MTC traffic models, these schemes can be designed in an efficient manner. Moreover, since MTC traffic is uplink dominant and the rigid QoS support framework of LTE designed for voice and data services may not be capable of addressing specific QoS requirements of MTC traffic in terms of latency, jitter and packet loss, suitable transmission scheduling techniques need to be investigated to support a large number of MTC devices while fulfilling their specific QoS requirements. Besides, to investigate suitable traffic management schemes such as peak traffic reduction in wireless IoT networks, it is essential to understand and characterize the traffic models applicable for a particular IoT application \cite{Sharmacommletter2018}. In addition, the traffic characteristics depend on the application scenarios and the MTC devices usually have heterogeneous traffic patterns in terms of their amplitudes, starting times and activation periods \cite{Kim2014M2M}.

Existing traffic models in telecommunication systems can be  categorized into: (i) source traffic models, mainly applicable for video, data and voice transmissions, and (ii) aggregated traffic model applicable for Internet, high-speed links and backbone networks \cite{Laner2013traffic}. Since an IoT network consists of a large number of sensors or MTC devices which are usually controlled by a gateway/server, IoT traffic at the gateway usually fits into the aggregated traffic model. In this regard, the 3GPP has defined the following two types of aggregated MTC traffic models \cite{3GPPRAN}.
\begin{enumerate}
\item \textbf{Model 1}: Uniform distribution over a duration $T$ in which MTC devices access the network uniformly over a period of time, i.e., in a non-synchronized manner. This model does not take account of the correlation between the transmissions of the devices.
\item \textbf{Model 2}: Beta distribution over $T$ in which  a large amount of MTC devices access the network in a highly synchronized manner. This model generates correlated traffic in a specific time interval.
\end{enumerate}

For the first model, the following function can be considered.
\begin{equation}
f_i(t)=\left\{\begin{array}{ll}
         P_i, \hspace{5 pt} 0 \leq t \leq \tau_i, \\
         0, \hspace{5 pt} \tau_i \leq t \leq T_i,
         \end{array}
              \right.
\label{eqn:trafficprofile}
\end{equation} where $P_i$ denotes the amplitude of the traffic profile for the $i$th device and $\tau_i/T_i$ represents the corresponding duty cycle. Similarly, for the second model, the probability density function of the Beta distribution, is given by \cite{3GPPRAN}
\begin{equation}
p(t,\alpha,\beta)=\frac{1}{B(\alpha,\beta)} t^{(\alpha-1)} (1-t)^{(\beta-1)},
\label{eqn:betaPDF}
\end{equation} where $t$ denotes a single realization over the time axis, $\alpha>0, \beta>0$ are the scale parameters, and $B(\alpha,\beta)$ denotes the Beta function, which is a normalization constant to ensure that the total probability equals to $1$ and can be defined as $B(\alpha,\beta)=\int_{0}^{1}x^{\alpha-1} (1-x)^{\beta-1} dx$ \cite{Sharmacommletter2018}.

\begin{figure}
\begin{center}
\includegraphics[width=3.2 in]{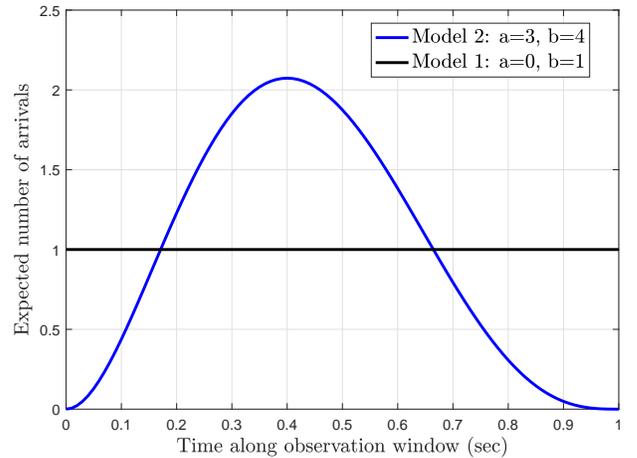}
\caption{\footnotesize{Access intensity for two 3GPP based MTC traffic models.}}
\label{fig: accessintensity3GPP}
\end{center}
\vspace{-15 pt}
\end{figure}

Considering $M$ number of MTC devices and their activation periods between $t=0$ and $t=T$, the RA intensity for the $m$th device is given by the probability distributions either in (\ref{eqn:trafficprofile}) or (\ref{eqn:betaPDF}) depending on the employed model. The number of arrivals in the $i$th access slot is given by \cite{3GPPRAN}
\begin{equation}
I_{\mathrm{access}}(i)=M \int_{t_i}^{t_{i+1}}p(t) dt,
\label{eqn:accessintensity}
\end{equation} where $t_i$ denotes the time of the $i$th access opportunity and $p(t)$ given by (\ref{eqn:trafficprofile}) or (\ref{eqn:betaPDF}). The distribution of access attempts should be limited over a certain observation period $T$ in such way that $\int_{0}^{T}p(t)dt=1$. Figure \ref{fig: accessintensity3GPP} illustrates the RA intensity of two 3GPP-based MTC traffic models with $T=1$.

Although aggregated traffic modeling is suitable for scenarios involving a large number of devices and is less complex to realize, it is less precise than the source traffic modeling since it is not able to capture the real traffic features at the source level \cite{Laner2013traffic,SmiljkovicM2M}. On the other hand, source traffic modeling treats the traffic for every devices separately, and hence is more precise. However, modeling source traffic becomes complex for a large number of source devices. Therefore, it is crucial to investigate suitable traffic models which can combine the benefits of both the source and aggregate traffic models. In this regard, coupled Markov modulated Poisson processes \cite{Laner2013traffic} seems a promising approach, which has higher accuracy than the aggregated modeling and has lower complexity than the conventional source traffic modeling.

Considering a variety of applications, the MTC traffic can be categorized into the following three traffic patterns: (i) Periodic Update (PU), (ii) Event-Driven (ED) and (iii) payload exchange \cite{Nikaein2013}. The PU traffic has a regular pattern, constant data size and is non-real time type (example: smart meter reading) while the ED traffic has a variable pattern, varying data size and is a real time traffic (example: health emergency alarming). On the other hand, payload exchange traffic follows either of the above traffic types and it may be of constant size or variable size, real time or non-real time depending on the application scenario. In addition, there exist three main types of traffic shaping policies \cite{AlAdwani2013traffic}: (i) traffic shaping for bulk applications where each flow is assigned a fixed bandwidth, (ii) traffic shaping for the aggregate traffic, and (iii) time-based traffic shaping which is applied only at the peak-time to reduce congestion and cost.

The uplink traffic generated from the sensors in most of MTC applications is heterogeneous and can be classified into \cite{KumarMilcom2016}: (i) non real-time with no task completion deadline, (ii) soft real-time with the decreased utility if the deadline not met and (iii) firm real-time having zero utility if the deadline is not met. As an example, industrial M2M traffic has very low latency requirements in the order of a few milliseconds \cite{Kumar2016online}. In general, the PU traffic is periodic with tight service deadline while the ED traffic is random with all three traffic categories, i.e., non real-time, firm or soft real-time. From the scheduler designer perspective, it is crucial to maximize a system utility metric in order to maximally satisfy the delay requirements of all the classes \cite{KumarMilcom2016}.

Moreover, possible network applications in wireless IoT networks can be classified into the following \cite{Zhengradio2012}.
\begin{enumerate}
\item \textbf{Elastic applications}: This category corresponds to more traditional HTC applications such as electronic email, file transfer as well as the downloading of remote data from the MTC servers. These applications are mostly delay tolerant in nature and the user utility usually has diminishing marginal improvements with the incremental increase in the achievable data rate.
\item \textbf{Hard real-time applications}: These applications have a desired delay constraint with hard real-time requirements. Beyond the desired time frame, there is no additional utility gain while increasing the data rate and the user utility becomes the step function of the achievable data rate.
\item \textbf{Delay adaptive applications}: Some delay sensitive applications can occasionally tolerate a small delay with a certain delay-bound violation and the packet dropping probability. The user utility in these applications (such as remote monitoring of e-Health services) deteriorates rapidly when the achievable data rate becomes less than the required intrinsic data rate.
\item \textbf{Rate-adaptive applications}: These applications try to adjust their transmission rates based on the available radio resources with the moderate delays. A highly efficient scheduler is needed to enhance the performance of these applications in time-varying channel conditions.
\end{enumerate}

Traffic shaping, also called packet shaping, delays certain types of data packets in order to optimize the overall performance of a network. To achieve the optimized network performance, Internet traffic thesedays is intentionally shaped into ON/OFF pattern \cite{Zhaotraffic}. Also, ON/OFF pattern is generated due to some inherent characteristics of applications such as HTTP web browsing and MapReduce operation at the data center/server. The main benefits in performing ON/OFF traffic shaping include: (i) reduction of computing overhead at the server-side, (ii) energy saving at the wireless terminals and (iii) minimizing the bandwidth waste while delivering streaming services. However, this On/OFF traffic shaping faces several challenges such as the impact on packet drop probability, harmful effect on other real-time applications and weakening the congestion control function of the transmission control protocol \cite{Zhaotraffic}. To address these issues, it is crucial to design suitable models to characterize the relation among the associated parameters of ON/OFF traffic such as the ratio of ON/OFF duration, burst size, and burst transmission rate, and also the models for packet loss probability and temporary congestion caused by the bursty transmissions.

Another approach to manage the peak-traffic is to employ demand-side management, which adopts suitable measures at the customer-side/sensor-side to optimize the overall network performance \cite{AlAdwani2013traffic}. On one hand, the demand profile can be flattened to limit the amplitude fluctuations while simultaneously accommodating the same amount of traffic volume. On the other hand, the utilization profile of network resources can be flattened by rescheduling or delaying the services \cite{AlAdwani2012}. Some approaches for traffic scheduling include limiting the queue size and setting a bandwidth limit so that aggregate traffic size in the active queue does not exceed the limit. However, scheduling techniques cause delay for processing the network demand and it is crucial to investigate suitable techniques to minimize the delay introduced by traffic scheduling mechanisms. Furthermore, a suitable bandwidth limit should be applied to balance the trade-off between latency and energy saving \cite{AlAdwani2013traffic}.

\section{Random Access Procedure in Cellular IoT Networks}
\label{sec:_sec3}
An MTC device must go through the access procedure to establish a connection to the Base Station (BS)/eNodeB/access point mainly in the following situations \cite{Layarandomaccess2014}: (i) while establishing an initial access to the network, (ii) while receiving/transmitting new data and MTC device is not synchronized to the network, (iii) during the transmission of new data when no scheduling resources are allocated on the uplink control channel, (iv) to perform a seamless handover, (v) in order to re-connect to the network in the case of radio link failure.

The RA methods in the LTE-based cellular systems can be categorized into contention-based (for delay-tolerant access requests) and contention-free (for delay-sensitive requests) schemes \cite{Kim2017random}. Out of these, the contention-based scheme is of the main interest here due to the limitation in the number of available Resource Blocks (RBs) as compared to the massive number of access requests to be supported. In the contention-based RA approach, a huge number of MTC devices have to select the same preambles because of the limitation in the available preambles  in the existing LTE-based cellular systems,  and this results in significantly high number of collisions in the access network and subsequently leads to the RAN overload or radio access congestion problem in ultra-dense IoT networks. In this direction, one important question to be answered is how to concurrently support the massive number of MTC devices in ultra-dense cellular IoT networks without affecting the performance of the existing cellular devices by using the current communication technologies/standards.

In the following, we briefly describe the RA procedure in the legacy LTE systems and its inefficiency in handling the massive number of devices in the mMTC environment, then present some adaptations made to support MTC devices in cellular networks, and subsequently highlight the main features and access mechanisms in emerging cellular IoT standards, namely, LTE-M and NB-IoT.

\subsection{RA Procedure in Legacy LTE Systems}
\begin{figure}
	\begin{center}
		\includegraphics[width=3.2 in]{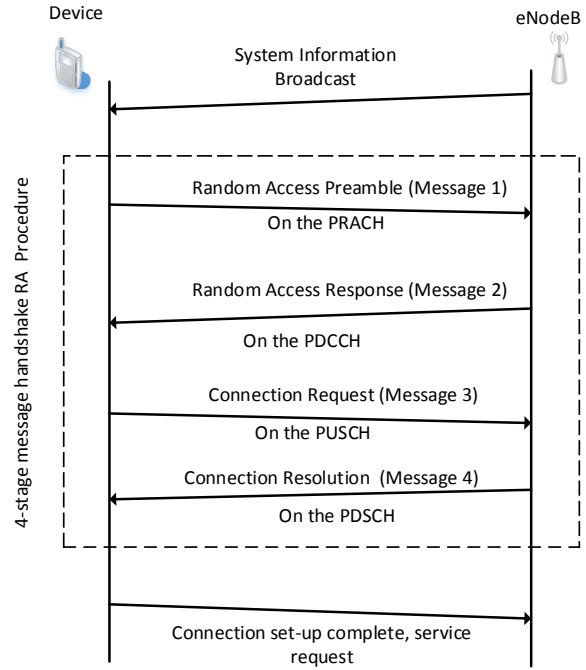}
		\caption{\footnotesize{Illustration of four-stage message handshake-based RA procedure in LTE-based cellular systems (PRACH: Physical Random Access Channel, PDCCH: Physical Downlink Control Channel, PUSCH: Physical Uplink Shared Channel, PDSCH: Physical Downlink Shared Channel).}}
	\label{fig: RAillus}
	\end{center}
\vspace{-15 pt}
\end{figure}

After an eNodeB broadcasts the system information to the devices, the contention-based RA procedure follows a four-stage message handshake procedure as depicted in Fig. \ref{fig: RAillus} \cite{Duan2016adaptive,Layarandomaccess2014}, which mainly involves the following four stages: (i) RA preamble transmission from the device to the eNodeB (Message 1), (ii) RA Response (RAR) from the eNodeB to the device (Message 2), (iii) connection request message from the device to the eNodeB (Message 3), and (iv) connection resolution message from the eNodeB to the device (Message 4).

In the first stage, each device randomly selects an RA preamble from the set of available preambles broadcasted by the eNodeB during the initial network synchronization phase and sends the RA request (Message 1) by transmitting thus selected preamble in an RACH. At this stage, the User Equipments (UEs) just transmit the selected preambles and not the device IDs. In the second stage, the eNodeB acknowledges the received distinct preambles with an RA response (Message 2) which includes the preamble index being acknowledged, instructions for the timing alignment and the command for the RB allocation. Subsequently, in the third stage, the UE recognizes the RA response addressed to it by noting the preamble index it has used for the RA request in the first step and utilizes the dedicated RB on the Physical Uplink Shared Channel (PUSCH). The devices which made current transmissions of the RA request with the same preamble in the first stage will be instructed to use the same RB in the step 3 and such transmissions will go through collisions. On the other hand, for the packets (which contain the corresponding device IDs) which are successfully decoded in step 3, the eNodeB sends a contention resolution message (Message 4) to the corresponding devices.

In this RA procedure, after sending the preamble in the RA request (Message 1), the device
sets an RAR window and waits for the eNodeB's response with an uplink grant (Message 2) in the RAR message. If the UE successfully receives its Message 2 within the defined RAR window, the UE sends the Radio Resource Control (RRC) connection request (Message 3) to the eNodeB. At this stage, the device starts the Message 4 timer and waits to receive its own RRC connection setup message (Message 4) from the eNodeB \cite{Lin2016estimation}.

In the above-mentioned RA procedure, the physical-layer mapping of RACHs is called Physical RACHs (PRACHs) which are time-frequency blocks specified by the eNodeB. The periodicity of RA slots is broadcasted by the eNodeB in terms of the PRACH configuration index, which varies between every $1$ ms (i.e., a maximum of $1$ RA slot per $1$ subframe) to $20$ ms (i.e., the minimum of $1$ RA slot every $2$ frames) \cite{Layarandomaccess2014}. The transmission scheduling in terms of time and frequency depends on the configuration of the PRACH. For example, for the PRACH configuration index of $6$, there will be RACHs in every $5$ ms within a bandwidth of $180$ kHz, with a duration ranging from $1$ ms to $3$ ms.

\subsection{Failure of RA Procedure and Its Inefficiency in mMTC systems}
The aforementioned RACH procedure in LTE/LTE-A based networks may fail due to the following reasons \cite{Lin2016estimation}.
\begin{enumerate}
\item \textbf{Failure of preamble transmission}: This occurs either due to the collision of RACH preambles (caused due to concurrent transmission of the same set of preambles by more than one devices) or insufficient preamble transmission power.
\item \textbf{Failure of Message 2 reception}: This occurs mainly due to the lack of downlink radio resources (i.e., PDCCH) to send the RA response (Message 2) to all the received preambles within the devices' RAR windows.
\item \textbf{Failure of Message 3 transmission}: This occurs due to the failure in transmitting Message 3 to the
eNodeB by employing the Hybrid Automatic Repeat Request (HARQ) process at the device.
\item \textbf{Failure of Message 4 reception}: This occurs due to the failure in receiving Message 4 by the devices within the Message 4 expiration time while using HARQ transmission from the eNodeB either due to insufficient PDCCH resources or an imperfect channel condition.
\end{enumerate}

Out of $64$ preambles used in LTE-A networks, $54$ preambles are used for the contention-based access, while the remaining $10$ preambles are reserved for the contention-free access which is needed for high-priority services such as handover. In every $5$ ms, there arises an access opportunity and $200$ access opportunities per second. This corresponds to the absolute maximum capacity of $10,800$ preambles per seconds in the absence of access collisions \cite{Layarandomaccess2014}. However, due to ALOHA type access protocol and random backoffs, performance becomes much lower than this maximum limit in practical cellular systems. Furthermore, the situation becomes worse while supporting the MTC devices. Besides the problem of supporting higher number of nodes in ultra-dense IoT networks, other performance metrics such as access delay and energy consumption are important to be considered.

To study the performance of the aforementioned $4$-stage message handshake RA procedure in LTE systems, authors in \cite{Osti2014analysis} studied the stability limit of this legacy RA procedure, which indicates the probability of failure of the RA procedure and the maximum achievable throughput. It has been shown that the performance of the RA procedure deteriorates rapidly while sharing the Physical Downlink Control Channel (PDCCH) resources between Messages 2 and 4 with different priorities and the overall RA performance can be enhanced by increasing the size of the PDCCH resource.

Furthermore, in mMTC systems, the system performance may severely degrade in the presence of concurrent massive access requests due to high probability of collision caused by the signaling and traffic load spikes since the contention-based operation of the RACH in LTE-A networks is based on ALOHA-type access \cite{BIRAL20151}. One of the possibilities to reduce the load of physical RACH is to increase the number of access opportunities scheduled in a frame, however, this approach will reduce the amount of resources needed for data transmission. In this regard, it is crucial to balance the tradeoff between the amount of resources available for data transmission and the amount of access opportunities to be scheduled per frame while designing an uplink scheduler for MTC applications by taking into account of limited available bandwidth. Besides, the main performance metrics to be improved include access success probability, preamble collision rate, access delay and device energy consumption \cite{Layarandomaccess2014}.

The LTE RA procedure employed in legacy LTE systems is not efficient to support MTC devices mainly due to the following main reasons \cite{Centenaro2017comparison}.
\begin{enumerate}
\item Because of the limited number of available preambles for the contention-based RA procedure, the massive number of concurrent transmissions of the same preambles would cause the overload of the RA procedure both in the uplink and the downlink and this will result in high collision probability, access failure rate and the access delay.
\item To support a huge number of access requests, additional downlink resources need to be allocated since each RAR message for one MTC device consists of $56$ bits.
\item Even after an MTC device becomes successful in the RA procedure, the signalling overhead degrades the overall system efficiency since the size of the upload data payload from the MTC device is significantly smaller than the traditional cellular terminals.
\end{enumerate}

\subsection{Adaptation of RA Procedure for MTC devices}
It should be noted that the RACH in the RA procedure is related to two different channels, namely,  PRACH and PDCCH as illustrated in \cite{Kim2017random}. A single PRACH consists of six physical RBs and has a bandwidth of $1.08$ MHz. Over the whole system bandwidth, a maximum of $6$ RACHs can be deployed for time-division multiplexing with one RACH for frequency-division multiplexing. However, while sending RAR in the downlink channel, only a single PDCCH is responsible for handling multiple PRACHs. Although this is not a problem in the conventional HTC devices, this becomes a serious problem in MTC devices due to their hardware limitations in terms of their capacity to listen to the wideband PDCCH. In general, a low-cost MTC device consists of a single RF interface operating with $1.4$ MHz bandwidth. To address this issue, the 3GPP has proposed Enhanced-PDCCH (EPDCCH) with the narrow bandwidth of $1.4$ MHz for low-cost MTC devices and  each PRACH has a dedicated NB EPDCCH. In this modified RACH structure adapted for low-cost MTC devices, the RA requests from the devices are distributed across multiple NB channels, thus reducing the congestion caused due to wideband nature of PDCCH in the conventional RA structure. Despite this enhanced RACH structure designed for low-cost MTC devices, the capacity of this RACH structure is not sufficient to handle the massive number of RA requests coming from the ever-increasing number of devices.

\begin{table*}
\caption{\small{Comparison of two main cellular IoT (LTE-M and NB-IoT) technologies}}
	\centering
\begin{tabular}{|l|l|l|}
\hline
Feature/parameter & LTE-M &  NB-IoT  \\
\hline
Channel bandwidth & 1.4 MHz    & 180 KHz \\
Transmission mode & HD-FDD/FDD/TDD  & HD-FDD \\
Peak data rate & 375 kbps (HD-FDD), 1 Mbps (TDD)  & $~$50 kbps (HD-FDD) \\
Latency & 50-100 ms & 1.5-10 seconds \\
Noise figure & 9 dB (uplink), 5 dB (downlink) & 5 dB (uplink), 3 dB (downlink) \\
Maximum coupling loss (MCL)  & 155.7 dB & 164 dB \\
Modes of operation & In-band & Inband, guard-band and standalone \\
Power consumption & Best at medium data rates  & Best at very low data rates \\
Mobility support & Full mobility   & No connected mobility \\
Voice over LTE support & Yes    & No \\
\hline
\end{tabular}
	\vspace{-15 pt}
\label{tab: diffLTEMnarribandIoT}
\end{table*}

Towards addressing the problem of RACH overload in the cellular IoT systems, several methods have been proposed in the literature \cite{Kim2017random}. From the perspective that whether the device or the eNodeB employs the solution, the existing schemes can be broadly categorized into push-based and pull-based. In the first approach, the RA requests are controlled from the device-side while in the pull-based approach, the contention in the RA procedure is controlled from the eNodeB. Besides, there are some strict separation schemes and soft separation schemes to concurrently support both the HTC and MTC traffic in LTE-A networks \cite{BIRAL20151,Layarandomaccess2014}. The strict separation schemes mainly comprise of the following: (i) resource separation: orthogonal allocation of resources between HTC and MTC traffic and dynamic shifting of resources among two classes, (ii) slotted access methods which define access cycles including the RA slots dedicated to the MTC device access, and (iii) pull-based scheme in which the MTC devices are allowed to access the PRACH only upon being paged by the corresponding eNodeB. On the other hand, soft-separation schemes include the following: (i) backoff tuning which assigns longer back-off intervals to the MTC devices which do not succeed during the preamble transmission of the RA procedure and (ii) Access Class Barring (ACB) scheme. A brief description of various existing and emerging solutions for the RAN congestion problem is provided in Section \ref{sec:_sec5}.

Most of the existing MTC related works focus on the BS load balancing, radio resource management and the grouping of MTC devices, and only a few studies have been conducted in optimizing the access control of massive requests from the MTC devices \cite{Huang2017M2M}. The incoming requests from the MTC devices can be categorized into delay-sensitive and delay-tolerant based on the delay tolerance level of the underlying applications and the aggregator/BS can be equipped with two queues with one having higher priority over the other in order to deal with the two traffic classes. The criteria used for defining delay tolerant and delay sensitive may differ from one scenario to another \cite{Huang2017M2M}.

The RA delay is one of the important aspects to be considered while designing RA techniques for mMTC in the existing cellular networks. In this regard, authors in \cite{KoseogluLOWER} derived lower bounds for the LTE-A RA delay by considering uniformly distributed and Beta-distributed traffic arrivals and analyzed the effect of frequency of RA opportunities and the number of preambles. It has been shown that the RA delay can be reduced by several orders of magnitude by effectively tuning these system parameters.

As briefly highlighted in Section \ref{sec:_sec1}, cellular IoT standards mainly comprise of two categories, namely, LTE-M and NB-IoT, which are described in the following subsections. Also, in Table \ref{tab: diffLTEMnarribandIoT}, we highlight the key differences between these two technologies optimized to provide cellular connectivity to IoT devices \cite{DIGIcellular,Itayatraisemiconductor}. For the detailed differences among LTE-M, NB-IoT and legacy LTE in terms of supported features and functionalities for different uplink and downlink physical channels, interested readers may refer to \cite{Elsaadany2017cellularLTEA}.

\subsection{LTE-M: Key Features and Channel Access Mechanisms}
\label{sec:_sec31}
While looking at the history of MTC, the first generation of a full featured MTC device emerged in 3GPP Release R12. In this release R12, the 3GPP has defined the category 0, i.e., CAT-0 for the low-cost MTC operation \cite{Alvarino2016overview}. In the subsequent releases, the efforts to incorporate mMTC devices continued and LTE release 13 (R13) in 2016 introduced two special categories, namely, CAT-M (also called LTE-M) for MTC and CAT-N for the NB-IoT to support various features of MTC/IoT applications. LTE Rel-14 enhancements were completed in June 2017, and the improvements under the Rel-15 are ongoing and are expected to be released by June 2018.

With respect to Cat-1 category which was the lowest UE category in LTE Release 11 from the perspective of transmission capability (peak rate of $10$ Mbps in the downlink and $5$ Mbps in the uplink), Cat-0 devices have a reduced complexity of about $50$ \% and have a reduced transmission rate of $1$ Mbps for both the downlink and the uplink \cite{Elsaadany2017cellularLTEA}. Also, Cat-0 category enables the use of only one receiver antenna with a maximum receiver bandwidth of $20$ MHz and supports FDD half-duplex operation with relaxed switching time, eliminating the need of dual receiver chains and duplex filters for low cost MTC devices, respectively. In the subsequent LTE releases after the introduction of LTE-M in release $13$, several new features have been added. The key features of LTE-M in different releases are included in the following \cite{Ratasuk2017LTEM}.
\begin{enumerate}
\item \textbf{Release-13}:  The main features included in this release include Coverage Enhancement (CE) mode A/B, bandwidth limited operations ($1.4$ MHz), half-duplex support, in-band operation mode, RRC connection, data transmission via a control plane, mobility support and eDRX.
\item \textbf{Release-14}: The key features incorporated in this version include multi-cast support with single-cell
point-to-multipoint, positioning enhancements such as enhanced-cell ID requirements and observed time difference of arrival support, larger channel PDSCH/PUSCH bandwidth (up to $5$ \& $20$ MHz), Voice over LTE enhancements, and support for HARQ-ACK bundling and inter-frequency measurements.
\item \textbf{Release-15}: The main features of this release are reduced latency and power consumption, lower UE power class, improved spectral efficiency, improved load control of idle UEs, eDRX enhancements and support for higher UE velocity.
\end{enumerate}

The four-stage message handshake procedure followed in the current LTE standard results in very high overhead for most of the IoT devices since the packets transmitted by the resource-constrained IoT devices are quite short as compared to the conventional cellular packets \cite{Layarandomaccess2014}. In this regard, several approaches are being investigated to design efficient channel access mechanisms to support MTC in the existing cellular systems. One approach investigated in the literature is to follow ALOHA-like immediate access without any reservation \cite{Andreev2013efficient}. Although this scheme completely eliminates the channel reservation phase and provides very low latency, the system throughput is limited by the slotted ALOHA capacity of $1/e$. Besides, another approach is to utilize a preamble-initiated contention-based mechanism in which the nodes transmit a randomly selected preamble to reserve a time/frequency resource \cite{Koseoglu2017}. In contrast to the conventional RACH procedure followed in LTE,  this method eliminates Message 3 and Message 4 of the four-stage message handshake procedure and the data is transmitted on the RB specified in the RAR message, thus significantly lowering the delay. However, if two or more nodes choose the same set of preambles for the RA request, the collisions occur which are detected by the lack of Acknowledgement (ACK) message.

From the performance analysis carried out in \cite{Koseoglu2017}, it is shown that the preamble-initiated access achieves $86$ \% more capacity in comparison to both the conventional LTE access mechanism and ALOHA-like immediate transmission scheme for small data packet transmissions in IoT application scenarios. However, in terms of delay, the ALOHA-like scheme reduces the delay by about $62$ \% for low traffic loads in comparison to the preamble initiated access and by about $77$ \% as compared to the conventional LTE access mechanism.

To capture the signature of the LTE signal, an MTC device will need to receive the synchronization signals which occupy $6$ RBs of the eNodeB's bandwidth \cite{Elsaadany2017cellularLTEA}. Although decoding PDCCH becomes impossible due to the bandwidth limitation (only $1.4$ MHz) of MTC devices, the enhanced version EPDCCH, which uses only one RB, is a good candidate, but is not sufficient for the required coverage enhancement \cite{Huenhanced2015}. However, increasing its bandwidth to $6$ RBS in the $1.4$ MHz bandwidth on the MTC device along with the repetition will provide the good coverage of about $-14$ dB  and the EPDCCH also supports beamforming to enhance the coverage \cite{Wang2016cellular}. Therefore, $6$ RBs, i.e., one narrowband is usually used as the basic unit for the MTC bandwidth \cite{Elsaadany2017cellularLTEA}.

The main changes incorporated in the physical layer operation of LTE-M as compared to the legacy LTE are briefly described below \cite{Elsaadany2017cellularLTEA}.
\begin{enumerate}
\item \textbf{Frequency hopping}: Due to narrow-bandwidth and a single receiver chain at the MTC device, the benefits due to spatial diversity and frequency diversity are not available. To compensate for the performance loss caused due to frequency diversity, the concept of frequency hopping, which allows MTC transmissions to hop from one NB channel to another, is employed. The challenges associated with frequency hopping in MTC devices include the need of retuning the RF chain, and the prior knowledge of the hopping pattern at the eNodeB and the device.
\item \textbf{Repetitions}: To achieve sufficient link budget in the downlink for the coverage enhancement, repeated copies of the same signal are transmitted over time to boost the link performance via time diversity. The main issue involved with this repeated transmission strategy is the requirement of increased decoding time, i.e., latency, demanding for longer wake-up time for the MTC device.
\item \textbf{MTC Physical RACH (MPRACH) and Physical Downlink Control Channel (MPDCCH)}: To compensate for the additional path-loss caused due to the extended coverage for the MTC device, the PRACH of legacy LTE needs to be modified. For this, frequency diversity and repetitions need to applied to the MPRACH to achieve the required diversity. Furthermore, additional features such as defining downlink control formats and enhancing control channel assignment procedure need to be added to support frequency hopping and repetitions in MPDCCH.
\item \textbf{MTC search spaces}: In order to reduce the number of decoding trials by the devices in the LTE systems, each MTC device can be assigned only a defined search space area of the whole control region to be monitored. In contrast to the legacy Enhanced PDCCH, there are mainly two classes of search spaces in MPDCCH, namely, device-specific search space and common search space.
\item \textbf{MTC Downlink Control Information (DCI) formats}: To reduce blind decoding iterations, i.e., device complexity as well as to facilitate the use of frequency hopping, repetition and enhanced coverage, three different DCI formats have been defined for uplink grant, downlink scheduling and paging in MTC devices.
\end{enumerate}

\subsection{Narrowband IoT: Key Features and Channel Access Mechanisms}
\label{sec:_sec32}
To address various challenges of supporting MTC devices in cellular IoT networks specified in Section \ref{sec:_sec22}, the 3GPP has proposed the concept of NB-IoT in its Release 13 \cite{Maldonado2017NBIoT}. The main objectives behind the NB-IoT concept include providing better indoor coverage and support to a massive number of low-throughput devices, with low power consumption and relaxed delay requirements \cite{3GppTR2015}. To accomplish these objectives, the NB-IoT follows the procedures of optimizing control plane and user plane of Cellular-IoT (CIoT) evolved packet system towards reducing the signalling overhead for small data packet transmissions \cite{3GppTS2017GPRS}.

For both the uplink and downlink operations, NB-IoT can operate with an effective narrowband operation of $180$ kHz bandwidth corresponding to one RB in the LTE network. In the downlink of an NB-IoT system, Orthogonal Frequency-Division Multiple Access (OFDMA) is employed with the subcarrier spacing of $15$ kHz over $12$ sub-carriers while in the uplink, both single tone and multi-tones are supported (single-tone with the subcarrier spacing of either $3.75$ kHz or $15$ kHz) \cite{Yuuplinkscheduling}. The NB-IoT usually can be operated in the following three operation modes \cite{Boisguene2017survey,Maldonado2017NBIoT}.
\begin{enumerate}
\item \textbf{In-band operation}: This mode of operation utilizes the RBs within an LTE carrier by reserving one RB for the NB-IoT system.
\item \textbf{Guard band operation}: This mode uses the unused resources within the guard band of the LTE carriers while ensuring that this does not affect the normal capacity of the LTE carrier.
\item \textbf{Stand-alone operation}: This mode utilizes the re-farmed GSM low band already existing in several countries ($700$ MHz, $800$ MHz, and $900$ MHz) \cite{Boisguene2017survey}.
\end{enumerate}

The NB-IoT technology provides greater flexibility for the deployment of IoT devices in different applications such as smart city, smart home, smart metering and smart agriculture by reusing the existing network architectures. The main requirements for the NB-IoT system include the following \cite{Boisguene2017survey}.
\begin{enumerate}
\item \textbf{Low power consumption}: NB-IoT systems utilize the power saving mode and eDRX to maximize the battery life.
\item \textbf{Low channel bandwidth}: Due to low channel bandwidth of $200$ kHz ($180$ kHz plus guard bands), GSM channel re-farming is applicable for NB-IoT systems since a single NB-IoT channel can utilize one GSM/GPRS channel.
\item \textbf{Low cost for the end-device}: Due to low channel bandwidth of $200$ kHz, the front-end and digitizer of NB-IoT receivers are much simpler than that of the existing LTE-based systems operating on the bandwidth of $1.4$ MHz, thus leading to low-complexity (cheaper) devices.
\item \textbf{Low deployment cost}: Besides the device cost, due to the capability of reusing existing GSM bands, the deployment cost for the network operators is significantly reduced.
\item \textbf{Extended coverage}: NB-IoT can provide about ten times better coverage area compared to the legacy GPRS systems as it can be achieve the additional $20$ dB link budget gain.
\item \textbf{Support for massive number of connections}: Due to improved coverage and low channel bandwidth, it can support significantly higher number of MTC devices.
\end{enumerate}

The main signals and channels involved in the downlink of an NB-IoT system are Narrowband Primary Synchronization Signal (NPSS), Narrowband Secondary Synchronization Signal (NSSS), Narrowband Physical Broadcast Channel
(NPBCH), Narrowband Reference Signal (NRS), Narrowband Physical Downlink Control Channel
(NPDCCH) and Narrowband Physical Downlink Shared Channel (NPDSCH) \cite{Wangprimer}. Out of these, NPSS and NSSS are used by an NB-IoT device to carry out cell search procedure including cell identity detection, and frequency and time synchronization. The NPBCH includes the master information block while the NRS is used to provide phase reference required for the demodulation of downlink signals. Similarly, NPDCCH includes the scheduling information for both the uplink and downlink data channels while the NPDSCH carries various information such as system information, paging message, RAR message and also data from the higher layers.

Besides, the uplink transmission scheduling of devices in the NB-IoT mainly comprises of Narrowband PRACH (NPRACH) and Narrowband PUSCH (NPUSCH) \cite{Yuuplinkscheduling}. Out of these, NPRACH corresponds to the time-frequency resource used to transmit RA preambles and the NPUSCH is used for carrying the uplink data. The differences of the above-mentioned uplink and downlink channels from the legacy LTE systems are highlighted in \cite{Wangprimer}. The key technique employed by an NB-IoT system to obtain enhanced coverage with low complexity is repetition, which utilizes the repeated transmission of both data transmission and the involved control signalling transmission \cite{Yuuplinkscheduling}.

The RA procedure in the NB-IoT system is responsible for establishing a radio link during the initial access, for scheduling the transmission  requests and to achieve uplink synchronization among the NB-IoT devices \cite{Wangprimer}. Three different types of NPRACH resource can be configured by assigning separate repetition values for a basic RA preamble to serve the devices belonging to different coverage classes with different ranges of path loss. The device estimates its coverage level by measuring the downlink received signal power, and then the device transmits an RA preamble in the NPRACH resources configured for the estimated coverage level. The configuration of NPRACH resources is made flexible in a time-frequency resource grid to enable the deployment of NB-IoT systems in different scenarios.

\section{Transmission Scheduling for MTC Systems with QoS Support}
\label{sec:_sec4}
Most of the models used to analyze the capacity of wireless systems are based on physical layer models and they can not capture the link-layer QoS requirements such as bounds on the delay \cite{Dapeng2003}. Therefore, physical-layer only models are not suitable for QoS support mechanisms such as resource reservation and admission control.  Furthermore, in contrast to the wired links, it is challenging to guarantee QoS requirements in wireless systems due to low reliability, multi-path fading, co-channel interference and time-varying capacities. In order to incorporate complex QoS requirements into account, it is important to understand the queuing behavior of the connections and to capture the QoS requirements while characterizing the performance of packet-switching based wireless networks. For analyzing the queuing behavior, the characterization of source traffic as well as the services is an important aspect to be considered \cite{Dapeng2003}.

Due to distinct QoS requirements of MTC devices, it is crucial to provide QoS support for MTC devices in future wireless networks. For example, non-real time MTC applications such as data transmissions aim to enhance the reliability of transmission and do not have strict delay constraint. Whereas, real-time MTC applications such as video surveillance/demand, the important QoS metrics are strict latency and data rate requirements rather than high spectral efficiency \cite{Ghavimi2017uplink}. To meet the QoS requirements of different network applications highlighted in Section \ref{sec:_sec24}, it is crucial to design efficient radio resource allocation algorithms for the MTC devices in the uplink while considering the constraints on the available radio spectrum.

One of the potential candidate platforms to support MTC devices is the LTE-A standard and the 3GPP has been working on several enhancements of LTE-A standard towards this direction. The 3GPP uses Single Carrier Frequency Division Multiple Access (SC-FDMA) \cite{Myung2006} as the multiple access scheme in the uplink of LTE cellular networks due to its main advantage of low Peak-to-Average Power Ratio (PAPR) as compared to that of the OFDMA. Due to this feature, the reduced requirements on the processing power and battery are suitable for the resource-constrained MTC devices \cite{Ghavimi2017uplink}. However, the allocation of RBs in the SC-FDMA  becomes complex as compared to that in the OFDMA scheme due to the sequential transmission of the RBs in the SC-FDMA in contrast to the transmission of orthogonal RBs in the OFDMA-based systems.

The minimum resource unit used for scheduling downlink and uplink transmissions in the LTE-A based cellular systems is referred to as an RB. Each RB comprises of $12$ sub-carriers with each sub-carrier having the bandwidth of $180$ kHz in the frequency domain and one sub-frame of $1$ ms duration in the time domain  \cite{Hasanrandom2013}. The RB can be considered as a time-frequency resource in which an UE performs RA and each RA slot comprises of the bandwidth equivalent to the $6$ RBs, i.e., $1.08$ MHz and its duration in the time domain is $1$ ms. In LTE-based cellular networks, the eNodeB broadcasts the periodicity of the RA slots by means of a variable referred to as the Physical RACH (PRACH) configuration Index \cite{Layarandomaccess2014}, and subsequently, the MTC devices and the legacy cellular users can perform RA by using the PRACH channel. Even though the data size from the MTC devices is significantly small, the massive number of devices attempt to concurrently communicate over the same radio channel, thus leading to the network overload problem \cite{Hasanrandom2013}. In contrast to the conventional HTC services such as multimedia for which the packet arrival periods range from $10$ ms to $40$ ms, the packet arrival periods in MTC applications may range from $10$ ms to several minutes \cite{Lien2011massive}.

\subsection{Framework for Performance Analysis with QoS Support}
In this subsection, we present a mathematical framework to carry out the performance analysis of mMTC systems with QoS support in terms of the effective capacity, effective SNR and the estimated number of MTC devices. For this analysis, we consider the uplink in a single-cell of 3GPP LTE-A networks serving multiple MTC devices with the SC-FDMA scheme. This scenario can also be studied in conjunction with the legacy cellular/HTC users as in \cite{Aijaz2014uplink}, however, herein, we deal only the case of MTC devices since we are interested in providing QoS support for MTC devices while maximizing some network performance metric subject to the constraints on the available radio spectrum. In practice, these devices can be grouped based on the employed transmission protocols and QoS requirements, and can be deployed on the cluster basis by using different wireless technologies such as WiFi, Bluetooth and Zigbee \cite{Ghavimi2017uplink}. Furthermore, MTC devices can communicate to the eNodeB via an MTC gateway and the total available RBs can be divided between the access link (MTC devices to the MTC gateway) and the backhaul link (from the MTC gateway to the eNodeB) in the time domain as considered in \cite{Aijaz2014uplink}.

Let us assume that there are $M$ number of total MTC devices in the coverage area of the eNodeB, indexed by the set $\mathcal{M}=\{1, \ldots ,m, \ldots , M\}$ and there are $L$ number of available RBs, indexed by the set $\mathcal{L}=\{1, \ldots ,l, \ldots , L\}$. We assume Poisson distribution for the traffic arrival rate of the MTC devices and block fading wireless channel between MTC devices and the eNodeB/gateway as in \cite{Ghavimi2017uplink}. Also, we assume that channel coherence time is greater than the Transmission Time Interval (TTI) and the channel gain remains constant during a TTI.

To incorporate QoS requirements of MTC devices into the problem formulation, one way is to define a QoS exponent for each MTC device and to introduce this exponent in the definition of system capacity. Let $\theta_m$ denote the QoS exponent of the $m$th MTC device  indicating a steady-state delay violation probability of the $m$th M2M device. Considering a queue of infinite buffer size required due to a constant arrival rate $\lambda$, the delay violation probability is given by \cite{Ghavimi2017uplink}
\begin{equation}
\delta=\mathrm{Pr}(d_m>d_{\mathrm{max}}) \approx \phi_m (\lambda) e^{-\theta_m} d_{\mathrm{max}},
\label{delayvoilation}
\end{equation} where $\mathrm{Pr}(\ldotp)$ denotes the probability operation, $d_m$ represents the delay experienced by a source packet of the $m$th MTC device, $d_{\mathrm{max}}$ is a delay bound, and $\phi_m(\lambda)= \mathrm{Pr}(d_m>0)$ indicates the probability of non-empty buffer. In this formulation, the pair of ($\phi_m(\lambda), \theta_m(\lambda)$) can be used to characterize the link from the $m$th device to the gateway/eNodeB.

In contrast to the conventional physical layer-based capacity, we define the effective capacity to take the link-layer QoS requirements into account. The effective capacity \cite{Dapeng2003} is defined as the maximum constant arrival rate that a given service process can support to guarantee a QoS requirement specified by $\theta$ and can be defined for the $m$th MTC device as
\begin{equation}
R_e^m(\theta_m)=-\frac{1}{\theta_m} \mathrm{ln} E[e^{-\theta_m R_m}],
\label{effcapacity}
\end{equation} where $R_e^m$ denotes the effective capacity for the $m$th MTC device, $\theta_m$ represents the statistical QoS exponent of the $m$th MTC device, $E(\ldotp)$ denotes the expectation, and $R_m$ is the data rate of the $m$th MTC device. In order to guarantee a QoS requirement of $\theta_m$ for the $m$th MTC device, the following condition should be satisfied \cite{Ghavimi2017uplink}
\begin{equation}
R_e^m(\theta_m) \geq \lambda_m,
\label{QoSguarantee}
\end{equation} where $\lambda_m$ is the traffic arrival rate for the $m$th MTC device. By solving (\ref{QoSguarantee}), one can obtain $\theta_m$. Subsequently, by using the Shannon's capacity formula, the maximum achievable transmission rate for the $m$th MTC device can be expressed as
\begin{equation}
R_m=B \mathrm{log}_2 (1+\gamma_m)=B \mathrm{log}_2 \left(1+\frac{P_m |h_m|^2}{\sigma_n^2} \right),
\label{achievablerate}
\end{equation} where $B$ is the bandwidth of each RB, $P_m$ is transmission power of the $m$th MTC device, $|h_m|^2$ is the channel gain, $\sigma_n^2$ is the Additive White Gaussian Noise (AWGN) power, and $\gamma_m=\frac{P_m |h_m|^2}{\sigma_n^2}$ is the SNR for the $m$th MTC device.

Although the value of $\theta_m$ can be derived by finding the probability density function of $\lambda_m$ corresponding to (\ref{achievablerate}) and by subsequently solving (\ref{QoSguarantee}), the evaluation process is quite complex. Another approach is to use an intuition from \cite{Dapeng2003} and to obtain the QoS exponent $\theta(\lambda)$ in the following way
\begin{equation}
\theta(\lambda)=\frac{\phi(\lambda)\lambda}{\lambda \tau_s(\lambda)+E[Q(t)]},
\label{QoSguarcal}
\end{equation} where $Q(t)$ is the length of a queue at time $t$ and $\tau_s$ denotes the
the average remaining service time of a packet being served. The detailed methodology to estimate the value of $\theta(\lambda)$ in (\ref{QoSguarcal}) for the considered MTC scenario has been illustrated in \cite{Ghavimi2017uplink}.

Despite the significant benefits of SC-FDMA in terms of power and battery requirements, there arise some restrictions for uplink resource allocation (RB and power allocations) while employing SC-FDMA in the uplink \cite{Myung2006}. The main aspects to be considered include: (i) a single RB can only be allocated to at most one user, (ii) multiple RBs allocated to a single user should be adjacent, and (iii) the transmit power on all the RBs allocated to a user should be equal. Let us assume that the set of RBs $\mathcal{L}_m$ is allocated to the $m$th MTC device in the current TTI, then the achievable rate (upper bound) from (\ref{achievablerate}) in terms of effective SNR can be written as
\begin{equation}
R_m=B. L_m \mathrm{log}_2 (1+\gamma_{\mathrm{eff},m}),
\label{achievablerateeff}
\end{equation} where $L_m=|\mathcal{L}_m|$ denotes the cardinality of the set $\mathcal{L}_m$ and $\gamma_{\mathrm{eff},m}$ denotes the effective SNR for the $m$th MTC device. Since each data symbol is spread over the whole bandwidth in SC-FDMA transmission, the effective SNR can be computed as an average of SNRs over the allocated set of RBs to a particular MTC device as follows
\begin{equation}
\gamma_{\mathrm{eff},m}=\frac{1}{L_m}\sum_{l \in \mathcal{L}_m} \gamma_{m,l},
\label{effectiverate}
\end{equation} where $\gamma_{m,l}$ is the SNR of the $m$th device for the $l$th RB.

Another aspect to be considered is how to effectively design the medium access scheme to support the massive number of devices. One approach is to determine the optimal size of the Random Access Window (RAW) based on the estimated number of MTC devices in the following way \cite{Parkenhancement2014}. If there are idle slots available at the RAW, the eNodeB/access point can estimate the number of devices for the uplink access by using suitable estimation techniques such as maximum likelihood estimation method. Let $˜I$ be the measured number of idle slots in the uplink RAW, $L_{\mathrm{UL}}$ be the number of slots of the uplink RAW and $N_{\mathrm{UL}}$ be the number of devices for the uplink access. When $N_{\mathrm{UL}}$ devices contend in $L_{\mathrm{UL}}$, the probability of selecting a slot by a device for the uplink access becomes $\frac{1} {L_{\mathrm{UL}}}$ and the corresponding complementary probability is $(1-\frac{1}{L_{\mathrm{UL}}})$. Thus, the idle probability that no devices for the uplink access transmit the power save poll message, by which the device requests for the downlink data or the ACK frame from the eNodeB, is $p_{\mathrm{idle}} = (1-1/L_{\mathrm{UL}})^{N_{\mathrm{UL}}}$ and the probability $p_{\mathrm{idle}}$ is estimated as: $\hat{p}_{\mathrm{idle}}=\frac{I}{L_{\mathrm{UL}}}$. Subsequently, the estimated number of devices for the uplink access by utilizing the aforementioned idle probability can be calculated as
\begin{equation}
{N}_{\mathrm{UL}}=\frac{\mathrm{log}(\hat{p}_{\mathrm{idle}})}{\mathrm{log}(1-\frac{1}{L_{\mathrm{UL}}})}.
\label{eqn:numberuplink}
\end{equation}

On the other hand, the existing packet schedulers are mainly designed for a specific wireless system such as LTE and do not fully capture the heterogeneous characteristics of ultra-dense IoT networks. In this regard, authors in \cite{KumarMilcom2016} proposed delay-efficient joint packet scheduling and subcarrier assignment by considering the classification of the uplink MTC traffic aggregated at the MTC aggregator into multiple classes based on traffic features such as packet size, arrival rate and delay requirements. By employing an MTC specific traffic model, the incoming data from the sensors at the aggregator is categorized either as ED or PU types and the delay requirements of these PU and ED traffic types are mapped onto sigmoidal and step utility functions, respectively. In addition, in order to ensure that the packets transmitted by an MTC device are within the delay budget, authors in \cite{Afrin2013performance} introduced a new MAC element, called Packet Age, with which the device informs the scheduler about the waiting time of the oldest packet in the device buffer along with the buffer size specified in the buffer status report.

\begin{table*}
\caption{\small{Recent research works towards supporting small data packet transmissions in wireless networks}}
\centering
\begin{tabular}{|l|l|l|}
\hline
\textbf{Main Theme} & \textbf{Applicable systems} &  \textbf{References}  \\
\hline
Optimization of pilot overhead & IoT Sensor networks & \cite{Mousaei2017pilot} \\
Design of air interface and waveforms & Multicarrier 5G systems  & \cite{Schaich2014waveform} \\
Coding and modulation schemes & IoT Sensor networks & \cite{Yoomodcoding} \\
Non-orthogonal multiple access & MTC scenarios & \cite{Sun2018short} \\
Minimization of the core network signalling & 5G cellular network & \cite{Aziz2016IEEE} \\
Joint encoding of grouped messages & Wireless broadcast channel & \cite{Trillingsgaard2017} \\
Autonomous transmission mode & Delay tolerant IoT/MTC scenarios & \cite{Balachandran2016delay} \\
Receiver algorithms to enhance the reception quality & 5G wireless networks & \cite{Leepacketstructure} \\
Energy and information outage performance analysis & Wireless powered network & \cite{Yang2018wireless} \\
Exploitation of frequency diversity to enhance reliability & Tactile Internet & \cite{Sheglobecom2016} \\
\hline
\end{tabular}
	\vspace{-15 pt}
\label{tab: shortpackettable}
\end{table*}

\subsection{Short Data Packet Transmission and Associated Issues}
\label{sec:_sec41}
In this subsection, we briefly discuss various issues related to short data packet transmission in IoT systems along with its information theoretic perspective. One of the emerging areas in the MTC systems is ultra-reliable and low-latency communications, also known as mission-critical MTC. Some of the applications of mission-critical MTC are industrial control, intelligent transportation systems and smart grids for power distribution automation \cite{Dahlman2014}. As an example, industrial control applications may need to transmit about $100$ bits within $100$ $\mu$seconds
with $10^{-9}$ PER \cite{Johansson2015}.

Existing wireless systems are designed to support the conventional HTC traffic having long packet sizes and each packet consists of information payload and the control information (metadata) which usually contains various information about logical addresses, packet initiation and termination, synchronization and security. As compared to the transmission of long packets, the transmission of short packets in the wireless IoT systems differs mainly in the following two ways \cite{Durisi2016IEEEproc}. First, existing transmission techniques are based on the assumption that the metadata is negligible as compared to the size of the information payload. However, this assumption does not apply to the transmission of short packets since the metadata size becomes no longer negligible, resulting in the need of highly efficient encoding schemes. Secondly, for the case of long packets, there exist channel codes which enable the reconstruction of information payload with high probability. The thermal noise and channel distortions average out for the case of long packets due to the law of large numbers, however, this averaging does not occur for the case of short packets and the classical law of large numbers is not applicable for mMTC applications, resulting in the need of new information theoretic principles. In this regard, authors in \cite{Durisi2016IEEEproc} discussed various information theoretic approaches to characterize the transmission of short packets in wireless communication systems and applied these principles on the transmission of short packets in various channels such as a two-way channel, a downlink broadcast channel and the uplink RACH. In addition, authors in \cite{Durisi2016shortpacket} investigated the tradeoffs among reliability, throughput, and latency for the transmission of information over multiple-antenna Rayleigh block-fading channels.

In the above context, several recent works have investigated different physical layer approaches to support small packet transmissions in mMTC/IoT environment by considering their specific characteristics, which are briefly reviewed in the following paragraphs. Also, in Table \ref{tab: shortpackettable}, we list the recent research works towards supporting small data packet transmission in wireless networks with their main themes and applicable systems.

In short data packet transmissions, one effective way of enhancing the packet transmission efficiency is to optimize the pilot overhead \cite{Mousaei2017pilot}. However, most of the existing pilot overhead optimization works considering the objective of ergodic channel capacity maximization are based on the assumption of sufficiently large packet length resulting in small packet error probability, which is not suitable for short-packet transmission. In this regard, authors in \cite{Mousaei2017pilot} formulated the optimization of approximate achievable rate as a function of block length, pilot length and error probability, and illustrated the importance of considering packet size and error probability while optimizing pilot overhead via numerical results.

Another potential enabling approach to support short packet transmissions in IoT/mMTC environments is to design suitable transmit waveforms. In the IoT environment, there are some applications with very small packet sizes such as the data transmitted from sensors like temperature sensors while some other applications such as car to car and car to infrastructure communications demand very fast response time. In order to support these diverse set of applications, the 5G and beyond air interface should be able to support transmissions with very small air interface latency enabled by very short transmission frames \cite{Schaich2014waveform}. In this regard, it is important to investigate suitable waveforms for supporting diverse applications in an IoT environment. Among potential multi-carrier waveform contenders such as filtered Cyclic Prefix-OFDM, Filter bank multi-carrier and Universal Filtered Multi-Carrier (UFMC), authors in \cite{Schaich2014waveform} concluded UFMC as the best choice for IoT systems with short burst transmissions due to its several benefits in terms of supporting fast Time Division Duplex (TDD) switching, low latency modes, low energy consumption and small packet transmission.

In addition, investigating suitable coding and modulation schemes is crucial to achieve high energy efficiency for short-packet transmissions having low-duty cycles. Due to lower duty cycle, time synchronization and phase coherency for short-packet transmissions become non-trivial. Furthermore, because of short packet length, a large coding can not be achieved as in the conventional voice or data networks. Moreover, the overhead required to maintain time synchronization and phase coherency becomes significantly large while using the conventional coherent modulation schemes \cite{Yoomodcoding}. In this regard, the time synchronization overhead can be reduced by employing either non-coherent modulation/demodulation schemes such as Phase-Shift Keying (PSK) with differential encoding or orthogonal modulations. To this end, authors in \cite{Yoomodcoding} analyzed the tradeoff between energy efficiency and bandwidth in non-coherent short packet transmission systems.

Towards addressing the problem of scalability and efficient connectivity to the massive number of MTC devices with short packets, the NOMA scheme is considered as one candidate multiple access solution \cite{Sun2018short}. Due to its benefit of improving fairness and spectral efficiency for low-latency transmission with respect to the orthogonal multiple access technique, it is considered promising for IoT applications. In this regard, authors in \cite{Sun2018short} analyzed a trade-off among the transmission
rate, transmission delay (in terms of block-length) and decoding error probability by considering a two user downlink NoMA system with finite block-length constraints.

Furthermore, towards minimizing the signalling overhead for small data packet transmissions, authors in \cite{Aziz2016IEEE} proposed a framework based on 5G RAN controlled user-centric mobility, in which an anchor node is allocated and updated for each end-device and it maintains the connection of the device to the core network within its coverage area. In this approach, an user centric area is dynamically allocated so that an user/device can move freely and communicate with the network without any state transitions signaling required in the existing connection management schemes with RRC protocols \cite{Aziz2016IEEE}.

Moreover, while implementing multiple-antenna based interference suppression in IoT systems with small data packet structures, the insufficient training period may result in severe degradation in the estimation of the desired channel and interference covariance matrix. The main challenge here is to obtain the reliable channel estimation without significantly affecting the data transmission duration, i.e., to balance the trade-off between the pilot training period and the data transmission period. In this regard, authors in \cite{Leepacketstructure} investigated an efficient receiver structure which can exploit information received during the data transmission period to enhance the reception quality for the short packet transmissions. Moreover, in the context of energy harvesting networks, authors in \cite{Yang2018wireless} provided a comprehensive analysis of the backscatter wireless powered communication with sporadic short data packets by using a stochastic geometry framework.

\section{Solutions for RAN Congestion Problem in Cellular IoT networks}
\label{sec:_sec5}
In this section, we first review several existing techniques towards addressing RAN congestion problem in cellular IoT networks, and then discuss some emerging solutions.
\subsection{Existing Techniques}
Towards addressing the RAN congestion problem in LTE-based cellular networks, 3GPP has specified the following six different solutions of LTE RA congestion \cite{3GPPRAN}:: (i) ACB, (ii) MTC-Specific backoff, (iii) dynamic resource allocation, (iv) Slotted random access, (v) separate RA resources and (vi) pull-based RA. In the following, we briefly describe the principles of these techniques along with other related solutions in the literature \cite{Ali2017LTE,Fawal2017overload,3GPPRAN}. Also, in Table \ref{tab: summaryRAschmesMTC}, we provide the list of these schemes along with their main principles and the corresponding references.
\begin{table*}
\caption{\small{Summary of existing solutions for RAN congestion problem in cellular IoT networks}}
	\centering
\begin{tabular}{|l|l|l|}
\hline
\textbf{RA Scheme} & \textbf{Main principle} &  \textbf{References}  \\
\hline
Back-off based scheme & In the occurrence of collisions, devices retransmit after an MTC-specific backoff period.   & \cite{Wu2013FASA,3GPPRAN} \\ \hline
Access Class Barring (ACB)   & Multiple access classes of devices are assigned different access probabilities.   &  \cite{BIRAL20151}  \\ \hline
Extended Access Barrier (EAB)  &  A certain access class of devices is barred from the channel access.  &  \cite{Cheng2015modeling} \\ \hline
Cooperative ACB scheme & ACB parameter is designed by many BSs in a collaborative way. & \cite{Lien2012cooperative} \\ \hline
Dynamic ACB   & ACB parameter is updated dynamically based on previous collisions.   & \cite{Duan2013IEEE} \\ \hline
Prioritized RA with dynamic ACB & Utilizes class-dependent back-offs and dynamic ACB. & \cite{Cheng2011IEEE,LINPARDA2014} \\ \hline
Dynamic resource allocation & Congestion level is predicted at the BS and additional RACH resources are allocated dynamically. & \cite{Ali2017LTE} \\ \hline
Slotted random access & Each device is assigned a dedicated RA slot and is allowed to perform RA only in that slot. & \cite{3GPPRAN} \\ \hline
Separation of RA Resources & Available preambles or RA slots are divided between MTC and HTC devices. & \cite{3GPPRAN,Ali2017LTE}  \\ \hline
Pull-based/Paging-based  & Devices perform RA attempts only after receiving paging messages from the BS.  & \cite{3GPPRAN}  \\ \hline
Group-based RA  & RACH resources are allocated on the basis of groups formed based on some defined criterion.    & \cite{Farhadi2013group,Chuang2015group} \\ \hline
Code-Expanded RA  & RA codewords are generated and each device utilizes a set of preambles in each RA slot. & \cite{Pratascode2012} \\
\hline
\end{tabular}
	\vspace{-15 pt}
\label{tab: summaryRAschmesMTC}
\end{table*}

\begin{enumerate}
\item \textbf{Back-off based scheme}: In this scheme, the devices retransmit after a backoff time if they encounter a collision. This scheme
    can enhance the network performance under a low congestion level, however, becomes problematic in high-level congestions \cite{Wu2013FASA}
    This is the conventional approach followed in contention-based wireless networks and 3GPP has suggested several improvements to solve the
    RAN overload problem. To support MTC devices in the existing cellular networks, 3GPP has suggested the use of MTC-specific backoff scheme in which MTC devices are subject to a larger backoff interval than the HTC devices \cite{3GPPRAN}.

\item \textbf{Access Class Barring (ACB) scheme}:  This scheme classifies the contending devices into multiple access classes with different access probabilities and each class is assigned to an ACB parameter and an access barring timer \cite{3GPPRAN}. The working principle of the ACB scheme can be summarized in the following way. First, the BS broadcasts the ACB parameter, i.e., $0 \leq p \leq 1$ to the MTC devices and each MTC device trying to connect to the BS generates a random number $0 \leq r \leq 1$ uniformly. Then, the MTC
    device is allowed to start the RA procedure if $r < p$ and otherwise, the access to that particular device is barred and the device has to wait for a random backoff time determined based on the barring duration of that class. Therefore, by controlling the ACB parameter $p$, the BS can control the stabilization of RA to optimize some network performance metrics such as throughput \cite{BIRAL20151}.

    However, in the presence of severe congestion caused by the presence of massive number of IoT devices, the value of $p$ may be set to be extremely low, thus leading to the intolerable delay. Also, the ACB scheme is not suitable for event-driven applications in which the contention may arise within a short time duration \cite{Layarandomaccess2014}. Furthermore, the operating parameters such as transmission probability should be adjusted based on the network status and estimating the number of devices/network status becomes challenging due to highly bursty traffic in event-driven MTC communications \cite{Wu2013FASA}.

    To address the above drawbacks of the ACB scheme, there have been some attempts in the literature. Some of the important ones include the following.
    \begin{enumerate}
    \item Extended Access Barrier (EAB) scheme \cite{Cheng2015modeling}: In this scheme, devices belonging to a certain access class are
        barred from the channel access to provide some form of service differentiation \cite{Cheng2015modeling}. The operation of
        this scheme depends mainly on the following two factors: (i) the sets of barred access classes and (ii) the time of turning EAB on or off. The larger the set of barred access classes, the higher will be the access success probability which comes at the cost of increased mean access
        delay. Furthermore, the timing of turning EAB on or off relies on the input network load which is proportional to the number of
        devices concurrently accessing the network.

    \item Cooperative ACB scheme \cite{Lien2012cooperative}: In this scheme, ACB parameters are determined across the network jointly by
        many BSs interconnected via the $X_2$ interface \cite{Lien2012cooperative} rather than individually calculated at each BS. This scheme
        aims to balance the traffic load among the BSs in a heterogeneous multi-tier cellular network with the objective of reducing the
        congestion level and also improving the access delay.

    \item Dynamic ACB scheme \cite{Duan2013IEEE}: In this approach, the ACB parameters are updated dynamically based on the information
        about the number of collisions in the previous time slots.

    \item Prioritized RA with dynamic ACB \cite{Cheng2011IEEE,LINPARDA2014}: This scheme pre-allocates the RACH
        resources for different classes of MTC devices with class-dependent backoff procedures and reduces the number of concurrent requests for the RACH by employing the dynamic ACB method.
\end{enumerate}

\item \textbf{Dynamic Resource Allocation}: In this scheme, the BS predicts the congestion level of the access network overload caused due to MTC devices and allocates additional RACH resources dynamically in the time domain or frequency domain or both for the MTC devices \cite{Ali2017LTE,3GPPRAN}. However, the allocation of more radio resources for RACH will reduce the radio resources available for the traffic channels and this trade-off needs to considered while implementing this solution.

\item \textbf{Slotted Random Access}: In this method, a dedicated RA opportunity is provided to each MTC device and is allowed to perform RA only in the access slot allocated to it \cite{3GPPRAN}. However, in ultra-dense IoT scenarios, this method will result in very high access delay since the duration for each RA cycle will be significantly large.

\item \textbf{Separation of RA Resources}: In this approach, different RACHs are allocated to MTC devices and HTC devices to avoid the impact of RA congestion on HTC devices. The separation of RA resources can be done either by splitting the available preambles into MTC and HTC subsets or by allocating different RA slots for MTC and HTC devices \cite{3GPPRAN,Ali2017LTE}.

\item \textbf{Pull-based/Paging-based scheme}: All the schemes described above fall under the category of push-based approach in which RA attempts are done randomly by the devices. However, in the pull-based method, the devices perform RA attempts only after receiving paging messages from the BS. To reduce the number of paging load in this approach, a number of MTC devices can be paged together by following a group paging method \cite{3GPPRAN}.

\item \textbf{Group-based RA Scheme} \cite{Farhadi2013group,Chuang2015group}: The MTC devices can be grouped based on some criterion such as having similar QoS/delay requirements and being deployed in a specific geographical region, and RACH resources can be allocated on the group-basis to reduce the access network congestion. In a group-based RA scheme proposed in \cite{Farhadi2013group}, the devices within one paging group are partitioned into different access groups based on some criterion and only one device within each access group, called group delegate/header, is made responsible for communicating with the BS. The group delegate can be decided by the BS based on some suitable metrics such as transmission power and channel conditions.

   Another grouping approach is to divide the cell coverage area into a different spatial groups and to enable the use of same preambles at the same RA slot by the MTC devices located in different groups if the minimum distance of these MTC devices is larger than the multi-path delay spread \cite{Jang2014comm,Kim2017random}. While sending the RAR message, the BS sends distinct RARs to all the detected devices having different Timing Alignment (TA) values even if they use the same preamble during the RA preamble transmission phase.

\item \textbf{Code-Expanded RA Scheme} \cite{Pratascode2012}: In this approach, the contention space is expanded to the code domain by creating the RA codewords. While initiating an RA attempt, each device sends a set of preambles over the given RA slots instead of transmitting only a single preamble at any random RA slot, thus creating a set of preambles in each RA slot.

\item \textbf{Tree-based RA scheme} \cite{Gursu2017hybrid,Madueno1433}: This category of RA schemes utilizes the tree-based algorithms such as q-ary tree splitting technique \cite{Janssen2000analysis}, which rely on the
    utilization of feedback obtained after each contention attempt \cite{Madueno1433}. This RA scheme is mostly used to address the contention problem caused due to synchronized arrivals of the traffic from a large number of MTC devices. Furthermore, the combination of collision avoidance techniques such as access barring can be used in combination with tree-based collision resolution in order to form a hybrid RA scheme \cite{Gursu2017hybrid}.
\end{enumerate}

\subsection{Emerging Solutions}
\label{sec:_sec51}
In the following, we provide some of the emerging research directions to address RAN congestion problem in wireless IoT networks.
\subsubsection{Learning-based Techniques}
Recently, learning-based techniques have received important attention in addressing the RAN congestion problem in cellular IoT networks. In this direction, an RL scheme has been applied in \cite{Hasanrandom2013} for the selection of an appropriate BS for the MTC devices  with the objective of avoiding access network congestion and minimizing the packet delay. In addition, a Q-learning based access scheme has been studied in \cite{Belloeuropean2014} to support MTC traffic in the existing cellular networks. In this Q-learning based approach, MTC devices learn to avoid collisions among each other without involving a central entity and after the learning convergence, each MTC device gets a unique RACH slot. Furthermore, authors in \cite{Mohammed2015base} applied a Q-learning based unsupervised algorithm in order to select an appropriate BS for MTC devices on the basis of QoS parameters in dynamic network traffic conditions. Moreover, a hierarchical stochastic learning algorithm has been applied in \cite{Ruandelay2017} to enable each device to make the access decision with the assistance
of common control information broadcasted from the BS. In addition, in \cite{Moon2017access}, a Q-learning algorithm has been applied to dynamically adjust the value of a barring factor to be allocated to the MTC device in the ACB scheme.

\subsubsection{Distributed Queueing}
The existing approaches to enhance the RACH performance are mainly based on the ALOHA-type mechanisms which suffer from some level of inefficiency, instability and uncertainty in the outcome of the access opportunities to be assigned to the devices \cite{Layarandomaccess2014}. In this regard, one promising approach is Distributed Queuing Collision Avoidance (DQCA) \cite{Alonso2005} which is a distributed and always-stable high performance protocol. This MAC protocol behaves as an RA mechanism for low traffic load and switches automatically and smoothly to a reservation scheme when the traffic volume increases \cite{Alonso2005,KartsakliIEEE2008}. More specifically, the DCQA protocol utilizes two distributed queues which operate in parallel \cite{KartsakliIEEE2008}. The first queue, called collision resolution queue, deals with the resolution of access-request signal collisions, while the other queue, called data transmission queue, helps to manage the data transmission. The main features of this protocol are the following \cite{Alonso2005}.
\begin{enumerate}
\item It can eliminate back-off periods and avoid collisions in data packet transmissions.
\item Its performance is independent of the number of transmitting nodes.
\item It is stable independently of the traffic conditions.
\item As compared to other centralized or distributed MAC, it utilizes very few bits for signaling operation purposes.
\end{enumerate}

Furthermore, authors in \cite{Samir2016} proposed a distributed queuing-based access protocol for LTE with the objective of improving the RA performance for MTC systems without altering the existing frame structure of LTE systems.
The original version of distributed queueing protocol envisions orthogonal mini-slots as access opportunities and its implementation requires a change in the LTE frame structure since the preambles in LTE are not orthogonal in time domain.
To address this issue, the authors in \cite{Samir2016} considered the distribution of allocated preambles for MTC devices among $N_g$ virtual groups, with each virtual group having $N_p$ number of preambles and each preamble being equivalent to one mini-slot considered in the original distributed queueing protocol.

\subsubsection{SDN and Virtualization for RAN Management}
To support differentiated MTC services with diverse QoS requirements, a physical wireless network can be abstracted and sliced into multiple virtual networks by employing suitable network function virtualization techniques \cite{Lirandomaccess2017}. On the other hand, Software Defined Networking (SDN) enables the separation of a data plane and a control plane, and provides the capability of programming a network via a centralized controller. Due to the global view of the underlying network, an SDN controller enables the efficient management of radio resources in dynamic network traffic and channel conditions. In cellular IoT networks, a hypervisor can divide the physical network into different IoT networks based on device classes and functionalities, and the SDN controller can dynamically allocate the available radio resources among these virtual networks to meet the QoS requirements of different IoT networks. Among these virtual networks, each MTC device can select one of the virtual networks to access to the physical network while meeting its connection requirements \cite{Li2017energy}. In addition to radio resources, it is also possible to enhance the utilization of other network resources such as computing, caching and networking resources \cite{Lirandomaccess2017}.

\section{Learning-Assisted Solutions for RAN Congestion Problem in Cellular IoT Networks}
\label{sec:_sec6}
\subsection{Advantages of Learning Techniques in Wireless Communications}
\label{sec:_sec61}
\begin{table*}
\caption{\small{Use cases for the applications of learning techniques in ultra-dense cellular systems}}
	\centering
\renewcommand{\arraystretch}{0.95}
\begin{tabular}{|l|l|l|}
\hline
\textbf{Use Case} & \textbf{Sub-cases/related description} & \textbf{Reference} \\ \hline
RACH congestion minimization  & Learning to find the unique access slot for each MTC device & \cite{Belloeuropean2014} \\
& Learning to adapt an ACB parameter & \cite{Moon2017access}  \\
& Learning to associate MTC devices with the best eNodeBs & \cite{Hasanrandom2013,Mohammed2015base} \\ \hline
Autonomous adaptive resource allocation & Learning to find the existence of the critical delay-sensitive messages  & \cite{Park2016resource} \\
 \hline
Dynamic spectrum sharing & Learning to predict the occupancy status of radio channels & \cite{Liglobecomlearning}  \\  \hline
Learning-assisted edge-side processing & Extracting the useful information from the raw sensor data   & \cite{Portelli2017leveraging} \\
& at the edge devices to reduce the communication burden &  \cite{Portelli2017leveraging} \\ \hline
Selection of a suitable RAT &  Learning to select a suitable RAT among different RAT technologies & \cite{Rakovic2010novel}\\
& under network conditions and user preferences constraints &  \\ \hline
Network traffic control  & Learning to control network traffic to enhance computational efficiency and scalability  &  \cite{Katodeeplearning2017} \\ \hline
Adaptation of transmission parameters & Learning link quality/reliability to adapt parameters such as MCS and transmission slot &
\cite{DanielsGlobecom2008,Daniels2009,Yun2010reinforcement,Desai2016energy,Saishankar2017} \\
\hline
Data analytics & To extract user activity/mobility patterns, temporal, spatial and social correlations & \cite{Mohammadienabling} \\
\hline
Provisioning of personalized services & To learn network contexts for a global view of communications, computing   & \cite{Chen2018datafriven} \\
& and caching resources & \\
\hline
\end{tabular}
	\vspace{-15 pt}
\label{tab: UsecasesMLappl}
\end{table*}

The main questions this section attempts to answer are why learning techniques are important in wireless communication systems, which parameters to learn and for what purposes. First, we present the main advantages of learning techniques in wireless communications systems in general, and then discuss why learning techniques are needed on the top of the conventional link adaptation techniques. Subsequently, we discuss various parameters which can be learnt by using learning techniques in different application scenarios.

As highlighted earlier in Section \ref{sec:_sec13}, the number of configurable system parameters has increased significantly from one cellular generation to the next one. For example, the number of configurable parameters has increased to about $1500$ in a 4G node from about $500$ in a 2G node and from about $1000$ in a 3G node, and it is predicted to be around $2000$ in a 5G node \cite{Imran2014challenges,Liintelligent2017}. In this regard, the process of optimizing these reconfigurable parameters in 5G and beyond systems becomes extremely complex and performing self-configuration, self-optimization and self-healing operations will be challenging. Also, emerging ultra-dense networks will need to observe environmental variations, learn uncertainties, plan response actions and configure the network parameters effectively to handle these operations. To this end,  emerging ML techniques could bring potential benefits in efficient handling of these operations. The main role of learning techniques include learning the system variations/parameter uncertainties, classifying the involved cases/issues, predicting the future results/challenges and investigating potential solutions/actions \cite{Liintelligent2017}.

Wireless systems utilize link adaptation techniques to adapt the physical layer parameters such as modulation and coding scheme based on the reliability/quality of the communication link. In practice, different applications such as
wireless video broadcasting and VoIP demand for different reliability constraints. Before performing this link adaptation, the reliability of a wireless link in the form of some metrics such as PER is predicted for each set of physical layer parameters, needs to be predicted \cite{Daniels2009}. During the link adaptation process, there arises a tradeoff between data rate and reliability since the PER calculated for a set of physical layer parameters is in general inversely related to the data rate. In order to predict the reliability, existing systems form explicit input/output models of a wireless channel and then analyze the performance of physical layer for each set of the parameters.

However, due to the increasing trend of using multiple antennas, wideband signals and a number of advanced signal processing algorithms, the above-mentioned reliability prediction process becomes extremely complex \cite{Daniels2009} and the prediction of PER with good accuracy becomes difficult in practice \cite{Blankenshiplink2004}. Furthermore, due to a significantly large number of environmental parameters such as channel state information, signal power, noise variance, non-Gaussian noise effect, transceiver hardware impairments such as power amplifier non-linearity and quantization error, it becomes challenging to provide the near-optimal/optimal tuning of the transmission parameters to achieve the efficient link adaptation \cite{Yun2010reinforcement}. The severity of this problem greatly increases in ultra-dense networks due to the involvement of various agents and system parameters such as Signal to Interference plus Noise Ratio (SINR) mismatch in ultra-dense small cell networks \cite{Park2017broadband}, and therefore, the link adaptation in emerging ultra-dense networks becomes extremely challenging.  Also, the existing link adaptation systems are localized to individual links and small coverage areas, and do not take into account of the consequences on other systems from the system-level perspective.

In order to make the link/system adaptation more flexible and efficient, existing works have applied ML techniques in different settings \cite{DanielsGlobecom2008,Daniels2009,Yun2010reinforcement,Desai2016energy,Saishankar2017}. The contribution in \cite{Daniels2009} investigated an online learning framework for the link adaptation by using a modified $k$ nearest neighbor (kNN) algorithm to learn the mappings between the channel conditions and PER values for all possible Modulation and Coding Schemes (MCS) supported by the system. In this online learning framework, when a new packet is delivered, the predicted PER associated with each MCS is calculated by using the kNN algorithm and the best MCS is selected. After a packet is transmitted, the packet is stored as a prior data of the selected MCS for future prediction purpose. Although the accuracy of PER prediction becomes more accurate with the increase in the number of packet transmissions, this kNN-based approach requires to store all the previous samples and has higher time complexity, not suitable for a real time operation \cite{Daniels2009}. In this regard, the authors in \cite{Yun2010reinforcement} proposed an online kernelized support vector regression method which can work with the minimal memory size and has low computational complexity while providing a comparable performance to those of the existing algorithms.

Learning techniques are expected to provide significant benefits by adaptively learning numerous parameters in various application scenarios as listed below. Also, we provide a brief summary of these use cases along with the corresponding references in Table \ref{tab: UsecasesMLappl}.
\begin{enumerate}
\item Learning to exploit an unique RA slot for each MTC device within the considered transmission frame in a way that concurrent transmissions in the same RACH opportunity can be avoided \cite{Belloeuropean2014}.
\item Learning to adapt an access control parameter, i.e., access barring factor for the RACH congestion \cite{Moon2017access}.
\item Learning to associate MTC devices with suitable BSs/eNodeBs with the objective minimizing overall access network congestion \cite{Hasanrandom2013,Mohammed2015base}.
\item Learning the existence of delay sensitive/critical messages by IoT devices in heterogeneous ultra-dense IoT networks so that enough resources can be dynamically allocated and critical information can be successfully transmitted to the BS/eNodeB/aggregator as soon as they are generated \cite{Park2016resource}. Based on the learned information about critical messages, IoT devices can collectively adjust their uplink transmission parameters such as orthogonal codes, transmission slot period, periodicity of transmission and the received power for performing autonomous resource allocation and the coordination for the usage of available codes.
\item Learning the radio spectrum by dynamic spectrum sharing among the systems/nodes in a collaborative manner to predict the occupancy status of radio channels \cite{Liglobecomlearning}.
\item Learning the relationship of the contextual information (related to the surrounding radio environment) collected from IoT sensors to extract knowledge and to predict the future context at the edge devices \cite{Portelli2017leveraging}. Instead of transferring all the raw contextual data to the network/cloud-centre, only the inferred knowledge can be transferred, thus reducing the communication burden. This approach deals with pushing learning intelligence from the network to the distributed edge devices having heterogeneous computing abilities.
\item Learning for the selection of Radio Access Technology (RAT) while performing vertical handovers among heterogeneous networks having different RAT technologies under the constraints of network conditions and user preferences, which can be designed on the device-side, network side or in a hybrid manner \cite{Rakovic2010novel}.
\item Learning for network traffic control (such as routing) in ultra-dense heterogeneous networks to alleviate the issues of computational efficiency and scalability of the existing approaches \cite{Katodeeplearning2017}.
\item Learning link quality/reliability to adapt the transmission parameters (such as MCS, transmission slot, received power etc.) of a wireless link. \cite{DanielsGlobecom2008,Daniels2009,Yun2010reinforcement,Desai2016energy,Saishankar2017}
\item Learning to extract user activity/mobility patterns, temporal, spatial and social correlations from raw unstructured/semi-structured data coming from massive number of sensors.
\end{enumerate}

\begin{figure*}
	\begin{center}
		\includegraphics[width=6.0 in]{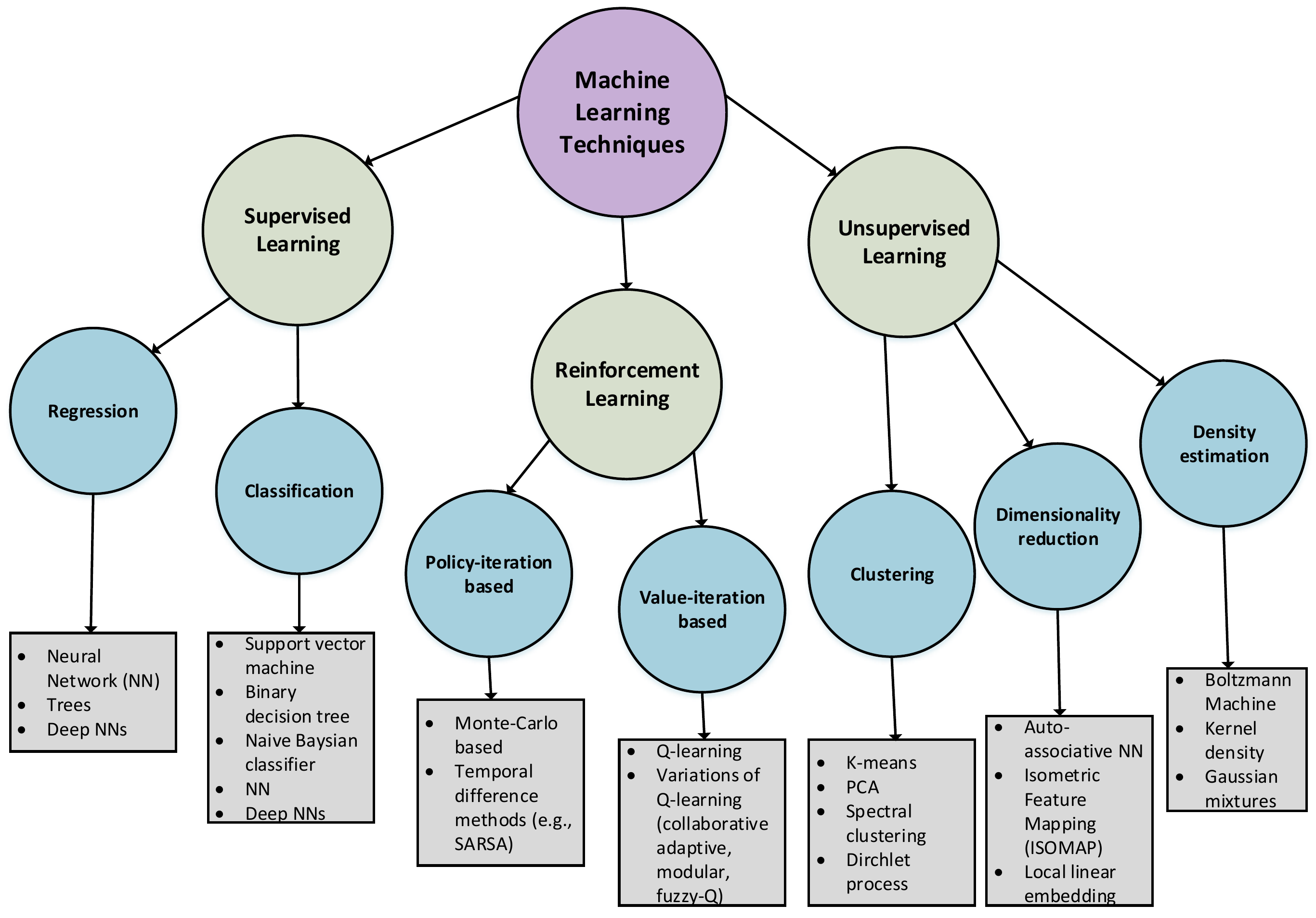}
		\caption{\footnotesize{Classification of existing machine learning techniques.}}
	\label{fig: MLclass}
	\end{center}
\vspace{-15 pt}
\end{figure*}

\subsection{Learning Techniques for IoT/MTC Systems}
\label{sec:_sec62}
The main challenges of applying learning techniques in an IoT environment include the following \cite{ParkIEEE2016}.
\begin{enumerate}
\item MTC devices have low computational capability, however, the widely-used ML techniques such as RL and decision trees can be computationally complex to implement.
\item Because of the distributed nature of IoT devices and high energy required to maintain constant communication with the BS/centralized aggregator, distributed learning needs to be investigated for an IoT environment.
\item Due to limited radio resources and energy constraints, only the limited amount of information is available at the IoT devices, and therefore, it is necessary to adapt the learning mechanisms based on the limited amount of information.
\item In some critical applications such as eHealthCare and industrial control, IoT devices need to learn quickly in order to satisfy the ultra-reliable and low-latency requirements. For this purpose, learning time should be as small as possible to quickly adjust the performance parameters.
\item To enable the harmonious coexistence of MTC and HTC systems, learning techniques should consider both the existing traffic as well as the new traffic from the MTC devices.
\end{enumerate}

Existing works have applied learning techniques in the context of MTC/IoT in the following ways: (i) adaptation of an access control parameter, i.e., access barring factor to minimize the RACH overload \cite{Moon2017access}, (ii) learning a dedicated slot within the MTC transmission frame by using an intelligent slot assignment strategy to avoid the collisions of access requests \cite{Belloeuropean2014}, (iii) BS/eNodeB selection by using Q-learning/RL techniques to minimize the access network overload \cite{Hasanrandom2013,Mohammed2015base}, and (iv) sequential learning with finite memory in order to learn transmission parameters under stringent memory and computational constraints \cite{Park2016resource,Park2016learning}.

\subsection{Overview of Existing Machine Learning Techniques}
\label{sec:_sec63}
\begin{figure}
\begin{center}
\subfigure[]{
\includegraphics[scale=.40]{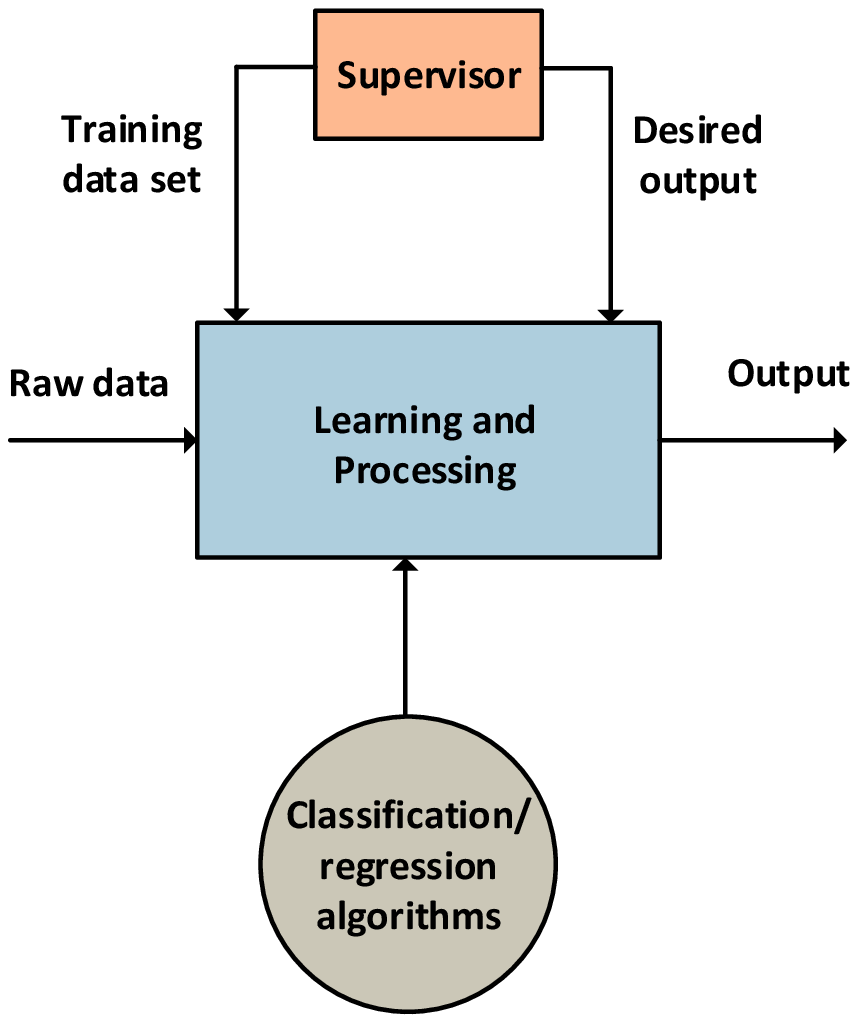}
} \subfigure[]{
\includegraphics[scale=.40]{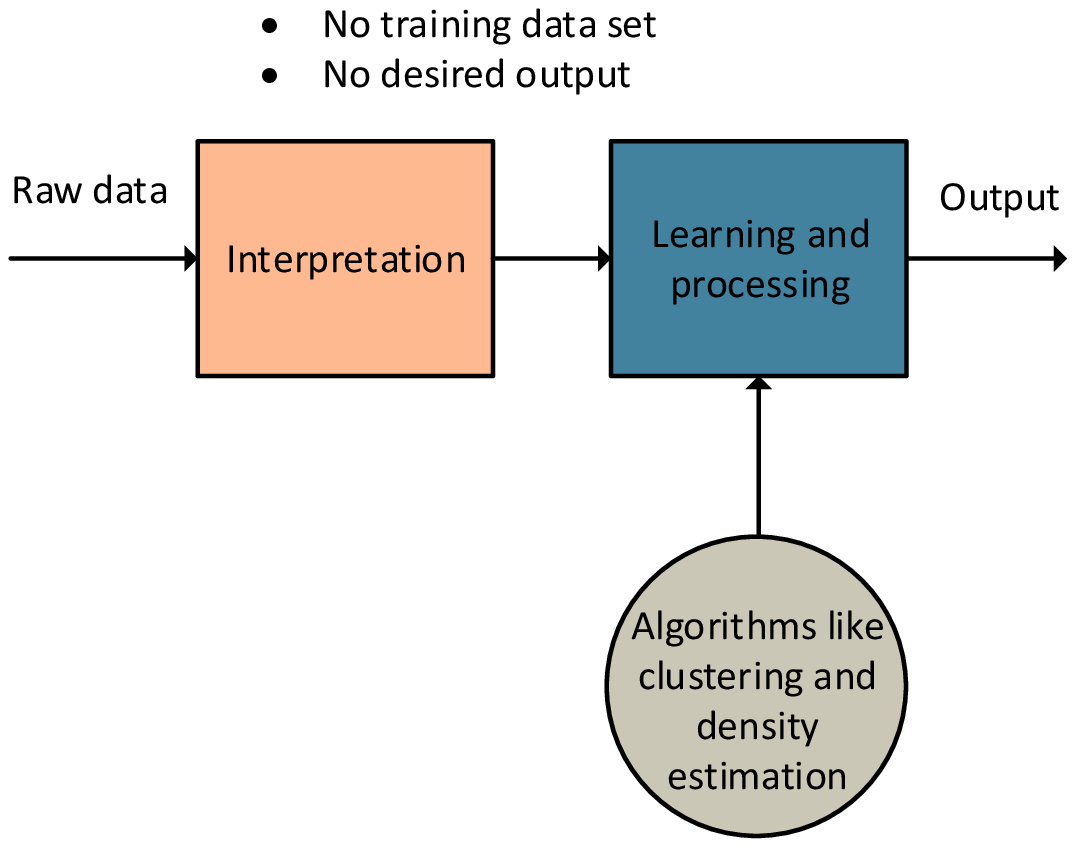}
}
\end{center}
\caption{\footnotesize{Illustrations of the principles of supervised and unsupervised learning.}}
\vspace{-10 pt}
\label{fig: superandunsuperlearning}
\end{figure}

\begin{figure}
	\begin{center}
		\includegraphics[width=3.0 in]{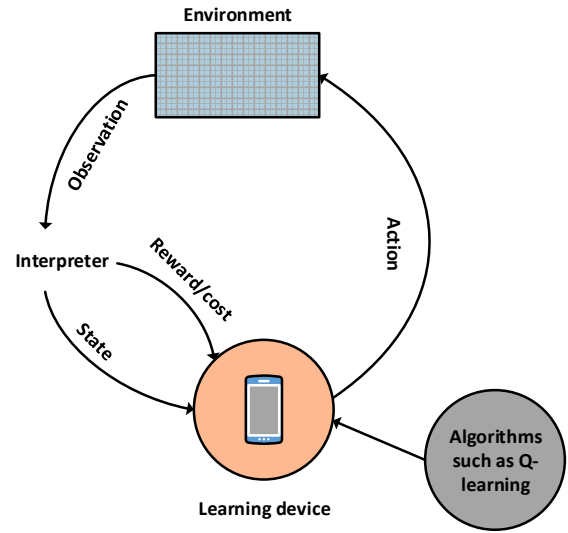}
		\caption{\footnotesize{Illustration of the principles of a reinforcement learning technique.}}
	\label{fig: Rlillus}
	\end{center}
\vspace{-15 pt}
\end{figure}

ML techniques learn necessary information either from the available data-sets or by interactions with the surrounding environments and make suitable decisions on future actions to be followed by the agents either based on some learned models or in a model-free manner. The classification of existing ML techniques in terms of different bases such as learning principles, objectives and the employed algorithm is provided in Fig. \ref{fig: MLclass} \cite{Katodeeplearning2017}. On the basis of the employed learning mechanisms, the existing ML techniques can be broadly categorized into the following \cite{Katodeeplearning2017,Alsheikh2014ML,ParkIEEE2016}: (i) supervised learning, (ii) unsupervised learning, and (iii) reinforcement learning, which are briefly described below. Figures \ref{fig: superandunsuperlearning} and \ref{fig: Rlillus} provide the illustrations of the principles of these three categories of ML techniques \cite{Ronald2018}.

The first category of ML techniques, i.e., supervised learning requires the need of training data to be labelled and the output of the algorithm needs to be already fed to the machine. Being aware of the output, the learning agent builds a model to move from the input to the output guided by the input training set. Based on the employed learning algorithms, the supervised learning techniques can be classified into \cite{Katodeeplearning2017,Sezer2018context}: Artificial Neural Networks (ANNs), Deep NNs, Bayesian Networks (BNs), Support Vector Machine (SVM), Deep Learning (DL), Case-based Reasoning (CBR), Decision Trees (DTs), K-Nearest Neighbor (KNN), Instance-based Reasoning (IBR) and Naive Bayesian Classifier. The main difficulty of applying supervised ML techniques in IoT scenarios is that they require the processing of extensive data-set to learn from the dynamic environment but IoT devices are limited in terms of computing and caching/memory resources \cite{ParkIEEE2016}.

On the other hand, the second category of ML techniques, i.e., unsupervised learning does not require the need of labels of data-sets and is more complex than supervised learning techniques in terms of the computational cost \cite{Katodeeplearning2017}. Although this learning class has not been widely used as compared to the supervised learning in the current context, it can be considered as a promising future ML paradigm since the main objective of ML is to make the learning agent capable of learning without any supervision or human intervention. Since the learning agent does not have any training data-set and the knowledge of the output, the learning process is quite complex as compared to the supervised counterpart. This learning approach divides the unlabelled heterogeneous data into smaller homogeneous sub-sets which can be easily understood and managed \cite{Sezer2018context}. Therefore, unsupervised learning techniques are mainly based on following three objectives \cite{Katodeeplearning2017}: (i) clustering, (ii) dimensionality reduction, and (iii) density estimation. For clustering-based unsupervised learning, different ML algorithms such as K-means, spectral clustering, Principal Component Analysis (PCA) and Dirchlet processes can be utilized. Similarly, unsupervised learning for dimensionality reduction can be employed by using ML algorithms like auto-associative NN, local linear embedding and Isometric Feature Mapping (ISOMAP). Furthermore, the density estimation-based unsupervised learning can be realized by using ML algorithms such as Kernel density, Gaussian mixtures, Boltzmann machine and Deep Boltzmann machine.

In order to take the advantages of supervised and unsupervised learning, there is another intermediate type of ML techniques, called semi-supervised learning technique \cite{Yoosemi2017} which uses both the labelled and unlabelled data for learning. The main objective of semi-supervised learning scheme to improve the learning accuracy by utilizing a small amount of the labelled data along with the large amount of the unlabelled data. The above-mentioned types of ML techniques have been applied mainly in the centralized framework \cite{Alsheikh2014ML}. For example, cloud-based processing can enable the operation of big data analytics to handle the massive amount of data gathered from the heterogeneous IoT sensors/devices  \cite{SKSIEEE2017}.

The third category of learning techniques, i.e., RL enables a number of agents to interact with the environment and involves an environment of states, actions to be taken by the agents, state transition functions, an immediate reward and an initial observation function \cite{ParkIEEE2016}. In this approach, an agent learns from the previous experience in the absence of the training data set. This learning mechanism deals with finding a proper balance between exploration for the random actions and exploitation of current knowledge, i.e., exploration-exploitation trade-off. The role of exploration phase in RL learning is to attempt some random actions towards searching better rewarding actions, while the exploitation phase tries to utilize the previously learned utility to maximize the reward for the agent \cite{Shah2007distributed}. Although an RL technique requires the knowledge of state transition function and it has slow convergence, its unique feature of action-reward feedback to the agents makes this suitable for the applications in IoT systems.

Besides the above-discussed three categories of ML techniques, some researchers have also considered another category of learning techniques, called as Sequential Learning (SL) \cite{ParkIEEE2016}, which helps the autonomous agents to learn the true underlying state of the environment having binary states. In this approach, the agents learn the system state in the sequence by following a given order while observing the environment and the actions or observations of previous agents, and then eventually converge to a true underlying state with repeated hypothesis testing \cite{ParkIEEE2016}. The main advantage of employing SL techniques in the IoT systems is its flexibility in terms of memory requirements since it enables the convergence of finite memory SL in the resource-constrained IoT devices \cite{Park2016learning}. However, the main drawback of the SL approach is that it relies on direct communication links among MTC devices since the information required for SL comes from other agents, thus leading to the requirement of additional network resources.

Among several RL techniques, Q-learning requires low computational resources for its implementation and does not require the knowledge of the model of the environment, thus being a suitable learning technique for the resource-constrained IoT devices \cite{Shah2007distributed}. Furthermore, it is possible to implement this technique in a distributed way. Therefore, in the following subsection, we utilize the Q-learning technique to address the problem of RACH congestion in cellular IoT networks.

\subsection{Q-learning for RACH Congestion Problem}
\label{sec:_sec64}
The main problem with the existing contention-based RA schemes is that the occurrence of collisions is unavoidable and the achievable RACH throughput in the presence of massive access requests/loads gets significantly reduced \cite{Belloeuropean2014}. For example, the maximum throughput of the widely-used slotted ALOHA technique is $e^{-1}$ $(~37\%)$. Furthermore, because of low RACH throughput and the employed backoff strategies, the aggregated traffic including both the newly generated and the retransmitted exceeds the RACH channel capacity at a certain point, thus making the system unstable. Although Slotted ALOHA can work well with the conventional cellular/HTC traffic despite the instability issue, the support of M2M traffic becomes problematic due to infrequent and massive number of access requests, thus causing the problem of RACH overload.

Learning techniques can be employed at the MTC devices in order to enable them to learn to avoid concurrent transmissions during the RACH contention period without any assistance from the central entity. After a learning technique achieves its convergence, each MTC device can get a unique dedicated slot, thus avoiding collisions among their transmissions. In the following, first, we present a framework for the RL with a single MTC device and then develop the formulation of Q-learning in the considered context.

The environment perceived by an MTC device can be usually described by a Markov Decision Process (MDP) and a finite MDP can be denoted by a tuple $<X,U,f,\rho>$, where $X$ represents the finite set of environment states, $U$ is the finite set of device actions, $f$ denotes the state transition probability function and captures the environmental dynamics, and $\rho$ is the reward function \cite{Busoniu2008Trans}. In this MDP modeling, a state parameter $x_t \in X$ indicates the characteristics of the environment at the $t$th time instance. At each time-step, the device can change its state by taking actions $u_t \in U$ and due to this action, the environmental state alters from the current state $x_t$ to the other some state $x_{t+1}$ based on the employed transition probability function $f(x_t,u_t,x_{t+1})$. During this transition, the device receives an instantaneous reward $r_{t+1} \in R$ with the defined function $\rho$, i.e., $r_{t+1}=\rho(x_t,u_t,x_{t+1})$.

Given a state, the device chooses its action based on its policy $\pi$ and the policy can be either stochastic or deterministic. Each time the device applies a policy, it accumulates the rewards from the environment, resulting in the return of $\sum_{l=0}^L \gamma^l r_{l+1}$, where $\gamma \in [0 \hspace {5 pt} 1]$ is the discount factor which provides more weights on the immediate rewards and $L$ denotes the length of one episode, which denotes the time period after which the state is reset for an episodic MDP \cite{deepRL2017}. For the case of non-episodic MDP, $L= \infty$. At each time-step, the learning device aims to maximize the expected discounted return in the long-term, i.e., long-term reward, given by \cite{deepRL2017,Busoniu2008Trans}
\begin{equation}\label{}
R_t=E\left(\sum_{l=0}^L \gamma^l r_{t+l+1}\right),
\end{equation} where $E$ denotes the expectation operator and this is taken over probability state transitions, i.e., dynamics of the considered environment.

In this RL process, the learning device attempts to maximize its long-term performance, while only receiving feedback about its immediate action, i.e., one-step performance. To achieve this, the device needs to compute an optimal action-value function, known as a Q-function. Given a certain policy $\pi$, the expected return of a state-action pair, $Q^{\pi}(x,u)$ is given by
\begin{equation}\label{}
Q^{\pi}(x,u)=E\left(\sum_{l=0}^L \gamma^l r_{t+l+1}|x_t=x,u_t=u,\pi \right).
\end{equation} Subsequently, the optimal Q-function can be written as: $Q^{*}(x,u)=\max_{\pi}Q^{\pi}(x,u)$ and it satisfies the well-known Bellman optimality equation in the following way.
\begin{eqnarray}\label{}
Q^{*}(x,u)=\sum_{x' \in X} f(x,u,x')[\rho(x,u,x') \nonumber \\
+\gamma \max_{u'}Q^{*}(x',u')], x \in X, u \in U.
\label{eq: Bellman}
\end{eqnarray}

One simplest way of choosing a future action by a device is to employ a greedy policy which selects the action with the highest Q-value at every state as follows:  $\pi(x)=\arg \max_u Q(x,u)$.

Among different ML techniques listed in Fig. \ref{fig: MLclass}, Q-learning is simple to implement and uses a look-up table with the Q-values representing the utilities for state-action pairs. The utility of taking an action `u' in a state `x' is denoted by $Q(x,u)$ and can be calculated as the expected value of the sum of immediate reward and discounted utility of the resulting state after executing the action `u'. In the Q-learning process, the current estimate of $Q^{*}$ value, i.e., $Q_t(x_t,u_t)$ is updated by using the estimated samples given by the right-hand side of (\ref{eq: Bellman}), which are computed by relating with the actual experience from the execution of the action, in the form of the pairs of subsequent states $(x_t,x_{t+1})$ and the rewards $r_{t+1}$. In this way, this Q-learning process transforms (\ref{eq: Bellman}) to the following iterative procedure \cite{Busoniu2008Trans}.
\begin{eqnarray}\label{}
Q_{t+1}(x_t,u_t)=Q_t(x_t,u_t)+\alpha_t[r_{t+1} \nonumber \\
+\gamma \max_{u'}Q_t(x_{t+1},u')-Q(x_t,u_t)].
\label{eq: Qlearning}
\end{eqnarray} where $\alpha_t$ denotes the learning rate applies at the $t$th time-step and the expression inside the square brackets indicates the difference between the estimates of $Q^{*}(x_t,u_t)$ at two successive time steps.

To provide an example framework for the application of Q-learning in an MTC scenario with $N$ number of devices, we consider a frame-based slotted ALOHA scheme as in \cite{ParkIEEE2016AI,Belloeuropean2014}, in which a frame is divided into $K$ number of access slots. Each MTC node has individual Q values corresponding to every slots in the frame and these values are updated based on the outcomes of transmission, i.e., success or failure. At the start, all the MTC nodes can start with zero or random Q values, learn gradually via their transmissions, and then finally reach to the optimal transmission strategy after finding unique RA slots for their transmissions.

Let $Q(i,k)$ indicate the preference of the $i$th node to transmit a packet in the $k$th RA slot. After every data transmission, the new Q value, i.e., $Q_{t+1} (i,k)$ is updated based on the previous Q value and the current reward based on the following relation
\begin{equation}\label{}
Q_{t+1}(i,k)=Q_t(i,k)+\alpha(R-Q_t(i,k)),
\end{equation} where $R$ is the current reward and $\alpha$ is the learning rate. The reward value of $R=+1$ is assigned if the transmission becomes successful, and otherwise, $R=-1$. At each instance of transmission, the node selects a slot with the highest $Q$ value and in case of two or more maximum values, a random selection approach can be applied. Regarding the selection of {$\alpha$}, the higher the value, the faster will be the convergence of $Q$ value towards the reward value $R$. However, a small value of $\alpha$ is usually preferred to enhance the robustness with respect to infrequent collisions caused by channel variations \cite{ParkIEEE2016AI}. Furthermore, although learning rate $\alpha$ is usually considered fixed in the existing works, it is typically time varying in nature, decreasing with time and each state-action pair may be associated with a different learning rate \cite{Busoniu2008Trans}.

\subsection{Exploration Strategies for Q-Learning}
\label{sec:_sec65}
Q-learning aims at finding an optimal policy to select an action in the current state and one of the design aspects for the Q-learning algorithm is to balance the exploration-exploitation tradeoff. The exploitation is performed by executing one of the actions which maximizes $Q(x,u)$ whereas the exploration is carried out by randomly selecting an action to build a better estimation of  the optimal Q-function.There are several strategies to create this balance and the widely-used three strategies are described below \cite{Tijsma2016comparing,Poole2017AI}.

\begin{enumerate}
\item \textbf{$\epsilon$-greedy strategy}: This is the most commonly used exploration strategy in which the Q-learning algorithm utilizes a parameter $0 \leq \epsilon \leq 1$ to decide on the action to follow. The algorithm chooses the action with the highest Q-value in the current state with $(1-\epsilon)$ probability and a random action with the probability $\epsilon$. The value of the parameter $\epsilon$ can be varied over time as the learning progresses. The main drawback of this approach is that it treats all the possible actions equivalently during its exploration by choosing an action uniformly from the set of possible actions.

\item \textbf{Soft-max strategy}: To address the drawback of $\epsilon$-greedy stage during the exploration phase, this soft-max strategy uses either a Gibbs or Boltzmann distribution, in which the learning device at a state $x$ selects an action $u$ with the following probability
\begin{equation}\label{}
\pi(x,u)=\frac{e^{\frac{Q(x,u)}{\tau}}}{\sum_u e^{\frac{Q(x,u)}{\tau}}},
\end{equation} where $\tau>0$ denotes the temperature parameter of the Boltzmann distribution and depicts how randomly the values are chosen. For example, $\tau=0$ case represents no exploration at all while $T \rightarrow \infty$ case reflects the scenario in which the learning device chooses the action values almost with equal probability.

\item \textbf{Optimism in the face of uncertainty}: This approach encourages exploration by assigning higher initial values to the Q-function. However, the convergence time will be increased since the estimated Q-values can be quite bad estimates of the actual Q-value and these bad estimates may last longer during the learning process. Also, in the presence of dynamic uncertainty, this technique is not useful since the exploration mostly occurs at the beginning of the learning process.
\end{enumerate}

\subsection{Performance Enhancement of Q-Learning}
\label{sec:_sec66}
In the following, we present several variants of Q-learning techniques which aim towards improving the performance of the ordinary Q-learning.
\subsubsection{Collaborative Q-learning}
In this subsection, we describe the need of collaborative Q-learning in ultra-dense IoT networks and review the related works. The traditional Q-learning algorithm which is based on single state-action may not be suitable for the multi-agent environment with multiple policies. In this regard, collaborative learning can be utilized by exploiting the overall/global objective/reward of the collective environment instead of the reward for a single learning device. Furthermore, the global reward can be utilized in addition to the individual reward to improve the performance of the existing learning schemes.

In a multi-agent environment like the case of ultra-dense IoT networks, if all agents keep mappings of their joint states and actions, this will require each learning device to maintain very large Q-value tables, whose sizes are exponential in the number of agents \cite{Limulti2011}. This becomes difficult even in the case of a single state case. For instance, when $M$ number of learning devices play the repeated game with only two actions, the size of the table becomes $2^M$. Therefore, it requires more state space, information space and action space. Furthermore, in the multi-agent case, state transitions, the instantaneous rewards and future expected return are based on the joint action of the multiple agents. In addition, instead of the individual policies, a joint policy formed by individual policies will impact the Q-function. Therefore, the design of state-action pairs or the optimal policy to maximize the overall reward becomes the joint/collaborative problem in a multi-agent environment, thus leading to the need of collaborative learning.

Some existing works have applied collaborative Q-learning in different settings \cite{Limulti2011,Portelli2017leveraging,Lamini2015oct,Kartransactions2013}. One way of realizing collaborative Q-learning is to estimate the belief of the opponent and the environment knowledge instead of the Q-value function in a way that the learning agent does not need to observe the opponents' reward and their Q learning
parameters \cite{Limulti2011}. In this regard, the authors in \cite{Rosyadi2016intelligent} applied the
collaborative Q-learning framework to optimize the waiting time in intelligent traffic control applications. Furthermore, the authors in \cite{Portelli2017leveraging} employed a collaborative ML technique at the edge computing devices with the objective of extracting the statistical relationships among the contextual information collected by the edge devices and constructing predictive models to maximize the communication efficiency. With this edge-centric learning approach, only the inferred knowledge can be transferred to the network instead of transmitting all the raw contextual information. Furthermore, the authors in \cite{Lamini2015oct} used collaborative Q-learning in finding an optimal path between any starting point and a target in a grid environment for a mobile robot. In contrast to the conventional approach of using a single Q-table, the work in \cite{Lamini2015oct} exploited the use of two Q-tables, i.e., one local Q table and another Master Q table.

Moreover, in a multi-agent environment, it is necessary for a learning agent  to keep the track of its environment, as well as other agents' actions. In such a multi-agent environment, rather than considering individual actions of the agents as in the ordinary Q-learning, the joint actions need to be considered while devising a learning strategy \cite{Ni2014multiagent}. Instead of the Q-value used in the ordinary Q-learning, a Nash Q-value function should be defined. For the $m$th learning device in a multi-agent environment, the Nash Q-value function can be defined as \cite{HuNash2003,Ni2014multiagent}
\begin{equation}\label{}
Q_m^{t+1}(x,u_1,...u_M)=(1-\alpha) Q_m^t(x,u_1,...u_M)+\alpha[r_m^t+\gamma \tilde{Q_m^t}(x')],
\label{eq: multiagent}
\end{equation} where $(u_1,...u_M)$ is a joint action, $r_m^t$ is an one-period reward for the $m$th learning device and $\tilde{Q}_m^t(x')$ denotes the payoff of the $m$th device in the state $x'$ for the chosen Nash equilibrium. The convergence of the Q-learning algorithm using the Q-value function defined in
(\ref{eq: multiagent}) becomes slower as the number of agents increases due to the resulted increase in the joint action set \cite{Ni2014multiagent}.

\subsubsection{Situation-Aware Adaptive Q-learning}
The collisions of the RA requests caused due to concurrent transmissions of multiple RA requests in one RACH sub-frame results in higher access delay since the devices have to retransmit their RA requests. Although the average access delay can be minimized by using a higher number of RACH sub-frames, the sub-frames available for data transmission will be reduced since the sub-frames allocated for RACH procedure can not be used for the data transmission purpose \cite{Choi2011automatic}. In this regard, it is crucial to balance the tradeoff between the radio resources allocated for RACH procedure and data transmission process. Furthermore, in many cases, the network usage pattern (the number of devices/users trying to connect to the network) is time varying in nature. In this context, a network should be capable of detecting the variation in the arrival rate of the access request and should adapt the number of sub-frames to be allocated for RACH accordingly \cite{Choi2011automatic}. In addition, it is important to balance a tradeoff between exploration and exploitation while adapting the Q-learning parameters to the dynamic uncertain environment  \cite{Rahimiyan2010adaptive}.

Although Q-learning has been shown to converge and has been used in many fields including mechatronics control and robotics, it has some issues such as how to improve the convergence rate and to avoid the convergence in the local optimum \cite{Hwang2004COOPERATIVE}. To address these issues, three different parameters of the Q-learning technique, namely, learning rate $\alpha$,  discount rate $\gamma$ and temperature parameter $\tau$ in Boltzmann distribution, should be dynamically adapted based on the dynamicity of the underlying learning environment. One of the widely used methods to adapt the learning parameters in various applications is fuzzy-logic based learning, which is briefly described in the following subsection along with the related literature.

\subsubsection{Fuzzy-logic based Adaptive Q-learning}
In the Q-learning process, the Q-values are usually stored in a look-up table but this storage process becomes infeasible in practice in the presence  of a large number of state-action spaces and with the continuous state space \cite{Glorennec1997fuzzy}. Although Q-values can be stored by using feed-forward neural networks or self-organizing maps, the learning process becomes slower. In contrast, incorporation of Q-learning into fuzzy environments seems promising since fuzzy interference systems being universal approximations can be considered as good candidates to store Q-values and the prior knowledge can be provided to the fuzzy rules in order to significantly reduce the training part.

The implementation of Q-learning becomes impractical and even impossible in continuous state spaces \cite{Glorennec1997fuzzy}. In such cases, fuzzy-logic based approach helps to discretize the continuous state or action spaces into finite states by employing suitable fuzzy rules and also the speed of fuzzy-logic based Q-learning can be increased by incorporating the prior knowledge via fuzzy rules \cite{Jamshidi2015self}. In other words, Fuzzy Q-learning discretizes continuous variables by using fuzzy labels and a fuzzy rule-based inference system is employed to find an action for these discretized states \cite{IslamcooperativeQ}.

Other drawbacks with the ordinary RL are that it is difficult with the continuous states and behaviors in the real world environment due to discrete set of actions and spaces, and it becomes complex to learn the problem with multiple objectives \cite{Maedafuzzysystems}. To solve these issues, fuzzy-logic based rules can be employed to tune the learning parameters of the Q-learning technique towards making it more adaptive.

In the context of cooperative fuzzy Q-learning, authors in \cite{IslamcooperativeQ} utilized this learning approach to optimize the coverage and capacity of cellular networks by adapting the tilting of vertical antennas. The employed cooperative Fuzzy Q-learning mechanisms enable cooperation among the learning agents during the exploration phase and is fully distributed in the exploitation phase. The cooperation in the exploration phase is employed by utilizing the global reward of all the considered cells instead of the local rewards belonging to individual cells. This cooperation with the help of global reward helps to speed up the exploration phase of the Q-learning process while also allowing the learning agents to exploit the learned knowledge independently while selecting their actions. Furthermore, a self-learning cooperative strategy is developed in \cite{Hwang2004COOPERATIVE} by combining adaptive Q-learning with the fuzzy method for its application in robot soccer systems.

Moreover, the contribution in \cite{Pervez2017fuzzy} recently proposed a fuzzy Q learning-based user centric backhaul-aware user association scheme in which the BSs broadcast their constraints and capabilities in terms of meeting the requirements of heterogeneous UEs in terms of the optimized bias factors. The employed fuzzy-logic based Q-learning scheme helps to dynamically adjust these bias factors based on network conditions and users' requirements in an automated and distributed manner.

Similarly, authors in  \cite{Mendil2016PIMRC} proposed a fuzzy Q-Learning based energy controller for a small cell powered by local renewable energy, local storage, and the smart grid to elongate the lifespan of the storage devices and  to minimize the electricity expenditures of the mobile operators. The employed fuzzy Q-learning based controlled can be utilized without the prior knowledge of the
mobile traffic demand, energy pricing and weather.

In order to avoid the inefficient and expensive manual tuning of cellular network parameters in 5G small cell networks, it is crucial to perform automatic configuration and optimization of the network  parameters including the handover parameters. In this regard, authors in \cite{Wudynamic2015} employed a fuzzy logic controller based dynamic fuzzy Q-Learning algorithm for mobility robustness optimization in a heterogeneous network. With the proposed dynamic fuzzy Q-learning algorithm, the system learns necessary parameter values towards optimizing the call dropping ratio and handover ratio, and it has been shown that the Q-Learning algorithm can lower the handover ratio while keeping the call-dropping ratio at the lower level. Besides, by considering various parameters such as link quality, the available bandwidth, link quality, and relative vehicle movement, authors in \cite{Wu2013flexible} proposed a fuzzy constraint Q-learning algorithm for vehicular ad-hoc networks in order to evaluate the quality of a wireless link towards finding the optimal route.

\subsubsection{Model-based Q-Learning}
The main drawback of model-free learning is the convergence time. By predicting a model about the transition state probabilities, the performance of the learning techniques could be improved. In time-varying dynamic scenarios, it becomes advantageous to predict the environmental dynamics in the centralized entities such as cloud-center and to utilize the corresponding model to enhance the performance of distributed Q-learning at the resource-constrained IoT devices. In such a  collaborative cloud-edge processing framework \cite{SKSIEEE2017}, the predicted model at the cloud-center can be communicated to the edge-side to improve the performance of distributed learning at the edge-side of the network.

Existing works have used model-based Q-learning in different applications such as robotic applications \cite{Dengcombining2017,Sharma2017model} and wireless channel allocation \cite{El-Alfy2001modelbased}.

\section{Research Challenges and Future Directions}
\label{sec:_sec7}
\subsection{Design Issues for Low-Power MTC Device Transmission Schemes}
Current protocols designed for cellular IoT such as NB-IoT and LTE-M are based on the assumption of the low-latency requirement. This requirement results in significant cost in terms of device price and system capacity \cite{Yangnarrwoband2017}. Besides, the main issues involved in providing cellular connectivity to the low-power MTC devices include the device battery life, system capacity, coverage and cost. Due to low cost and low capability of MTC devices, a significant compromise in the link performance needs to be made resulting in the shrinkage of the coverage area. One way of compensating this coverage loss is to utilize an extended transmission time interval, which, however, will lead to the increase in the battery consumption. Therefore, it is crucial to balance the tradeoff between energy consumption and coverage expansion while designing transmission technologies for the MTC devices.

Furthermore, transmission schemes should be able to support a significantly large number of devices within a given bandwidth while ensuring higher battery efficiency. In this regard, the concept of effective bandwidth \cite{Yangnarrwoband2017} could be utilized to achieve a good balance between the spectral efficiency for a given coverage and the corresponding transmission time. On one hand, the bandwidth allocated for a device should not be far less than the effective bandwidth to avoid the excessive transmission time. On the other hand, for a given sensitivity level of the device, it is preferred to have allocated bandwidth not more than the effective bandwidth in order to accommodate more number of devices in the saved bandwidth without significantly increasing the transmission time. Moreover, advanced transmission scheduling techniques and
low signalling overhead MAC protocols need to be investigated to effectively support the massive number of MTC devices in the upcoming 5G and beyond cellular networks.

\subsection{Spectrum Issues for mMTC Systems}
Since the available usable spectrum below $6$ GHz is limited, it is crucial to investigate suitable
spectrum sharing solutions for emerging 5G and beyond systems including eMBB, URLLC and mMTC.
The most commonly discussed dynamic spectrum sharing solutions for 5G and beyond systems include interweave cognitive communications, underlay cognitive communications, carrier aggregation, Licensed Assisted Access (LAA), Licensed Shared Access (LSA) and Spectrum Access System (SAS) \cite{Sharma2018DSS}. Another option is to explore higher frequency bands such as millimeter wave (mmWave) bands. However, due to propagation related and hardware imperfections issues in mmWave bands, it becomes challenging to operate mMTC devices in mmWave bands as they are constrained in terms of resources and are mostly deployed in indoor or underground environments where propagation issues could become problematic. Therefore, for mMTC applications, the frequency bands below $6$ GHz is of more importance. In this spectrum region, there arises the issue of whether licensed or unlicensed band becomes suitable for mMTC applications as pointed out in the following.

In the spectrum region below $6$ GHz, the operation in the unlicensed band by using carrier aggregation and LAA could be a cost-effective solution for mMTC applications since these frequencies are freely available to any device. However, uncontrolled interference conditions resulted from the free access from the massive number of devices and the lack of QoS guarantees may severely limit the ability of utilizing unlicensed bands for mMTC applications \cite{JayawickramaspectrumIoT}. On the other hand, emerging techniques such as LSA and SAS provide better interference characterization due to the centralized control and seem more suitable for mMTC applications. Besides, some of the mMTC applications based on periodic data reporting such as smart metering can effectively work with the shared spectrum bands in a demand-based manner instead of exclusively assigning the portion of a licensed spectrum. Due to low bit-rate requirements of mMTC applications, the bandwidth requirement may not be significant, however, the exclusive licensed spectrum can be allocated to more demanding applications such as eMBB and URLLC. To this end, it is an important future research direction to investigate the feasibility of utilizing shared spectrum for mMTC applications.

\subsection{Traffic Characterization Issues for mMTC Systems}
Since traffic characteristics in IoT sensory networks usually depend on the application scenarios, traffic characterization is considered to be an important issue for the design and optimization of the network infrastructure. The mMTC networks may generate various types of traffic patterns such as PU, ED and streaming, and these traffic patterns may have different amplitudes, activation periods and starting times \cite{Kim2014M2M}. Besides, the data packets generated by MTC devices can be of varying sizes and bandwidth requirements. For example, the data packets generated by the temperature and humidity sensors are usually of small size in the order of bytes, whereas video monitoring devices can generate data sizes in the order of megabytes.

As highlighted in Section \ref{sec:_sec24}, the 3GPP has defined two different models for the aggregated traffic models for the MTC traffic. The first model based on uniform distribution is for the non-synchronized type of traffic whereas the second model based on Beta distribution is for the highly synchronized traffic. This aggregated traffic modeling is simpler but is less precise than the source traffic modeling which models the traffic for each MTC device separately and is more complex. In this regard, investigating suitable low-complexity and precise traffic modeling is an important future research direction.

Furthermore,  the Transmission Control Protocol (TCP) employed at the transport layer of current LTE-A networks is not efficient for MTC traffic due to several issues associated with connection set up, congestion control, data buffering and real-time applications \cite{Kim2014M2M}. Therefore, it is crucial to develop an enhanced version of TCP over LTE/LTE-A in order to accommodate the MTC traffic.

\subsection{Issues with Machine Learning in the mMTC environment}
While employing ML techniques in an mMTC environment, an important aspect to be considered is how to make ML techniques more practically realizable for the resource-constrained MTC devices. To achieve this objective, the following issues need to be considered. First, the convergence rate/learning time of the employed learning algorithm should be as small as possible and there may arise the issue of a local minimum. Since the required learning time may reduce the time for data transmission purpose, their trade-off should be designed properly. Second, in mMTC environments, the distributed implementation of the algorithm needs to be considered across multiple learning devices. Third, different parameters associated with the learning algorithms such as learning rate $\alpha$, discount rate $\gamma$ and exploration-exploitation tradeoff parameter $\epsilon$ should be adapted dynamically to enhance the performance of Q-learning algorithm in dynamic environments. Furthermore, there may arise the fairness issue while applying learning algorithms in a multi-agent environment since different devices may reach to convergence at different time intervals, thus creating different learning times for different devices. Moreover, the  heterogeneity of MTC devices need to be considered in terms of different aspects such as learning capability, cache size, data rate and delay tolerance limit.

In the above context, future works should focus on addressing the aforementioned issues to employ the ML-assisted solutions towards enabling the incorporation of MTC devices in the upcoming 5G and beyond cellular networks. Various emerging techniques described in Section \ref{sec:_sec66} such as collaborative learning, situation-aware adaptive learning, fuzzy-logic based Q-learning and model-based learning could be exploited to enhance the performance of the ML techniques in ultra-dense IoT networks involving MTC devices.

\subsection{Distributed Resource Management in Ultra-Dense IoT Networks}
In contrast to the connection-oriented design approach of the existing wireless networks while considering only the communication resources, future content-oriented networks are expected to utilize other resources such as computing and caching resources. In ultra-dense IoT networks, these communications, computing and caching resources are usually distributed across the different entities of the network including the devices, aggregator/eNodeBs and core-network/cloud-center. The coordination of these distributed resources to enhance the performance of ultra-dense IoT networks involving low-end MTC devices is an important research issue. Furthermore, in the emerging cloud-assisted IoT networks, it is advantageous to handle computationally intensive task at the cloud-center due to the availability of huge computing and storage capabilities while it becomes beneficial to handle delay-sensitive applications at the edge-side of the network. In this regard, it is an important future research direction to investigate suitable techniques for edge-cloud collaborative processing in various applications including dynamic spectrum sharing, event detection, context-aware resource allocation, live data analytics and security enhancement \cite{SKSIEEE2017}.

Moreover, various features of MTC device transmissions such as time-controlled, time tolerant, priority alarm message, infrequent transmission, group-based policing and addressing can be useful to utilize the distributed cache embedded in MTC devices. The caching capabilities distributed across heterogeneous IoT devices can enable the scheduling of sporadic transmissions from the MTC devices in order to significantly reduce the peak traffic in an IoT access network as demonstrated in \cite{Sharmacommletter2018}. This will subsequently reduce the demand for the radio resources at the peak time, thus significantly saving the radio resource cost for the telecommunication operators. In addition, distributed caching can be exploited to enable aggregate transmission for various other applications such as saving energy for low-power IoT devices and reducing signalling/protocol overhead for MTC device transmissions. Also, the physical-level cache embedded in low-end distributed MTC devices can be exploited to facilitate the implementation of a cross-layer based transmission scheduling in pushing data from the physical layer to the MAC layer at the suitable intervals.

\subsection{Device Heterogeneity and Grouping-based Transmission Schemes}
The heterogeneity of MTC devices in terms of different aspects such as computing capability, cache size, battery power, data rate and latency requirements becomes problematic for the efficient QoS provisioning in ultra-dense IoT networks. Furthermore, RAN congestion, high signalling overhead and power consumption are critical issues to be addressed in ultra-dense IoT networks. In this regard, grouping-based features of MTC transmissions such as group-based policing and group-based addressing \cite{3GPPMTCservice} can be utilized to enable the group-based operation in the mMTC environment towards alleviating various problems such as signalling congestion and power consumption. Furthermore, by implementing a group-based access authentication technique, severe signalling congestion caused by the conventional independent access authentication scheme in the existing cellular systems can be avoided \cite{Caogroupauthenticcation} and the security of emerging MTC applications can be significantly enhanced.

By grouping MTC devices on the basis of either service requirements or the physical locations of MTC devices, group-based data gathering, aggregation and reporting can be utilized in ultra-dense IoT networks including IEEE 802.11ah based systems \cite{Sabin2018ICC}. In such group-based schemes, a group header/cluster head collects the access requests, uplink data packets, and device status information from the resource-constrained MTC devices belonging to that group, and then forwards the aggregated traffic to an eNodeB/aggregator \cite{Ghavimi2015M2M}. Also, the downlink data packets and signalling messages can be relayed to the MTC devices by the group header, thus significantly reducing the radio resources required for direct communications between the devices and the eNodeB. Moreover, as the jittering constraints become challenging while transmitting small data packets from a large number of devices, this issue can be addressed by employing grouping-based resource scheduling which allocate radio resources to the specific groups having similar QoS characteristics \cite{Lien2011ubiquitous}. However, the existing device grouping mechanisms are formed mostly in a traditional way by selecting the devices in a random fashion or in an uniform manner, and it is an important research direction to investigate efficient grouping mechanisms by exploiting the QoS characteristics and locations of heterogeneous MTC devices.

\subsection{Deep Learning for Emerging IoT Applications and Associated Issues}
In real-world IoT environments, one of the major issues is how to reliably extract the meaningful information out of the massive unstructured/semi-structured IoT data obtained from a complex and noisy environment. Conventional ML learning techniques may fail in such complex and dynamic environments and deep learning can be considered as a promising solution \cite{Verma2017survey}. Due to multi-layer structure of deep learning, it is considered as an effective approach for the edge-computing environment in order to accurately extract necessary information from the  raw IoT sensor data \cite{Liota2018deep}. Since it is possible to offload the parts of learning layers in the edge-side of the network and transfer the reduced intermediate data to the remote cloud-center, the deep learning model seems suitable for the emerging edge computing paradigms in IoT systems. Another interesting advantage of deep learning in edge-computing environment is that it can provide privacy preservation in transferring the intermediate data. In contrast to the traditional big data systems such as Spark or MapReduce in which intermediate data contains the user privacy information, the intermediate data generated in deep learning networks usually have different semantics than that in the original source data \cite{Verma2017survey}.

In practical IoT systems, the data is usually dynamic and unlabelled, and the conventional statistically trained models are not efficient to handle the large unlabelled and dynamic data-set. Furthermore, it is highly impractical to manually label all the IoT raw data. Due to this, the conventional supervised training-based learning techniques are not suitable for large-scale IoT/mMTC environments \cite{SongAItowards}. Moreover, the conventional cloud-based architecture requires the transfer of a huge amount of IoT data from the edge-devices to the cloud-center. To address these issues, the application of deep learning with collaborative cloud-edge/fog processing seems to be a promising future research direction \cite{SongAItowards,SKSIEEE2017}.

However, the application of deep learning in IoT systems faces the crucial challenge of meeting the computational requirements. The main associated issues include the high-speed training of large-scale IoT networks with the massive data-sets and embedding deep learning capability in low-power IoT devices \cite{Venkataramani2016}. This computational challenge caused due to the expected growth in the size of the data-sets and the algorithmic complexity of ML algorithms is demanding the need of improving the computational efficiency of existing computing platforms. Also, it is not feasible to offload all the data to the cloud-center for processing due to constraints on the bandwidth, privacy and battery life, and the computational efficiency at the device-side needs to be accelerated for deep learning applications. In this regard, some of the potential future solutions to improve computational efficiency of ML algorithms include deep-learning accelerators, approximate computing and post-CMOS device technologies \cite{Venkataramani2016}.

Many IoT products have already used the ML techniques to acquire and analyze the environmental data. For example, Google's Nest Learning Thermostat uses ML algorithms to understand the patterns of its users' temperature schedules and preferences by utilizing the temperature data recorded in a structure way. However, unstructured multimedia data such as visual images and audio signals are difficult to learn by using the conventional ML techniques. In this regard, some emerging IoT devices are already using sophisticated deep learning techniques to capture and understand the complex environments \cite{Tang2017computer}. As an example, the face-recognition security system from the Microsoft's Windows IoT team uses deep-learning technology to perform tasks such as unlocking a door by recognizing its users' faces.

Due to demanding real-time requirements of IoT applications in terms of latency and high cost of radio resources required in delivering information to the cloud-center, it is advantageous to implement deep learning techniques at the device-side. However, due to the limited computing power and low memory size of IoT devices, it is challenging to implement deep learning at the device-side. Therefore, most of the time, existing deep-learning applications require third-party libraries and it may be difficult to migrate them to the IoT devices \cite{Tang2017computer}. In this regard, it is an important future research direction to investigate suitable paradigms such as convolution neural networks based inference engine  \cite{Tang2017computer} to facilitate the implementation of deep learning at the IoT devices.

\vspace{- 5 pt}
\section{Conclusions}
\label{sec:_sec8}
Future cellular IoT networks are expected to support the massive number of resource-constrained MTC devices while satisfying their diverse QoS requirements and will need to deal with several challenges for enhancing the access latency, scalability, connection reliability, energy efficiency and  network throughput. To this end, this paper has discussed various challenges of mMTC systems such as QoS provisioning, mMTC traffic characterization, transmission scheduling with QoS support, small data packet transmission and RAN congestion, and has provided a detailed review on the existing studies attempting to address these issues. By considering machine learning as an important enabler to address some of these issues in ultra-dense cellular IoT networks, the paper has identified the potential advantages, research challenges and the application scenarios of ML-assisted solutions. Among potential ML techniques, the application of Q-learning in minimizing the RAN congestion has been presented along with some performance enhancement techniques. Finally, some important research issues and interesting directions for future research have been provided.

\begin{thebibliography}{100}
\providecommand{\url}[1]{#1}
\csname url@rmstyle\endcsname
\providecommand{\newblock}{\relax}
\providecommand{\bibinfo}[2]{#2}
\providecommand\BIBentrySTDinterwordspacing{\spaceskip=0pt\relax}
\providecommand\BIBentryALTinterwordstretchfactor{4}
\providecommand\BIBentryALTinterwordspacing{\spaceskip=\fontdimen2\font plus
\BIBentryALTinterwordstretchfactor\fontdimen3\font minus
  \fontdimen4\font\relax}
\providecommand\BIBforeignlanguage[2]{{%
\expandafter\ifx\csname l@#1\endcsname\relax
\typeout{** WARNING: IEEEtran.bst: No hyphenation pattern has been}%
\typeout{** loaded for the language `#1'. Using the pattern for}%
\typeout{** the default language instead.}%
\else
\language=\csname l@#1\endcsname
\fi
#2}}

\bibitem{Fuqaha2015IoT}
A.~Al-Fuqaha, \emph{et~al.}, ``Internet of things: A survey on enabling technologies, protocols, and applications,''
  \emph{IEEE Commun. Surveys Tuts.}, vol.~17, no.~4, pp. 2347--2376,
  Fourthquarter 2015.

\bibitem{Wang2017survey}
H.~Wang and A.~O. Fapojuwo, ``A survey of enabling technologies of low power
  and long range machine-to-machine communications,'' \emph{IEEE Commun. Surveys Tuts.}, vol.~19, no.~4, pp. 2621--2639, Fourthquarter 2017.

\bibitem{ITURvision2015}
{ITU-R}, ``{IMT} vision - framework and overall objectives of the future
  development of {IMT} for 2020 and beyond,'' ITU-R Rec. M.2083-0, Sept. 2015.

\bibitem{JayawickramaspectrumIoT}
B.~A. Jayawickrama, Y.~He, E.~Dutkiewicz, and M.~D. Mueck, ``Scalable spectrum
  access system for massive machine type communication,'' \emph{IEEE Network},
  vol.~PP, no.~99, pp. 1--7, 2018.

\bibitem{3GPPtechnical2017}
3GPP, ``Study on scenarios and requirements for next generation access
  technologies (Release 14),'' Technical Report 38.913, 2017.

\bibitem{Ghavimi2015M2M}
F.~Ghavimi and H.~H. Chen, ``M2M communications in 3GPP LTE/LTE-A networks:
  Architectures, service requirements, challenges, and applications,''
  \emph{IEEE Commun. Surveys Tuts.}, vol.~17, no.~2, pp. 525--549,
  Secondquarter 2015.

\bibitem{Koseoglu2017}
M.~Koseoglu, ``Performance analysis of small data transmission schemes for
  cellular M2M communications,'' in \emph{Proc. 16th Annual Mediterranean Ad Hoc
  Networking Workshop}, June 2017, pp. 1--6.

\bibitem{DawymMTC2017}
Z.~Dawy, \emph{et~al.}, ``Toward massive machine type cellular communications,'' \emph{IEEE Wireless Communications},
  vol.~24, no.~1, pp. 120--128, Feb. 2017.

\bibitem{3GPPMTCservice}
3GPP, ``Service requirements for machine-type communications (MTC), TS 22.368, V11.5.0, June 2012.

\bibitem{Belloeuropean2014}
L.~M. Bello, P.~Mitchell, and D.~Grace, ``Application of Q-learning for RACH
  access to support M2M traffic over a cellular network,'' in \emph{European
  Wireless 2014; 20th European Wireless Conference}, May 2014, pp. 1--6.

\bibitem{Durisi2016IEEEproc}
G.~Durisi, T.~Koch, and P.~Popovski, ``Toward massive, ultrareliable, and
  low-latency wireless communication with short packets,'' \emph{Proc. of
  the IEEE}, vol. 104, no.~9, pp. 1711--1726, Sept. 2016.

\bibitem{SingminMDPI}
S.~Oh and J.~W. Jang, ``A scheme to smooth aggregated traffic from sensors with
  periodic reports,'' \emph{MDPI Sensors}, vol.~17, no. 503, pp. 1--18, March
  2017.

\bibitem{Sharmacommletter2018}
S.~K. Sharma and X.~Wang, ``Distributed caching enabled peak traffic reduction
  in ultra-dense IoT networks,'' \emph{IEEE Commun. Letters}, vol. 22, no. 6, pp. 1252-1255, June 2018.

\bibitem{Bockelmann2016}
C.~Bockelmann, \emph{et~al.}, ``Massive machine-type communications in 5G: physical
  and MAC-layer solutions,'' \emph{IEEE Commun. Mag.}, vol.~54,
  no.~9, pp. 59--65, Sept. 2016.

\bibitem{Imran2014challenges}
A.~Imran, A.~Zoha, and A.~Abu-Dayya, ``Challenges in 5G: how to empower SoN
  with big data for enabling 5G,'' \emph{IEEE Network}, vol.~28, no.~6, pp.
  27--33, Nov 2014.

\bibitem{Liintelligent2017}
R.~Li, , ``Intelligent\emph{et~al.}, 5G: When cellular networks meet artificial intelligence,'' \emph{IEEE
  Wireless Commun.}, vol.~24, no.~5, pp. 175--183, Oct. 2017.

\bibitem{Daniels2009}
R.~C. Daniels and R.~W. Heath, ``An online learning framework for link
  adaptation in wireless networks,'' in \emph{Proc. Info. Theory and
  Applications Workshop}, Feb. 2009, pp. 138--140.

\bibitem{Yun2010reinforcement}
S.~Yun and C.~Caramanis, ``Reinforcement learning for link adaptation in
  MIMO-OFDM wireless systems,'' in \emph{Proc. IEEE GLOBECOM }, Dec. 2010, pp. 1--5.

\bibitem{SKSIEEE2017}
S.~K. Sharma and X.~Wang, ``Live data analytics with collaborative edge and
  cloud processing in wireless {IoT} networks,'' \emph{IEEE Access}, vol.~5,
  pp. 4621--4635, March 2017.

\bibitem{Wang2015IEEE}
X.~Wang, X.~Li, and V.~C.~M. Leung, ``Artificial intelligence-based techniques
  for emerging heterogeneous network: State of the arts, opportunities, and
  challenges,'' \emph{IEEE Access}, vol.~3, pp. 1379--1391, 2015.

\bibitem{Moon2017access}
J.~Moon and Y.~Lim, ``Access control of MTC devices using reinforcement
  learning approach,'' in \emph{Proc. Int. Conf. on Information
  Networking}, Jan. 2017, pp. 641--643.

\bibitem{Hasanrandom2013}
M.~Hasan, E.~Hossain, and D.~Niyato, ``Random access for machine-to-machine
  communication in {LTE}-advanced networks: issues and approaches,'' \emph{IEEE
  Commun. Mag.}, vol.~51, no.~6, pp. 86--93, June 2013.

\bibitem{Mohammed2015base}
A.~H. Mohammed, A.~S. Khwaja, A.~Anpalagan, and I.~Woungang, ``Base station
  selection in M2M communication using Q-learning algorithm in LTE-A
  networks,'' in \emph{Proc. IEEE 29th Int. Conf. Advanced
  Info. Networking and Applications}, March 2015, pp. 17--22.

\bibitem{Park2016resource}
T.~Park and W.~Saad, ``Resource allocation and coordination for critical
  messages using finite memory learning,'' in \emph{Prof. IEEE Globecom
  Workshops (GC Wkshps)}, Dec. 2016, pp. 1--6.

\bibitem{Portelli2017leveraging}
K.~Portelli and C.~Anagnostopoulos, ``Leveraging edge computing through
  collaborative machine learning,'' in \emph{Proc. 5th Int. Conf.
  on Future Internet of Things and Cloud Workshops (FiCloudW)}, Aug. 2017, pp.
  164--169.

\bibitem{ParkIEEE2016}
T.~Park, N.~Abuzainab, and W.~Saad, ``Learning how to communicate in the
  internet of things: Finite resources and heterogeneity,'' \emph{IEEE Access},
  vol.~4, pp. 7063--7073, 2016.

\bibitem{Tang2017computer}
J.~Tang, D.~Sun, S.~Liu, and J.~L. Gaudiot, ``Enabling deep learning on IoT
  devices,'' \emph{Computer}, vol.~50, no.~10, pp. 92--96, 2017.

\bibitem{Wang2016cellular}
M.~Wang, \emph{et~al.}, ``Cellular machine-type communications: physical challenges and solutions,'' \emph{IEEE
  Wireless Commun.}, vol.~23, no.~2, pp. 126--135, April 2016.

\bibitem{Dawy2017towards}
Z.~Dawy, \emph{et~al.}, ``Toward massive
  machine type cellular communications,'' \emph{IEEE Wireless Commun.},
  vol.~24, no.~1, pp. 120--128, 2017.

\bibitem{Alvarino2016overview}
A.~Rico-Alvarino, \emph{et~al.}, ``An overview of 3GPP enhancements on machine to
  machine communications,'' \emph{IEEE Commun. Magazine}, vol.~54,
  no.~6, pp. 14--21, June 2016.

\bibitem{Elsaadany2017cellularLTEA}
M.~Elsaadany, A.~Ali, and W.~Hamouda, ``Cellular LTE-A technologies for the
  future internet-of-things: Physical layer features and challenges,''
  \emph{IEEE Commun. Surveys Tuts.}, vol.~PP, no.~99, pp. 1--1,
  2017.

\bibitem{Hoglund2017overview}
A.~Hoglund, \emph{et~al.}, ``Overview of 3GPP release 14 enhanced NB-IoT,''
  \emph{IEEE Network}, vol.~31, no.~6, pp. 16--22, Nov. 2017.

\bibitem{Layarandomaccess2014}
A.~Laya, L.~Alonso, and J.~Alonso-Zarate, ``Is the random access channel of
  {LTE} and {LTE-A} suitable for {M2M} communications? a survey of
  alternatives,'' \emph{IEEE Commun. Surveys Tuts.}, vol.~16, no.~1,
  pp. 4--16, Firstquarter 2014.

\bibitem{Yangnarrwoband2017}
W.~Yang, \emph{et~al.}, ``Narrowband
  wireless access for low-power massive internet of things: A bandwidth
  perspective,'' \emph{IEEE Wireless Commun.}, vol.~24, no.~3, pp.
  138--145, 2017.

\bibitem{Ali2017LTE}
M.~S. Ali, E.~Hossain, and D.~I. Kim, ``LTE/LTE-A random access for massive
  machine-type communications in smart cities,'' \emph{IEEE Commun. Mag.}, vol.~55, no.~1, pp. 76--83, Jan. 2017.

\bibitem{Soltanmohammadi2016}
E.~Soltanmohammadi, K.~Ghavami, and M.~Naraghi-Pour, ``A survey of traffic
  issues in machine-to-machine communications over LTE,'' \emph{IEEE Internet
  of Things J.}, vol.~3, no.~6, pp. 865--884, Dec 2016.

\bibitem{Gotsis2012M2M}
A.~G. Gotsis, A.~S. Lioumpas, and A.~Alexiou, ``M2M scheduling over LTE:
  Challenges and new perspectives,'' \emph{IEEE Veh. Technol. Mag.},
  vol.~7, no.~3, pp. 34--39, Sept. 2012.

\bibitem{Mehaseb2016classification}
M.~A. Mehaseb, Y.~Gadallah, A.~Elhamy, and H.~Elhennawy, ``Classification of
  LTE uplink scheduling techniques: An M2M perspective,'' \emph{IEEE
  Commun. Surveys Tuts.}, vol.~18, no.~2, pp. 1310--1335, Secondquarter 2016.

\bibitem{Marjani2017bigdata}
M.~Marjani, \emph{et~al.}, ``Big IoT data analytics: Architecture, opportunities,
  and open research challenges,'' \emph{IEEE Access}, vol.~5, pp. 5247--5261,
  2017.

\bibitem{Verma2017survey}
S.~Verma, \emph{et~al.}, ``A survey
  on network methodologies for real-time analytics of massive IoT data and open
  research issues,'' \emph{IEEE Commun.. Surveys Tuts.}, vol.~19,
  no.~3, pp. 1457--1477, thirdquarter 2017.

\bibitem{Elhammouti2017self}
H.~Elhammouti, \emph{et~al.},
  ``Self-organized connected objects: Rethinking QoS provisioning for IoT
  services,'' \emph{IEEE Commun. Mag.}, vol.~55, no.~9, pp. 41--47,
  2017.

\bibitem{Mohammadienabling}
M.~Mohammadi and A.~Al-Fuqaha, ``Enabling cognitive smart cities using big data
  and machine learning: Approaches and challenges,'' \emph{IEEE Commun. Mag.}, vol.~56, no.~2, pp. 94--101, Feb. 2018.

\bibitem{Sezer2018context}
O.~B. Sezer, E.~Dogdu, and A.~M. Ozbayoglu, ``Context-aware computing,
  learning, and big data in internet of things: A survey,'' \emph{IEEE Internet
  of Things J.}, vol.~5, no.~1, pp. 1--27, Feb. 2018.

\bibitem{Busoniu2008Trans}
L.~Busoniu, \emph{et~al.}, ``A comprehensive survey of multiagent reinforcement
  learning,'' \emph{IEEE Trans. Systems, Man, and Cybernetics, Part C
  (Applications and Reviews)}, vol.~38, no.~2, pp. 156--172, March 2008.

\bibitem{deepRL2017}
K.~Arulkumaran, M.~P. Deisenroth, M.~Brundage, and A.~A. Bharath, ``Deep
  reinforcement learning: A brief survey,'' \emph{IEEE Signal Process. Mag.}, vol.~34, no.~6, pp. 26--38, Nov 2017.

\bibitem{Baldemair2015ultra}
R.~Baldemair, \emph{et~al.}, ``Ultra-dense networks in millimeter-wave
  frequencies,'' \emph{IEEE Commun. Mag.}, vol.~53, no.~1, pp. 202--208, Jan. 2015.

\bibitem{ChenM2Msurvey}
S.~Chen, \emph{et~al.}, ``Machine-to-machine
  communications in ultra-dense networks: A survey,'' \emph{IEEE Commun. Surveys Tuts.}, vol.~19, no.~3, pp. 1478--1503, thirdquarter 2017.

\bibitem{Kamel2016ultra}
M.~Kamel, W.~Hamouda, and A.~Youssef, ``Ultra-dense networks: A survey,''
  \emph{IEEE Commun. Surveys Tuts.}, vol.~18, no.~4, pp. 2522--2545,
  Fourthquarter 2016.

\bibitem{SmiljkovicM2M}
K.~Smiljkovic, V.~Atanasovski, and L.~Gavrilovska, ``Machine-to-machine traffic
  characterization: Models and case study on integration in {LTE},'' in
  \emph{Proc. 4th Int. Conf. Wireless Commun., Veh. Technol., Info. Th. and Aerospace Electronic Systems (VITAE)}, May
  2014, pp. 1--5.

\bibitem{LiIoTjournal}
X.~Li, J.~B. Rao, and H.~Zhang, ``Engineering machine-to-machine traffic in
  {5G},'' \emph{IEEE Internet of Things J.}, vol.~3, no.~4, pp. 609--618,
  Aug 2016.

\bibitem{Afrin2014adaptive}
N.~Afrin, J.~Brown, and J.~Y. Khan, ``An adaptive buffer based semi-persistent
  scheduling scheme for machine-to-machine communications over {LTE},'' in
  \emph{Proc. Eighth Int. Conf. Next Generation Mobile Apps, Services and Technol.}, Sept 2014, pp. 260--265.

\bibitem{Kim2014M2M}
J.~Kim, J.~Lee, J.~Kim, and J.~Yun, ``M2M service platforms: Survey, issues,
  and enabling technologies,'' \emph{IEEE Commun. Surveys Tuts.},
  vol.~16, no.~1, pp. 61--76, Firstquarter 2014.

\bibitem{FabiniM2M}
J.~Fabini and T.~Zseby, ``M2M communication delay challenges: Application and
  measurement perspectives,'' in \emph{Proc. IEEE Int. Instrumentation
  and Measurement Technol. Conf.}, May 2015, pp. 1859--1864.

\bibitem{Lien2011massive}
S.~Y. Lien and K.~C. Chen, ``Massive access management for QoS guarantees in
  3GPP machine-to-machine communications,'' \emph{IEEE Commun. Letters},
  vol.~15, no.~3, pp. 311--313, March 2011.

\bibitem{Saleemarxiv2017}
Y. Saleem, N. Crespi, M. H. Rehmani and R. Copeland, ``Internet of Things-aided Smart Grid: Technologies, Architectures, Applications, Prototypes, and Future Research Directions'', CoRR, abs/1704.08977,2017, online: http://arxiv.org/abs/1704.08977.

\bibitem{Zhengradio2012}
K.~Zheng, \emph{et~al.}, ``Radio resource allocation
  in {LTE}-advanced cellular networks with {M2M} communications,'' \emph{IEEE
  Commun. Mag.}, vol.~50, no.~7, pp. 184--192, July 2012.

\bibitem{Chengoverload}
M.~Y. Cheng, \emph{et~al.}, ``Overload control for
  machine-type-communications in LTE-advanced system,'' \emph{IEEE
  Commun. Mag.}, vol.~50, no.~6, pp. 38--45, June 2012.

\bibitem{Ratasuk2017LTEM}
R.~Ratasuk, N.~Mangalvedhe, D.~Bhatoolaul, and A.~Ghosh, ``LTE-M evolution
  towards 5G massive MTC,'' in \emph{Proc. IEEE Globecom Workshops (GC Wkshps)},
  Dec. 2017, pp. 1--6.

\bibitem{Wali2014}
P.~K. Wali and D.~Das, ``A novel access scheme for IoT communications in
  {LTE}-advanced network,'' in \emph{Proc. IEEE Int. Conf. Advanced Networks and Telecommun. Systems (ANTS)}, Dec 2014, pp. 1--6.

\bibitem{AlAdwani2012}
A.~M. AlAdwani, A.~Gawanmeh, and S.~Nicolas, ``A demand side management traffic
  shaping and scheduling algorithm,'' in \emph{Proc. Sixth Asia Modelling
  Symp.}, May 2012, pp. 205--210.

\bibitem{Lien2011ubiquitous}
S.~Y. Lien, K.~C. Chen, and Y.~Lin, ``Toward ubiquitous massive accesses in
  3GPP machine-to-machine communications,'' \emph{IEEE Commun. Mag.}, vol.~49, no.~4, pp. 66--74, April 2011.

\bibitem{Lienabling2017hetnet}
Z.~Li, \emph{et~al.}, ``Enabling heterogeneous MMTC by energy-efficient and connectivity-aware clustering and routing,'' in
  \emph{Proc. IEEE Globecom Workshops (GC Wkshps)}, Dec. 2017, pp. 1--6.

\bibitem{3GppTR2015}
3GPP, ``Cellular system support for ultralow complexity and low
  throughput Internet of Things (CIoT),'', TR 45.820, V13.1.0, Nov. 2015.

\bibitem{Maldonado2017NBIoT}
P.~Andres-Maldonado, \emph{et~al.}, ``Narrowband IoT data transmission procedures for massive
  machine-type communications,'' \emph{IEEE Network}, vol.~31, no.~6, pp.
  8--15, Nov. 2017.

\bibitem{Huang2017M2M}
J.~Huang, \emph{et~al.},, ``Optimizing {M2M} communications and quality of services in the {IoT} for sustainable smart
  cities,'' \emph{IEEE Trans. on Sustainable Computing}, vol.~PP, no.~99,
  pp. 1--1, 2017.

\bibitem{Centenaro2017comparison}
M.~Centenaro, \emph{et~al.},, ``Comparison of
  collision-free and contention-based radio access protocols for the internet
  of things,'' \emph{IEEE Trans. Commun.}, vol.~65, no.~9, pp.
  3832--3846, Sept. 2017.

\bibitem{UmaIBM}
U.~Devi, \emph{et~al.}, ``SERA: A scheduling framework for M2M transmission in
  cellular networks,'' online, web:
  http://researcher.ibm.com/researcher/files/in-umamadev/M2M-long.pdf.

\bibitem{Medjahdi2017road}
Y.~Medjahdi, \emph{et~al.}, ``On the road to 5G: Comparative study of physical layer
  in MTC context,'' \emph{IEEE Access}, vol.~5, pp. 26556--26581, 2017.

\bibitem{3GPPRAN}
3GPP, ``Study on {RAN} improvements for machine-type communications,'' Tech.
  Rep., TR 37.868, 2012.

\bibitem{Choi2016commlett}
J.~Choi, ``On the adaptive determination of the number of preambles in RACH for MTC,''
\emph{IEEE Commun. Letters}, vol.~20, no.~7, pp. 1385--1388, July 2016.

\bibitem{Lee2017MTC}
S.~J. Sung‐Hyung~Lee and J.~Kim, ``Dynamic resource allocation of random
  access for MTC devices,'' \emph{ETRI Journal}, vol.~39, no.~4, pp. 546--557,
  Aug. 2017.

\bibitem{Sharma2018DSS}
S.~K. Sharma, \emph{et~al.}, ``Dynamic spectrum sharing in 5G wireless networks with
  full-duplex technology: Recent advances and research challenges,'' \emph{IEEE
  Commun. Surveys Tuts.}, vol.~20, no.~1, pp. 674--707, Firstquarter
  2018.

\bibitem{Lirandomaccess2017}
M. Li, \emph{et al.}, ``Random access and
  virtual resource allocation in software-defined cellular networks with
  machine-to-machine communications,'' \emph{IEEE Trans. Veh. Technol.}, vol.~66, no.~7, pp. 6399--6414, July 2017.

\bibitem{Samir2016}
A.~Samir, \emph{et al.}, ``Partial contention free random access protocol for M2M communications in LTE
  networks,'' \emph{J. of Wireless Networking and Commun.}, vol.~6, no.~3, pp. 66--72, 2016.

\bibitem{Compressivecoded2015}
G.~Wunder, C.~Stefanovic, P.~Popovski, and L.~Thiele, ``Compressive coded
  random access for massive MTC traffic in 5G systems,'' in \emph{Proc. 49th
  Asilomar Conf. on Signals, Systems and Computers}, Nov. 2015, pp. 13--17.

\bibitem{Delgado2012european}
I. M. Delgado-Luque, \emph{et al.}, ``Evaluation of latency-aware scheduling techniques for M2M traffic over LTE,''
in \emph{Proc. 20th European Signal Processing Conference (EUSIPCO)}, Bucharest, 2012, pp. 989-993.

\bibitem{AliCoRR2018}
S. Ali, N. Rajatheva and W. Saad, ``Fast Uplink Grant for Machine Type Communications: Challenges and
Opportunities'', CoRR, abs/1801.04953, 2018, online: http://arxiv.org/abs/1801.04953.

\bibitem{Sabin2017Globecom}
S. Bhandari, S. K. Sharma and X. Wang, ``Latency Minimization in Wireless IoT Using Prioritized Channel Access and Data Aggregation,'' in \emph{Proc. IEEE Global Commun. Conf.}, Singapore, Dec. 2017, pp. 1-6.

\bibitem{Sharmacooperative2017}
S. K. Sharma and X. Wang, ``Cooperative sensing delay minimization in cloud-assisted DSA networks'' in \emph{Proc. IEEE 28th Annual Int. Symp. on Personal, Indoor, and Mobile Radio Commun. (PIMRC)}, Montreal, QC, Oct. 2017, pp. 1-6.

\bibitem{Andrade2015impact}
T. P. C. de Andrade, C. A. Astudillo and N. L. S. da Fonseca, ``Impact of M2M traffic on human-type communication users on the LTE uplink channel,'' in \emph{Proc. 7th IEEE Latin-American Conference on Communications (LATINCOM)}, Arequipa, 2015, pp. 1-6.

\bibitem{Laner2013traffic}
M.~Laner, P.~Svoboda, N.~Nikaein, and M.~Rupp, ``Traffic models for machine
  type communications,'' in \emph{Proc. The Tenth Int. Symp. Wireless Commun. Systems}, Aug. 2013, pp. 1--5.

\bibitem{Nikaein2013}
N.~Nikaein, \emph{et al.},
  ``Simple traffic modeling framework for machine type communication,'' in
  \emph{Proc. The Tenth Int. Symp. Wireless Commun. Systems},
  Aug 2013, pp. 1--5.

\bibitem{AlAdwani2013traffic}
A.~M. AlAdwani, A.~Gawanmeh, and S.~Nicolas, ``Traffic shaping and delay
  optimization in demand side management,'' in \emph{Proc. 15th Int. Conf. on Computer Modelling and Simulation}, April 2013, pp. 722--727.

\bibitem{KumarMilcom2016}
A.~Kumar, A.~Abdelhadi, and C.~Clancy, ``A delay efficient multiclass packet
  scheduler for heterogeneous M2M uplink,'' in \emph{Proc. IEEE MILCOM Conf.}, Nov 2016, pp. 313--318.

\bibitem{Kumar2016online}
------, ``An online delay efficient packet scheduler for M2M traffic in
  industrial automation,'' in \emph{Proc. 2016 Annual IEEE Systems Conf.}, April 2016, pp. 1--6.

\bibitem{Zhaotraffic}
Y.~Zhao, B.~Zhang, C.~Li, and C.~Chen, ``On/off traffic shaping in the
  internet: Motivation, challenges, and solutions,'' \emph{IEEE Network},
  vol.~31, no.~2, pp. 48--57, March 2017.

\bibitem{Kim2017random}
J.~S. Kim, S.~Lee, and M.~Y. Chung, ``Efficient random-access scheme for
  massive connectivity in 3GPP low-cost machine-type communications,''
  \emph{IEEE Trans. Veh. Technol.}, vol.~66, no.~7, pp.
  6280--6290, July 2017.

\bibitem{Duan2016adaptive}
S.~Duan, V.~Shah-Mansouri, Z.~Wang, and V.~W.~S. Wong, ``D-ACB: Adaptive
  congestion control algorithm for bursty M2M traffic in LTE networks,''
  \emph{IEEE Trans. Veh. Technol.}, vol.~65, no.~12, pp.
  9847--9861, Dec. 2016.

\bibitem{Lin2016estimation}
G.~Y. Lin, S.~R. Chang, and H.~Y. Wei, ``Estimation and adaptation for bursty
  LTE random access,'' \emph{IEEE Trans. Veh. Technol.},
  vol.~65, no.~4, pp. 2560--2577, April 2016.

\bibitem{Osti2014analysis}
P.~Osti, \emph{et al.}, ``Analysis of PDCCH performance for M2M traffic in LTE,''
\emph{IEEE Trans. Veh. Technol.}, vol.~63, no.~9, pp. 4357--4371, Nov. 2014.

\bibitem{BIRAL20151}
A.~Biral, \emph{et al.}, ``The challenges of M2M massive access in wireless
  cellular networks,'' \emph{Digital Commun. and Networks}, vol.~1,
  no.~1, pp. 1--19, 2015.

\bibitem{Lien2012cooperative}
S.~Y. Lien, \emph{et al.}, ``Cooperative access class
  barring for machine-to-machine communications,'' \emph{IEEE Trans. Wireless Commun.}, vol.~11, no.~1, pp. 27--32,
  Jan. 2012.


\bibitem{Duan2013IEEE}
S.~Duan, V.~Shah-Mansouri, and V.~W.~S. Wong, ``Dynamic access class barring
  for M2M communications in LTE networks,'' in \emph{Proc. IEEE Global Commun. Conf.}, Dec. 2013, pp. 4747--4752.

\bibitem{KoseogluLOWER}
M.~Koseoglu, ``Lower bounds on the LTE-A average random access delay under
  massive M2M arrivals,'' \emph{IEEE Trans. Commun.}, vol.~64, no.~5, pp. 2104--2115, May 2016.

\bibitem{DIGIcellular}
DIGI, ``What are the difference between LTE-M and NB-IoT Cellular Protocols,'' available online: https://www.digi.com/, last accessed 15th March 2018.

\bibitem{Itayatraisemiconductor}
I. Lusky, ``CAT-M1 vs NB-IoT-examining the real differences,'' available online: https://www.iot-now.com/, last accessed 16th March 2018.

\bibitem{Andreev2013efficient}
S.~Andreev, \emph{et al.}, ``Efficient small data access for
  machine-type communications in LTE,'' in \emph{Proc. IEEE Int. Conf. Commun. (ICC)}, June 2013, pp. 3569--3574.

\bibitem{Huenhanced2015}
J.~Hu, \emph{et al.}, ``Enhanced LTE physical downlink control channel design for
  machine-type communications,'' in \emph{Proc. 7th Int. Conf. New Technologies, Mobility and Security (NTMS)}, July 2015, pp. 1--5.

\bibitem{3GppTS2017GPRS}
3GPP, ``General packet radio service (GPRS) enhancements for evolved universal
  terrestrial radio access network (E-UTRAN) access,'' TS 23.401 V14.3.0, March
  2017.

\bibitem{Yuuplinkscheduling}
C.~Yu, \emph{et al.}, ``Uplink scheduling and link adaptation
  for narrowband internet of things systems,'' \emph{IEEE Access}, vol.~5, pp.
  1724--1734, 2017.

\bibitem{Boisguene2017survey}
R.~Boisguene, S.~C. Tseng, C.~W. Huang, and P.~Lin, ``A survey on NB-IoT
  downlink scheduling: Issues and potential solutions,'' in \emph{Proc. 13th
  Int. Wireless Commun. and Mobile Computing Conf. (IWCMC)}, June 2017, pp. 547--551.

\bibitem{Wangprimer}
Y.~P.~E. Wang, \emph{et al.}, ``A primer on 3GPP narrowband internet of things,''
  \emph{IEEE Commun. Mag.}, vol.~55, no.~3, pp. 117--123, March
  2017.

\bibitem{Dapeng2003}
D.~Wu and R.~Negi, ``Effective capacity: a wireless link model for support of
  quality of service,'' \emph{IEEE Trans. Wireless Commun.},
  vol.~2, no.~4, pp. 630--643, July 2003.

\bibitem{Ghavimi2017uplink}
F.~Ghavimi, Y.~W. Lu, and H.~H. Chen, ``Uplink scheduling and power allocation
  for {M2M} communications in {SC-FDMA}-based {LTE-A} networks with {QoS}
  guarantees,'' \emph{IEEE Trans. Veh. Technol.}, vol.~66,
  no.~7, pp. 6160--6170, July 2017.

\bibitem{Myung2006}
H.~G. Myung, J.~Lim, and D.~J. Goodman, ``Single carrier FDMA for uplink
  wireless transmission,'' \emph{IEEE Veh. Technol. Mag.}, vol.~1,
  no.~3, pp. 30--38, Sept. 2006.

\bibitem{Aijaz2014uplink}
A.~Aijaz, \emph{et al.}, ``Energy-efficient uplink resource allocation in lte networks with M2M/H2H
  co-existence under statistical QoS guarantees,'' \emph{IEEE Trans. Commun.}, vol.~62, no.~7, pp. 2353--2365, July 2014.

\bibitem{Parkenhancement2014}
C.~W. Park, D.~Hwang, and T.~J. Lee, ``Enhancement of IEEE 802.11ah MAC for M2M
  communications,'' \emph{IEEE Commun. Letters}, vol.~18, no.~7, pp. 1151--1154, July 2014.


\bibitem{Afrin2013performance}
N.~Afrin, J.~Brown, and J.~Y. Khan, ``Performance analysis of an enhanced delay
  sensitive lte uplink scheduler for M2M traffic,'' in \emph{Proc. 2013 Australasian Telecommun. Networks and Applications Conf.}, Nov. 2013, pp. 154--159.

\bibitem{Dahlman2014}
E.~Dahlman, \emph{et al.}, ``5G wireless access: requirements and realization,''
  \emph{IEEE Commun. Mag.}, vol.~52, no.~12, pp. 42--47, Dec. 2014.

\bibitem{Johansson2015}
N.~A. Johansson, Y.~P.~E. Wang, E.~Eriksson, and M.~Hessler, ``Radio access for
  ultra-reliable and low-latency 5G communications,'' in \emph{Proc. IEEE
  Int. Conf. Commun. Workshop (ICCW)}, June 2015, pp. 1184--1189.

\bibitem{Durisi2016shortpacket}
G.~Durisi, \emph{et al.}, ``Short-packet
  communications over multiple-antenna rayleigh-fading channels,'' \emph{IEEE
  Trans. Commun.}, vol.~64, no.~2, pp. 618--629, Feb. 2016.

\bibitem{Mousaei2017pilot}
M.~Mousaei and B.~Smida, ``Optimizing pilot overhead for ultra-reliable
  short-packet transmission,'' in \emph{Proc. IEEE Int. Conf. Commun. (ICC)}, May 2017, pp. 1--5.

\bibitem{Schaich2014waveform}
F.~Schaich, T.~Wild, and Y.~Chen, ``Waveform contenders for 5G-suitability
  for short packet and low latency transmissions,'' in \emph{Proc. IEEE 79th
  Veh. Technol. Conf. (VTC Spring)}, May 2014, pp. 1--5.

\bibitem{Yoomodcoding}
D.~S. Yoo, \emph{et al.}, ``Coding and modulation for
  short packet transmission,'' \emph{IEEE Trans. Veh. Technol.}, vol.~59, no.~4, pp. 2104--2109, May 2010.

\bibitem{Sun2018short}
X.~Sun, \emph{et al.}, ``Short-packet downlink transmission with non-orthogonal multiple access,'' \emph{IEEE
  Trans. Wireless Commun.}, pp. 1--1, 2018.

\bibitem{Aziz2016IEEE}
D.~Aziz, H.~Bakker, A.~Ambrosy, and Q.~Liao, ``Signalling minimization
  framework for short data packet transmission in 5G,'' in \emph{Proc. IEEE 84th
  Veh. Technol. Conf. (VTC-Fall)}, Sept. 2016, pp. 1--6.

\bibitem{Trillingsgaard2017}
K.~F. Trillingsgaard and P.~Popovski, ``Downlink transmission of short packets:
  Framing and control information revisited,'' \emph{IEEE Trans. Commun.},
  vol.~65, no.~5, pp. 2048--2061, May 2017.

\bibitem{Balachandran2016delay}
K.~Balachandran, J.~H. Kang, K.~Karakayali, and K.~M. Rege, ``Delay-tolerant
  autonomous transmissions for short packet communications,'' in \emph{Proc.
  IEEE Globecom Workshops (GC Wkshps)}, Dec. 2016, pp. 1--6.

\bibitem{Leepacketstructure}
B.~Lee, \emph{et al.}, ``Packet structure and
  receiver design for low latency wireless communications with ultra-short
  packets,'' \emph{IEEE Trans. Commun.}, vol.~66, no.~2, pp. 796--807, Feb. 2018.

\bibitem{Yang2018wireless}
Q.~Yang, \emph{et al.}, ``Wireless powered asynchronous backscatter networks with sporadic short packets: Performance
  analysis and optimization,'' \emph{IEEE Internet of Things J.}, vol.~5,
  no.~2, pp. 984--997, April 2018.

\bibitem{Sheglobecom2016}
C.~She, C.~Yang, and T.~Q.~S. Quek, ``Uplink transmission design with massive
  machine type devices in tactile internet,'' in \emph{Proc. IEEE Globecom
  Workshops (GC Wkshps)}, Dec. 2016, pp. 1--6.

\bibitem{Fawal2017overload}
A.~H.~E. Fawal, \emph{et al.}, ``RACH overload congestion mechanism for M2M communication in LTE-A: Issues and
  approaches,'' in \emph{Proc. Int. Symp. on Networks, Computers and Commun. (ISNCC)}, May 2017, pp. 1--6.

\bibitem{Wu2013FASA}
H.~Wu, \emph{et al.}, ``FASA: Accelerated S-ALOHA using access history for event-driven M2M communications,'' \emph{IEEE/ACM
  Trans. Network.}, vol.~21, no.~6, pp. 1904--1917, Dec. 2013.

\bibitem{Cheng2015modeling}
R.~G. Cheng, \emph{et al.}, ``Modeling and analysis of an
  extended access barring algorithm for machine-type communications in LTE-A
  networks,'' \emph{IEEE Trans. on Wireless Commun.}, vol.~14,
  no.~6, pp. 2956--2968, June 2015.

\bibitem{Cheng2011IEEE}
J.~P. Cheng, C.~h.~Lee, and T.~M. Lin, ``Prioritized random access with dynamic
  access barring for RAN overload in 3GPP LTE-A networks,'' in \emph{Proc. IEEE
  GLOBECOM Workshops (GC Wkshps)}, Dec. 2011, pp. 368--372.

\bibitem{LINPARDA2014}
T.~M. Lin, \emph{et al.}, ``PRADA: Prioritized random
  access with dynamic access barring for MTC in 3GPP LTE-A networks,''
  \emph{IEEE Trans. Veh. Technol.}, vol.~63, no.~5, pp. 2467--2472, June 2014.

\bibitem{Farhadi2013group}
G.~Farhadi and A.~Ito, ``Group-based signaling and access control for cellular
  machine-to-machine communication,'' in \emph{Proc. IEEE 78th Veh. Technol. Conf.}, Sept. 2013, pp. 1--6.

\bibitem{Chuang2015group}
T.~H. Chuang, M.~H. Tsai, and C.~Y. Chuang, ``Group-based uplink scheduling for
  machine-type communications in LTE-advanced networks,'' in \emph{Proc. IEEE Int.
  Conf. Advanced Info. Networking and Applications Workshops}, March 2015, pp. 652--657.

\bibitem{Pratascode2012}
N.~K. Pratas, H.~Thomsen, C.~Stefanović, and P.~Popovski, ``Code-expanded
  random access for machine-type communications,'' in \emph{Proc. 2012 IEEE Globecom
  Workshops}, Dec 2012, pp. 1681--1686.

\bibitem{Jang2014comm}
H.~S. Jang, \emph{et al.}, ``Spatial group based
  random access for M2M communications,'' \emph{IEEE Communications Letters},
  vol.~18, no.~6, pp. 961--964, June 2014.

\bibitem{Gursu2017hybrid}
H.~M. Gürsu, M.~Vilgelm, W.~Kellerer, and M.~Reisslein, ``Hybrid collision
  avoidance-tree resolution for M2M random access,'' \emph{IEEE Trans. Aerospace and Electronic Systems},
  vol.~53, no.~4, pp. 1974--1987, Aug. 2017.

\bibitem{Madueno1433}
G.~C. Madueno, C.~Stefanovic, and P.~Popovski, ``Efficient LTE access with
  collision resolution for massive M2M communications,'' in \emph{Proc. 2014 IEEE
  Globecom Workshops (GC Wkshps)}, Dec. 2014, pp. 1433--1438.

\bibitem{Janssen2000analysis}
A.~J. E.~M. Janssen and M.~J. de~Jong, ``Analysis of contention tree
  algorithms,'' \emph{IEEE Trans. Info. Th.}, vol.~46, no.~6,
  pp. 2163--2172, Sept. 2000.

\bibitem{Ruandelay2017}
Y.~Ruan, W.~Wang, Z.~Zhang, and V.~K.~N. Lau, ``Delay-aware massive random
  access for machine-type communications via hierarchical stochastic
  learning,'' in \emph{Proc. IEEE Int. Conf. Commun. (ICC)}, May 2017, pp. 1--6.

\bibitem{Alonso2005}
L.~Alonso, R.~Ferrus, and R.~Agusti, ``WLAN throughput improvement via
  distributed queuing MAC,'' \emph{IEEE Commun. Letters}, vol.~9, no.~4,
  pp. 310--312, April 2005.

\bibitem{KartsakliIEEE2008}
E.~Kartsakli, \emph{et al.},
  ``Cross-layer enhancement for WLAN systems with heterogeneous traffic based
  on DQCA,'' \emph{IEEE Commun. Mag.}, vol.~46, no.~6, pp. 60--66, June 2008.

\bibitem{Li2017energy}
M.~Li, \emph{et al.}, ``Energy-efficient M2M communications with mobile edge computing in virtualized cellular networks,''
  in \emph{Proc. IEEE Int. Conf. Commun. (ICC)}, May 2017, pp. 1--6.

\bibitem{Liglobecomlearning}
L.~Li, X.~He, and H.~Li, ``Learning the spectrum using collaborative filtering
  in directional millimeter wave networks,'' in \emph{Proc. IEEE GLOBECOM}, Dec. 2017, pp. 1--7.

\bibitem{Rakovic2010novel}
V.~Rakovic and L.~Gavrilovska, ``Novel RAT selection mechanism based on
  hopfield neural networks,'' in \emph{Proc. Int. Congress on Ultra Modern
  Telecommun. and Control Systems}, Oct. 2010, pp. 210--217.

\bibitem{Katodeeplearning2017}
N.~Kato, \emph{et al.}, ``The deep learning vision for heterogeneous network traffic
  control: Proposal, challenges, and future perspective,'' \emph{IEEE Wireless
  Commun.}, vol.~24, no.~3, pp. 146--153, June 2017.

\bibitem{DanielsGlobecom2008}
R.~C. Daniels, C.~Caramanis, and R.~W.~Heath, ``A supervised learning approach
  to adaptation in practical MIMO-OFDM wireless systems,'' in \emph{Proc. IEEE GLOBECOM}, Nov. 2008, pp. 1--5.

\bibitem{Desai2016energy}
T.~Desai and H.~Shah, ``Energy efficient link adaptation using machine learning
  techniques for wireless OFDM,'' in \emph{Proc. Int. Conf. Inventive Computation Technologies (ICICT)}, vol.~3, Aug. 2016, pp. 1--4.

\bibitem{Saishankar2017}
S.~K. Pulliyakode and S.~Kalyani, ``Reinforcement learning techniques for outer
  loop link adaptation in 4G/5G systems,'' \emph{ArXiv e-prints}, 2017.

\bibitem{Chen2018datafriven}
M.~Chen, \emph{et al.}, ``Data-driven computing and
  caching in 5G networks: Architecture and delay analysis,'' \emph{IEEE
  Wireless Communications}, vol.~25, no.~1, pp. 70--75, Feb. 2018.

\bibitem{Blankenshiplink2004}
Y.~W. Blankenship, \emph{et al.}, ``Link error prediction methods for multicarrier systems,'' in \emph{IEEE
  60h Veh. Technol. Conf.}, vol.~6, Sept. 2004, pp. 4175--4179.

\bibitem{Park2017broadband}
Y.~Park, \emph{et al.}, ``A new link adaptation method to mitigate sinr mismatch in ultra-dense small cell LTE
  networks,'' \emph{IEEE Trans. Wireless Commun.}, vol.~PP, no.~99, pp. 1--1, 2017.

\bibitem{Park2016learning}
T.~Park and W.~Saad, ``Learning with finite memory for machine type
  communication,'' in \emph{Proc. Annual Conf. on Information Science and Systems (CISS)}, March 2016, pp. 608--613.

\bibitem{Alsheikh2014ML}
M.~A. Alsheikh, S.~Lin, D.~Niyato, and H.~P. Tan, ``Machine learning in
  wireless sensor networks: Algorithms, strategies, and applications,''
  \emph{IEEE Commun. Surveys Tuts.}, vol.~16, no.~4, pp. 1996--2018,
  Fourthquarter 2014.

\bibitem{Ronald2018}
R.~van Loon, ``{Machine Learning Explained: Understanding Supervised,
  Unsupervised, and Reinforcement Learning},''
  https://www.datasciencecentral.com/profiles/blogs/machine-learning-explained-understanding-supervised-unsupervised,
  online; accessed 5th March 2018.

\bibitem{Yoosemi2017}
J.~Yoo and K.~H. Johansson, ``Semi-supervised learning for mobile robot
  localization using wireless signal strengths,'' in \emph{Proc. Int. Conf. on Indoor Positioning and Indoor Navigation (IPIN)}, Sept 2017, pp. 1--8.

\bibitem{Shah2007distributed}
K.~Shah and M.~Kumar, ``Distributed independent reinforcement learning (DIRL)
  approach to resource management in wireless sensor networks,'' in \emph{IEEE
  Int. Conf. Mobile Adhoc and Sensor Systems}, Oct. 2007, pp. 1--9.

\bibitem{ParkIEEE2016AI}
Y.~Chu, \emph{et al.}, ``Application of reinforcement learning to medium access
  control for wireless sensor networks,'' \emph{Engineering Applications of
  Artificial Intelligence}, pp. 23--32, Nov. 2015.

\bibitem{Tijsma2016comparing}
A.~D. Tijsma, M.~M. Drugan, and M.~A. Wiering, ``Comparing exploration
  strategies for Q-learning in random stochastic mazes,'' in \emph{IEEE
  Symp. Series on Computational Intelligence (SSCI)}, Dec. 2016, pp. 1--8.

\bibitem{Poole2017AI}
D. L. Pooole and Alan K. Mackworth, ``Artificial Intelligence: Foundations of Computational Agents,'' in \emph{Cambridge University Press}, Second Edition, 2017.

\bibitem{Limulti2011}
C.~g.~Li, \emph{et al.}, ``Multi-intersections traffic
  signal intelligent control using collaborative Q-learning algorithm,'' in
  \emph{Proc. Seventh Int. Conf. on Natural Computation}, vol.~1, July 2011, pp. 185--188.

\bibitem{Rosyadi2016intelligent}
A.~R. Rosyadi, T.~A.~B. Wirayuda, and S.~Al-Faraby, ``Intelligent traffic light
  control using collaborative Q-learning algorithms,'' in \emph{Proc. 4th
  Int. Conf. Info. and Commun. Technol.}, May 2016, pp. 1--6.

\bibitem{Lamini2015oct}
C.~Lamini, Y.~Fathi, and S.~Benhlima, ``Collaborative Q-learning path planning
  for autonomous robots based on holonic multi-agent system,'' in \emph{Proc.
  10th Int. Conf. on Intelligent Systems: Theories and Applications}, Oct. 2015, pp. 1--6.

\bibitem{Kartransactions2013}
S.~Kar, J.~M.~F. Moura, and H.~V. Poor, ``QD-learning: A collaborative
  distributed strategy for multi-agent reinforcement learning through
  consensus+innovations,'' \emph{IEEE Trans. Signal Process.},
  vol.~61, no.~7, pp. 1848--1862, April 2013.

\bibitem{Ni2014multiagent}
J.~Ni, M.~Liu, L.~Ren, and S.~X. Yang, ``A multiagent Q-learning-based optimal
  allocation approach for urban water resource management system,'' \emph{IEEE
  Trans. Automation Science and Engg.}, vol.~11, no.~1, pp. 204--214, Jan. 2014.

\bibitem{HuNash2003}
J.~Hu and M.~P. Wellman, ``NASH Q-learning for general-sum stochastic games,''
  \emph{J. Mach. Learning Res.}, vol.~4, no.~6, pp. 1039-1069, 2003.

\bibitem{Choi2011automatic}
S.~Choi, \emph{et al.}, ``Automatic configuration of random access channel parameters in LTE systems,'' in
  \emph{Proc. IFIP Wireless Days (WD)}, Oct. 2011, pp. 1--6.

\bibitem{Rahimiyan2010adaptive}
M.~Rahimiyan and H.~R. Mashhadi, ``An adaptive $Q$-learning algorithm developed
  for agent-based computational modeling of electricity market,'' \emph{IEEE
  Trans. Systems, Man, and Cybernetics, Part C (Applications and Reviews)}, vol.~40, no.~5, pp. 547--556, Sept. 2010.

\bibitem{Hwang2004COOPERATIVE}
K.-S. Hwang, S.-W. Tan, and C.-C. Chen, ``Cooperative strategy based on
  adaptive Q-learning for robot soccer systems,'' \emph{IEEE Trans. Fuzzy Systems}, vol.~12, no.~4, pp. 569--576, Aug. 2004.

\bibitem{Glorennec1997fuzzy}
P.~Y. Glorennec and L.~Jouffe, ``Fuzzy Q-learning,'' in \emph{Proc. Int. Fuzzy Systems Conf.}, vol.~2, Jul 1997, pp. 659--662.

\bibitem{Jamshidi2015self}
P.~Jamshidi, \emph{et al.},
  ``Self-learning cloud controllers: Fuzzy Q-learning for knowledge
  evolution,'' in \emph{Proc. Int. Conf. Cloud and Autonomic Computing}, Sept 2015, pp. 208--211.

\bibitem{IslamcooperativeQ}
M.~N. ul~Islam and A.~Mitschele-Thiel, ``Cooperative fuzzy Q-learning for
  self-organized coverage and capacity optimization,'' in \emph{Proc. IEEE 23rd
  Int. Symp. (PIMRC)}, Sept. 2012, pp. 1406--1411.

\bibitem{Maedafuzzysystems}
Y.~Maeda, ``Modified Q-learning method with fuzzy state division and adaptive
  rewards,'' in \emph{Proc. IEEE Int. Conf. Fuzzy Systems}, vol.~2, 2002, pp. 1556--1561.

\bibitem{Pervez2017fuzzy}
F.~Pervez, \emph{et al.}, ``Fuzzy Q-learning-based user-centric backhaul-aware user cell association scheme,''
  in \emph{Proc. 13th Int. Wireless Commun. and Mobile Computing Conf.}, June 2017, pp. 1840--1845.

\bibitem{Mendil2016PIMRC}
M.~Mendil, \emph{et al.}, ``Fuzzy Q-learning based energy management of small cells powered by the smart
  grid,'' in \emph{Proc. IEEE 27th Annual Int. Symp. PIMRC}, Sept. 2016, pp. 1--6.

\bibitem{Wudynamic2015}
J.~Wu, J.~Liu, Z.~Huang, and S.~Zheng, ``Dynamic fuzzy Q-learning for handover
  parameters optimization in 5G multi-tier networks,'' in \emph{Proc.
  Int. Conf.  Wireless Commun. Signal Process. (WCSP)}, Oct 2015, pp. 1--5.

\bibitem{Wu2013flexible}
C.~Wu, S.~Ohzahata, and T.~Kato, ``Flexible, portable, and practicable solution
  for routing in vanets: A fuzzy constraint Q-learning approach,'' \emph{IEEE
Trans. Veh. Technol.}, vol.~62, no.~9, pp. 4251--4263, Nov. 2013.

\bibitem{Dengcombining2017}
Z.~Deng, \emph{et al.}, ``Combining
  model-based Q-learning with structural knowledge transfer for robot skill
  learning,'' \emph{IEEE Trans. Cognitive and Developmental Systems}, pp. 1--1, 2017.

\bibitem{Sharma2017model}
A.~Sharma, \emph{et al.}, ``Model based path
  planning using Q-learning,'' in \emph{Proc. IEEE Int. Conf. Industrial Technology (ICIT)},
  March 2017, pp. 837--842.

\bibitem{El-Alfy2001modelbased}
E.~S. El-Alfy, Y.-D. Yao, and H.~Heffes, ``A model-based Q-learning scheme for
  wireless channel allocation with prioritized handoff,'' in \emph{Proc. IEEE
  GLOBECOM}, vol.~6, 2001, pp. 3668--3672.

\bibitem{Caogroupauthenticcation}
J. Cao, M. Ma and H. Li, ``A group-based authentication and key agreement for MTC in LTE networks,''
in {Proc. IEEE Global Communications Conference (GLOBECOM)}, Anaheim, CA, 2012, pp. 1017-1022.

\bibitem{Sabin2018ICC}
S. Bhandari, S. K. Sharma and X. Wang, ``Device Grouping for Fast and Efficient Channel Access in IEEE 802.11ah based IoT Networks,'' in \emph{Proc. IEEE Int. Conf. Commun. (ICC) Workshops}, June 2018.

\bibitem{Liota2018deep}
H.~Li, K.~Ota, and M.~Dong, ``Learning IoT in edge: Deep learning for the
  internet of things with edge computing,'' \emph{IEEE Network}, vol.~32,
  no.~1, pp. 96--101, Jan. 2018.

\bibitem{SongAItowards}
M. Song, \emph{et al.}, ``In-Situ AI: Towards Autonomous and Incremental Deep Learning for IoT Systems,''
in \emph{2018 Proc. IEEE Int. Symp. High Performance Computer Architecture (HPCA)}, Vienna, Feb. 2018, pp. 92-103.

\bibitem{Venkataramani2016}
S.~Venkataramani, K.~Roy, and A.~Raghunathan, ``Efficient embedded learning for
  IoT devices,'' in \emph{Proc. 21st Asia and South Pacific Design Automation
  Conference (ASP-DAC)}, Jan. 2016, pp. 308--311.

\end{thebibliography}

\vspace{- 5 pt}

\end{document}